\pdfoutput=1

\documentclass[a4paper,11pt]{article}
\usepackage{jheppub}
\usepackage[section]{placeins}
\usepackage{bm}

\usepackage{float}
\usepackage{subfig}
\usepackage{cancel}
\usepackage{lscape}

\usepackage{color}
\usepackage[usenames,dvipsnames,svgnames,table]{xcolor}

\usepackage{amsmath}
\usepackage{amssymb}
\usepackage{graphicx}
\usepackage{slashed}
\usepackage{soul}
\usepackage{multirow}

\def\lsim{\mathrel{\raise.3ex\hbox{$<$\kern-.75em\lower1ex\hbox{$\sim$}}}}
\def\gsim{\mathrel{\raise.3ex\hbox{$>$\kern-.75em\lower1ex\hbox{$\sim$}}}}

\newcommand{\bea}{\begin{eqnarray}}
\newcommand{\eea}{\end{eqnarray}}

\newcommand{\be}{\begin{equation}}
\newcommand{\ee}{\end{equation}}

\title{Benchmark Suggestions for Resonant Double Higgs Production at the LHC for Extended Higgs Sectors}
\author[a,b]{Sebastian~Baum,}
\author[c]{Nausheen~R.~Shah}

\affiliation[a]{The Oskar Klein Centre for Cosmoparticle Physics, Department of Physics, Stockholm University, Alba Nova, 10691 Stockholm, Sweden}
\affiliation[b]{Nordita, KTH Royal Institute of Technology and Stockholm University, Roslagstullsbacken 23, 10691 Stockholm, Sweden}
\affiliation[c]{Department of Physics \& Astronomy, Wayne State University, Detroit, MI 48201, USA}

\emailAdd{sbaum@fysik.su.se}
\emailAdd{nausheen.shah@wayne.edu}

\preprint{NORDITA-2019-037
\\\phantom{0} \hfill WSU-HEP-1903}

\abstract{
In this note we present benchmark scenarios for resonant double Higgs production at the $13\,$TeV LHC in the 2HDM+S model and the $Z_3$ Next-to-Minimal Supersymmetric Standard Model~(NMSSM), which may be accessible with 300\,fb$^{-1}$ of data. The NMSSM Higgs sector can be mapped onto the 2HDM+S. We show benchmark points and relevant parameter planes in the 2HDM+S for three sets of signatures: ($A\to h_{125} a, ~H\to h_{125} h)$, ($A\to a h, H\to hh, H\to aa)$, and ($H\to h_{125}h_{125}$). The first two signatures are optimized in what we call {\it $Z$-Phobic} scenarios where $H/A$ decays into final states with $Z$ bosons are suppressed. The last signature, $h_{125}$ pair production, is directly proportional to the misalignment of $h_{125}$ with the interaction states sharing the couplings of a SM Higgs boson, and hence is presented in the {\it Max Misalignment} scenario. We also present two NMSSM benchmark points for the ($A\to h_{125} a, ~H\to h_{125} h)$ signatures. The benchmark scenarios presented here are based on Refs.~\cite{Baum:2018zhf, Baum:2019uzg}.
}

\begin{document}

\maketitle

\section{Introduction} \label{sec:intro}
In this note we present benchmark scenarios for resonant double Higgs production at the 13\,TeV LHC in the 2HDM+S and the Next-to-Minimal Supersymmetric Standard Model (NMSSM). We refer the reader to Ref.~\cite{Baum:2018zhf} for a detailed description of the 2HDM+S, in Sec.~\ref{sec:general} we give only a brief summary of the most important aspects of the 2HDM+S before presenting benchmark scenarios for resonant double Higgs production. Some recent work on resonant pair production of SM-like Higgs bosons in 2HDMs and the pMSSM can also be be found in Refs.~\cite{Babu:2018uik,Adhikary:2018ise}. 

In Sec.~\ref{sec:NMSSM} we present benchmark scenarios for double Higgs production in the $Z_3$ NMSSM, previously published in Ref.~\cite{Baum:2019uzg}. The 2HDM+S constitutes a generalized version of the NMSSM's Higgs sector, see Ref.~\cite{Baum:2018zhf} for a mapping of both the general and the $Z_3$ NMSSM to the 2HDM+S. Additional benchmark scenarios for double Higgs production in the NMSSM have e.g. been published in Ref.~\cite{Basler:2018dac}.

\section{The 2HDM+S} \label{sec:general}
We consider an extension of the Standard Model (SM) with a Higgs sector comprised of two $SU(2)$-doublets and a complex gauge singlet scalar, dubbed the 2HDM+S. Such an extended Higgs sector is of high interest since it constitutes a generalized version of the Next-to-Minimal Supersymmetric Standard Model's~(NMSSM's) Higgs Sector. For a detailed description of the 2HDM+S, its phenomenology, and the mapping to both the general and the $Z_3$ NMSSM, see Ref.~\cite{Baum:2018zhf}. We will only reproduce the portion of the analysis from Refs.~\cite{Baum:2018zhf,Baum:2019uzg} relevant for the description of the benchmark scenarios presented here. 

We will assume a type-II Yukawa structure for the coupling of the Higgs bosons to SM fermions, as is expected in generic supersymmetric theories. Note that for the purposes of this note, we consider the low $\tan\beta$ regime (in particular, we present benchmarks for fixed $\tan\beta = 2.5$), where a different choice for the Yukawa structure would yield similar numerical results. 

The scalar sector of the 2HDM+S is comprised of 6 physical states,
\begin{equation}
   \{h_{125}\;, \quad h\;, \quad H\}\;, \qquad \{a\;, \quad A\}\;, \qquad H^\pm \;.
\end{equation}
The first three are the neutral CP-even states, $a$ and $A$ are the two CP-odd neutral states, and $H^\pm$ is the charged Higgs. Of the CP-even states, $h_{125}$ is identified with the SM-like 125\,GeV state observed at the LHC, and the remaining states are order by their masses, $m_h < m_H$. Likewise, $m_a < m_A$.

\begin{figure}
   \centering
   \includegraphics[width=0.5\linewidth]{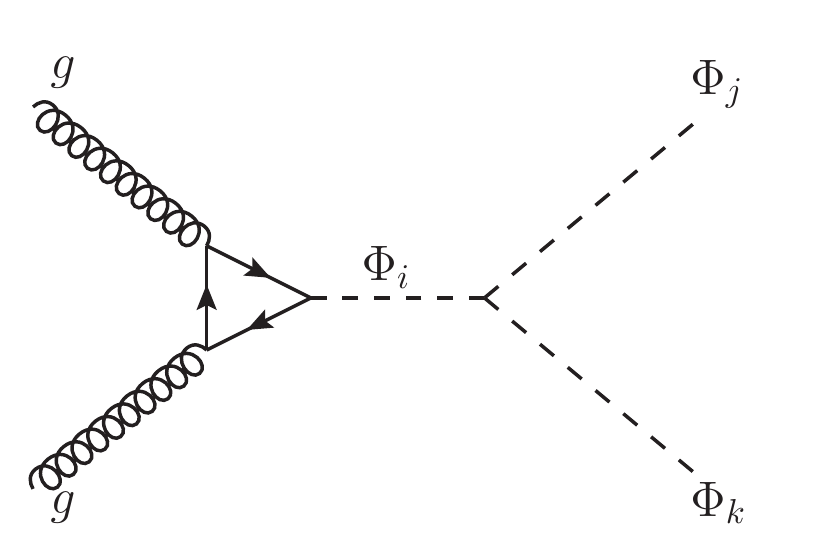}
   \caption{Illustration of resonant double Higgs production in the 2HDM+S. The $\Phi_i$ stand for any of the neutral Higgs bosons. Note that CP conservation requires either all three states to be CP-even (e.g. $H \to h_{125} h$) or two of the states to be CP-odd and the third to be CP-even (e.g. $A \to h_{125} a$).}
   \label{fig:diagram}
\end{figure}

It is useful to write the neutral Higgs states in terms of the interaction states of the extended Higgs basis~\cite{Georgi:1978ri, Donoghue:1978cj, gunion2008higgs, Lavoura:1994fv, Botella:1994cs, Branco99, Gunion:2002zf,Carena:2015moc}
\begin{equation}
   h_i = S_{h_i}^{\rm SM} H^{\rm SM} + S_{h_i}^{\rm NSM} + S_{h_i}^{\rm S}\;,
\end{equation}
for the CP-even states $h_i = \{ h_{125}, h, H\}$, and
\begin{equation}
   a_i = P_{a_i}^{\rm NSM} A^{\rm NSM} + P_{a_i}^{\rm S} A^{\rm S} \;,
\end{equation}
for the CP-odd states $a_i = \{ a, A \}$. In a type-II Yukawa structure, the coupling of the Higgs basis eigenstates to pairs of SM particles can be written as
\begin{align}
   H^{\rm SM} \left( {\rm up}, {\rm down}, {\rm VV} \right) &= \left( g_{\rm SM}, \; g_{\rm SM}, \; g_{\rm SM} \right), 
   \\ H^{\rm NSM} \left( {\rm up}, {\rm down}, {\rm VV} \right) &= \left( - g_{\rm SM} / \tan\beta, \; g_{\rm SM} \tan\beta, \; 0 \right),
   \\ H^{\rm S} \left( {\rm up}, {\rm down}, {\rm VV} \right) &= \left( 0, \; 0, \; 0 \right),
   \\ A^{\rm NSM} \left( {\rm up}, {\rm down}, {\rm VV} \right) &= \left( g_{\rm SM} / \tan\beta, \; g_{\rm SM} \tan\beta, \; 0 \right), 
   \\ A^{\rm S} \left( {\rm up}, {\rm down}, {\rm VV} \right) &= \left( 0, \; 0, \; 0 \right), 
\end{align}
where ``up'' (``down'') stands for pairs of up-type (down-type) SM fermions, ``VV'' for pairs of massive vector bosons, and $g_{\rm SM}$ is the coupling of a SM Higgs boson to such particles.

The parameter space of the 2HDM+S can be intuitively described in terms of the physical masses
\begin{equation}
   m_{h_{125}}\;, \quad m_H\;, \quad m_h\;, \quad m_A\;, \quad m_a\;, \quad m_{H^\pm}\;,
\end{equation}
the mixing angles
\begin{equation}
   S_{h_{125}}^{\rm NSM}\;, \quad S_{h_{125}}^{\rm S}\;, \quad S_H^{\rm S}\;, \quad P_A^{\rm S} \;,
\end{equation}
the vacuum expectation values (vevs) of the 2 doublets and the singlet, which we parameterize by 
\begin{equation} 
   v\;, \quad \tan\beta\;, \quad v_S\;,
\end{equation}
10 independent trilinear couplings between the interaction states of the extended Higgs basis 
\begin{equation}\label{eq:trilin} \begin{split}
   &\{ g_{H^{\rm SM} H^{\rm NSM} H^{\rm NSM}}, \quad g_{H^{\rm SM} H^{\rm S} H^{\rm S}}, \quad g_{H^{\rm SM} A^{\rm S} A^{\rm S}} \}, \\
   &\{ g_{H^{\rm NSM} H^{\rm NSM} H^{\rm NSM}}, \quad g_{H^{\rm NSM} H^{\rm NSM} H^{\rm S}}, \quad g_{H^{\rm NSM} H^{\rm S} H^{\rm S}}, \quad g_{H^{\rm NSM} A^{\rm S} A^{\rm S}} \}, \\
   &\{ g_{H^{\rm S} H^{\rm S} H^{\rm S}}, \quad g_{H^{\rm S} A^{\rm NSM} A^{\rm S}}, \quad g_{H^{\rm S} A^{\rm S} A^{\rm S}} \} \;.
\end{split} \end{equation}
and 4 independent quartic couplings
\begin{equation}\label{eq:quartic}
   \{ \lambda_{H^{\rm NSM} H^{\rm NSM} H^{\rm S} H^{\rm S}}, \quad \lambda_{H^{\rm NSM} H^{\rm NSM} A^{\rm S} A^{\rm S}}, \quad \lambda_{H^{\rm S} H^{\rm S} A^{\rm S} A^{\rm S}} , \quad \lambda_{A^{\rm S} A^{\rm S} A^{\rm S} A^{\rm S}} \} .
\end{equation}
All couplings, in particular the remaining trilinear couplings which play an important role for resonant double Higgs production, are fixed in terms of the above parameters. 

Let us comment on the sign conventions we use for the mixing angles. The conventions are particularly relevant when reconstructing the full mixing matrix of the CP-even states from our input parameters $\{ S_{h_{125}}^{\rm NSM}, S_{h_{125}}^{\rm S}, S_H^{\rm S}$\}. Assuming no CP violation, the mixing matrix can be written as
\begin{equation}
   \begin{pmatrix} h_{125} \\ H \\ h \end{pmatrix} = \begin{pmatrix} c_{12} c_{13} & s_{12} c_{13} & s_{13} \\ -s_{12} c_{23} - c_{12} s_{23} s_{13} & c_{12} c_{23} - s_{12} s_{23} s_{13} & s_{23} c_{13} \\ s_{12} s_{23} - c_{12} c_{23} s_{13} & - c_{12} s_{23} - s_{12} c_{23} s_{13} & c_{23} c_{13} \end{pmatrix} \begin{pmatrix} H^{\rm SM} \\ H^{\rm NSM} \\ H^{\rm S} \end{pmatrix},
\end{equation} 
where $s_{12} \equiv \sin\theta_{12}$, etc. In order to obtain all of the entries of the mixing matrix, we use the following identifications:
\begin{itemize}
   \item $s_{13} = S_{h_{125}}^{\rm S}$\;,
   \item $c_{13} = \sqrt{1-s_{13}^2}$ \;,
   \item $s_{12} = S_{h_{125}}^{\rm NSM}/c_{13}$\;,
   \item $c_{12} = \sqrt{1 - s_{12}^2}$\;,
   \item $s_{23} = S_H^{\rm S}/c_{13}$\;,
   \item $c_{23} = \sqrt{1-s_{23}^2}$\;.
\end{itemize}
Thus, we choose sgn$(c_{ij}) \geq 0$ everywhere. However, since the $s_{ij}$ can take both signs (depending on the signs of the inputs $\{ S_{h_{125}}^{\rm NSM}, S_{h_{125}}^{\rm S}, S_H^{\rm S}$\}), we can explore all possible configurations of the mixing angles by varying the signs of the input values. 

The SM-like nature of the observed 125\,GeV state is achieved in the 2HDM+S by enforcing approximate alignment
\begin{equation}
   \{ (S_{h_{125}}^{\rm NSM})^2, (S_{h_{125}}^{\rm S})^2 \} \ll 1 \;,
\end{equation}
and fixing $m_{h_{\rm 125}} \approx 125\,$GeV. Of particular interest for LHC phenomenology is the alignment-without-decoupling limit, where $A$ and $H$ remain relatively light. Then, the structure of the potential enforces the couplings between pairs of SM-like states and a non SM-like Higgs, $g_{h_{125} h_{125} H}$, to vanish, since in this limit~\cite{Baum:2018zhf}
\begin{equation}
   \{ g_{H^{\rm SM} H^{\rm SM} H^{\rm NSM}}\;, g_{H^{\rm SM} H^{\rm SM} H^{\rm S}} \} \to 0 \;
\end{equation}

For the purposes of this note, we are interested in resonant double Higgs production channels such as ($gg \to \Phi_i \to \Phi_j + \Phi_k$), cf. Fig.~\ref{fig:diagram}, where the $\Phi_i$ stand for any of the neutral Higgs bosons. Thus, we consider the charged Higgs to be sufficiently heavy such that it will not affect the decays of the neutral states, e.g. $m_{H^\pm} > 1\,$TeV. Further, we note that ($\Phi_i \to Z \Phi_j$) decays give rise to search channels for the 2HDM+S complimentary to ($\Phi_i \to \Phi_j + \Phi_k$) channels, cf. Ref.~\cite{Baum:2018zhf}. CP-conservation allows decays into pairs of Higgs bosons only if they are of the type ($h_i \to h_j h_k$), ($h_i \to a_j a_k$), or ($a_i \to h_j a_k$), while decays into a $Z$ and a Higgs boson must be of the type ($h_i \to Z a_j$) or ($a_i \to Z h_j$). Further, decays into pairs of either $Z$ or $W$ gauge bosons are only allowed for the CP-even states, ($h_i \to ZZ/W^+W^-$). 
The corresponding partial widths are given by 
\begin{align} \label{eq:width_hh}
   \Gamma\left( \Phi_i \to \Phi_j \Phi_k \right) &= \frac{g_{\Phi_i \Phi_j \Phi_k}^2}{16 \pi m_{\Phi_i}} \left(\frac{1}{1+\delta_{jk}}\right) \sqrt{1- 2\frac{m_{\Phi_j}^2 + m_{\Phi_k}^2}{m_{\Phi_i}^2} + \frac{\left(m_{\Phi_j}^2 - m_{\Phi_k}^2\right)^2}{m_{\Phi_i}^4}} \;, \\
	\Gamma(\Phi_i \to Z \Phi_j) &= \frac{\left(C_{\Phi_i}^{\rm NSM} C_{\Phi_j}^{\rm NSM}\right)^2}{32 \pi} \frac{m_Z^2}{m_{\Phi_i} v^2 } \left[ \frac{ \left( m_{\Phi_i}^2 - m_{\Phi_j}^2\right)^2 }{m_Z^2} - 2\left(m_{\Phi_i}^2 + m_{\Phi_j}^2 \right) + m_Z^2 \right] \nonumber
	\\ & \qquad \times \sqrt{1 - 2 \frac{m_{\Phi_j}^2 + m_Z^2}{m_{\Phi_i}^2} + \frac{\left( m_{\Phi_j}^2 - m_Z^2 \right)^2}{m_{\Phi_i}^4}} \;, \label{eq:width_Zh} \\
	\Gamma(h_i \to Z Z) &= \frac{\left(S_{h_i}^{\rm SM}\right)^2 m_Z^4}{16\pi m_{h_i} v^2} \left( 3 - \frac{m_{h_i}^2}{m_Z^2} + \frac{m_{h_i}^4}{4 m_Z^4} \right) \sqrt{1-4 \frac{m_Z^2}{m_{h_i}^2}} \;,
	\\ \Gamma(h_i \to W^+ W^-) &= \frac{\left( S_{h_i}^{\rm SM}\right)^2 m_W^4}{8 \pi m_{h_i} v^2} \left( 3 - \frac{m_{h_i}^2}{m_W^2} + \frac{m_{h_i}^4}{4 m_W^4} \right) \sqrt{1-4 \frac{m_W^2}{m_{h_i}^2}} \;, \label{eq:width_WW}
\end{align}
where we use $C_\Phi^J$ to refer to the mixing angles for both the CP-even and the CP-odd mass eigenstates $\Phi$ with mass $m_\Phi$; for example, $C_\Phi^{\rm NSM} = S_{h_i}^{\rm NSM}$ or $P_{a_i}^{\rm NSM}$ depending on context. For  partial widths into pairs of SM fermions, see Ref.~\cite{Baum:2018zhf}. The trilinear couplings between the Higgs mass eigenstates are given by
\begin{equation} \begin{split} \label{eq:tricoup_mass}
	g_{h_i h_j h_k} &= \sum_{H^l} \sum_{H^m} \sum_{H^n} S_{h_i}^{H^l} S_{h_j}^{H^m} S_{h_k}^{H^n} g_{H^l H^m H^n} \;, \\
	g_{h_i a_j a_k} &= \sum_{H^l} \sum_{A^m} \sum_{A^n} S_{h_i}^{H^l} P_{a_j}^{A^m} P_{a_k}^{A^n} g_{H^l A^m A^n}\;,
\end{split} \end{equation}
where the $g_{H^l H^m H^n}$ and $g_{H^l A^m A^n}$ are the couplings between the interaction states of the Higgs basis. The exact expressions for these trilinear couplings can be found in App.~B of Ref.~\cite{Baum:2018zhf}. Note that all trilinear couplings involving at least a pair of $H^{\rm SM}$ states, e.g. $g_{H^{\rm SM} H^{\rm SM} H^{\rm NSM}}$, as well as the couplings $g_{H^{\rm SM} H^{\rm NSM} H^{\rm S}}$ and $g_{H^{\rm SM} A^{\rm NSM} H^{\rm S}}$ are proportional to the entries of the scalar mass matrices, which are in turn determined by the physical masses and the mixing angles. The scale of most of the remaining couplings, in particular the couplings $g_{H^{\rm NSM} H^{\rm S} H^{\rm S}}$, $g_{H^{\rm NSM} A^{\rm S} A^{\rm S}}$, and $g_{H^{\rm S} A^{\rm NSM} A^{\rm S}}$, which will become important in Sec.~\ref{sec:SS}, is set by $v$. 

In the limit of perfect alignment, the relevant couplings for decays into one SM-like and one non SM-like state, i.e. ($H \to h_{125} h$) and ($A \to h_{125} a$), can be written as
\begin{align}\label{eq:freecoup_combination}
   g_{h_{125} H h} &= \frac{S_H^{\rm NSM} S_H^{\rm S}}{\sqrt{2} v} \left\{ \left[ 1 - 2(S_H^{\rm S})^2 \right] \left( m_H^2 - m_h^2 \right) + \sqrt{2} v~\widetilde{g}_H \right\} \;,\\
   g_{h_{125} A a} &= \frac{P_A^{\rm NSM} P_A^{\rm S}}{\sqrt{2}v} \left\{ \left[ 1 - 2(P_A^{\rm S})^2 \right] \left( m_A^2 - m_a^2 \right) + \sqrt{2} v~\widetilde{g}_A \right\} \;,
\end{align}
where we have defined
\begin{align} \label{eq:gHeff}
 \widetilde{g}_H &\equiv \left( g_{H^{\rm SM} H^{\rm S} H^{\rm S}} - g_{H^{\rm SM} H^{\rm NSM} H^{\rm NSM}} \right) \;,\\
 \widetilde{g}_A &\equiv \left( g_{H^{\rm SM} A^{\rm S} A^{\rm S}} - g_{H^{\rm SM} A^{\rm NSM} A^{\rm NSM}} \right)\;. \label{eq:gAeff}
\end{align}
The scale of the trilinear couplings entering $\tilde{g}_H$ and $\tilde{g}_A$ is set by $v$. Since ($H \to h_{125} h$) and ($A \to h_{125} a$) decays require sizable mass splittings between $H$ and $h$ or $A$ and $a$, respectively, we expect the contribution of the $\tilde{g}_H$ to $g_{h_{125}H h}$ and $\tilde{g}_A$ to $g_{h_{125} A a}$ to be sub-dominant, unless $(S_H^{\rm S})^2 \approx 0.5$ [$(P_A^{\rm S})^2 \approx 0.5$].

As mentioned above, decays into pairs of $h_{125}$'s (i.e. $H \to h_{125} h_{125}$) are suppressed by alignment. To leading order in misalignment, and assuming $S_{h_{125}}^{\rm NSM}\approx -S_H^{\rm SM}$, the relevant coupling can be approximated as
\begin{equation} \begin{split}
   g_{h_{125} h_{125} H} &\sim (S_{h_{125}}^{\rm SM})^2 S_H^{\rm NSM} g_{H^{\rm SM} H^{\rm SM} H^{\rm NSM}} + (S_{h_{125}}^{\rm SM})^2 S_H^{\rm S} g_{H^{\rm SM} H^{\rm SM} H^{\rm S}} \\
   & \quad - S_{h_{125}}^{\rm SM} S_{h_{125}}^{\rm NSM} \left[ S_{h_{125}}^{\rm SM} g_{H^{\rm SM} H^{\rm SM} H^{\rm SM}} - 2 S_H^{\rm NSM} g_{H^{\rm SM} H^{\rm NSM} H^{\rm NSM}} \right] \\
   & \quad + 2 S_{h_{125}}^{\rm SM} \left[ \left( S_{h_{125}}^{\rm NSM} S_H^{\rm S} + S_{h_{125}}^{\rm S} S_H^{\rm NSM} \right) g_{H^{\rm SM} H^{\rm NSM} H^{\rm S}} + S_{h_{125}}^{\rm S} S_H^{\rm S} g_{H^{\rm SM} H^{\rm S} H^{\rm S}} \right].
\end{split} \end{equation}
The couplings in the first line, $g_{H^{\rm SM} H^{\rm SM} H^{\rm NSM}}$ and $g_{H^{\rm SM} H^{\rm SM} H^{\rm S}}$, are proportional to the entries of the mass matrix corresponding to $H^{\rm SM} - H^{\rm NSM}$ and $H^{\rm SM} - H^{\rm S}$ mixing, respectively. Thus, they vanish in the alignment limit. The contributions in the second and third line are proportional to trilinear couplings which are not suppressed by alignment, but their respective contribution to $g_{h_{125} h_{125} H}$ all have coefficients of either $S_{h_{125}}^{\rm NSM}$ or $S_{h_{125}}^{\rm S}$, and are thus likewise suppressed by alignment. Hence, as noted above, $g_{h_{125} h_{125} H} = 0$ for perfect alignment.

Note that rather small departures from alignment can have considerable impact on the gluon fusion production cross section for $H$. At low to moderate values of $\tan\beta$, the gluon fusion production cross section is proportional to the coupling of $H$ to pairs of top quarks, $g_{H t\bar{t}} \propto (-S_H^{\rm NSM}/\tan\beta+S_H^{\rm SM})$. Thus, even relatively small variations of the value of $S_H^{\rm SM}$ can lead to large changes in $\sigma(ggH)$. 

The mixing angle between the non SM-like states, $S_H^{\rm S}$ and $P_A^{\rm S}$ are not experimentally constrained by the observed phenomenology of $h_{125}$. However, large values of $(S_H^{\rm S})^2$ and $(P_A^{\rm S})^2$ suppress the production cross sections of $H$ and $A$, respectively, and are thus challenging to probe at the LHC.

The trilinear couplings relevant for the decays ($H \to hh$), ($H \to aa$), and ($A \to ha$) are governed by free trilinear couplings between the Higgs basis interactions states and are thus not related to entries in the mass matrices. In our parameterization of the 2HDM+S, they are free parameters. 

In summary, the branching ratios of $(H \to h_{125} h_{125}/ZZ/W^+W^-)$ and $(A \to Z h_{125})$ decays are suppressed by the SM-like nature of $h_{125}$. Decays into a SM-like mode and a light non SM-like mode ($H \to h_{125} h$, $A \to h_{125} a$), or into two light non SM-like states ($H \to hh$, $H \to aa$, $H \to ha$) are not suppressed and can potentially have large branching ratios in the 2HDM+S. The detailed discussion of the various couplings and branching ratios can be found in Ref.~\cite{Baum:2018zhf}. In the following we will present benchmark scenarios optimizing all three types of decays. 

As discussed above, we consider the low to moderate $\tan\beta$ regime here, in particular we fix $\tan\beta = 2.5$ for all benchmark scenarios. The dominant production mechanism for the Higgs bosons at the LHC is then gluon fusion, which we will focus on in the following. Note in particular that vector-boson-fusion production of the non SM-like Higgs bosons is suppressed by approximate alignment. This is because of the Higgs basis interaction eigenstates, only $H^{\rm SM}$ couples to pairs of massive vector bosons at tree level. Further, the benchmarks presented here are meant to portray only the prospects of exploring the 2HDM+S parameter space using double Higgs production. The regions displayed do not take into account bounds/projections for direct searches for additional Higgs bosons beyond $h_{125}$, such as $(gg \to \Phi_i\to \tau\tau)$ etc. 

Finally, we observe that while in 2HDM's $A$ and $H$ are approximately mass degenerate, their masses are a priori less tightly correlated in the 2HDM+S. For example, large mass splittings can be achieved via different mixing angles $S_H^{\rm S}$ and $P_A^{\rm S}$. Hence when showing the planes for the benchmark scenarios, we show the effect of varying both $m_A$ and $m_H$ independently as labeled on the plots.

Computations of the cross sections for the benchmark values for the 2HDM+S make use of gluon
fusion Higgs production cross sections and SM Higgs branching ratios taken from Ref.~\cite{deFlorian:2016spz}, which we rescale by the appropriate mixing angles and $\tan\beta$.\footnote{We use the NNLO+NNLL  YR4 BSM 13\,TeV cross sections~\cite{deFlorian:2016spz} as input, which we rescale by the appropriate mixing angles and $\tan\beta$. We have cross-checked the numerical results with an implementation of the 2HDM+S in \texttt{SusHi-1.7.0}~\cite{Harlander:2012pb,Harlander:2016hcx,Liebler:2015bka,Harlander:2002wh,Aglietti:2004nj,Bonciani:2010ms,Degrassi:2010eu,Degrassi:2011vq,Degrassi:2012vt,Harlander:2005rq,Chetyrkin:2000yt}. For CP-even states our rescaled cross sections agree with the \texttt{SusHi} results within a few percent, an error much smaller than the renormalization and matching scale uncertainties. For CP-odd states, we found differences as large as $\sim 10 - 20\,\%$. This is due to our rescaling using leading-order loop factors for the conversion to the CP-odd gluon fusion cross section, cf. Ref.~\cite{Baum:2018zhf}.} For the NMSSM benchmarks, particle spectra, reduced couplings, and branching ratios are computed with \texttt{NMSSMTools}~\cite{NMSSMTools,Ellwanger:2004xm,Ellwanger:2005dv,Das:2011dg, Muhlleitner:2003vg}.

\subsection{$Z$-Phobic Scenarios:}
In this sections, we propose benchmark scenarios for ($gg \to \Phi_i \to \Phi_j \Phi_k$) production at the LHC. By choosing relatively small mixing angles between the non SM-like states, $\{ (S_H^{\rm S})^2, (P_A^{\rm S})^2\} \ll 1$, we consider a region of parameter space where ($\Phi_i \to Z \Phi_j$) decays are suppressed and hence $(\Phi_i \to \Phi_j \Phi_k)$ is enhanced. We dub these the {\it $Z$-Phobic scenarios}.

Further, we assume perfect alignment, forbidding decays such as ($H \to h_{125} h_{125}$) and ($A \to Z h_{125}$). Note that deviations from perfect alignment to a degree allowed by current data would have only minor impact on the decays of the heavy Higgs bosons into non SM-like states, but as mentioned previously could impact the gluon fusion production cross section of $H$ in a relevant way. We will show production cross sections at the LHC as well as projections for the reach with 300\,fb$^{-1}$ of data in various planes of interest in the parameter space.

The projected sensitivity of the LHC for the decays $(H \to h_{125} h)$ and $(A \to h_{125} a)$ depends on the final state decay of $h/a$. Assuming perfect alignment, the decays of $h/a$ into SM fermions proceed via their NSM components, and for low values of $\tan\beta$ are close to what is expected of a SM-like Higgs of the same mass. However, if there are additional Majorana fermions $\chi_i$, e.g. dark matter candidates, which couple to the Higgs bosons via a coupling $g_{\Phi_i\chi_j\chi_k}$
\begin{equation}\label{eq:int_chichi}
	\mathcal{L} \supset \frac{g_{h_i\chi_j\chi_k}}{2(1 + \delta_{jk})} h_i \bar{\chi}_j \chi_k + \frac{g_{a_i\chi_j\chi_k}}{2(1 + \delta_{jk})} a_i \bar{\chi}_j \gamma_5 \chi_k \;,
\end{equation}
$a$ and $h$ could decay invisibly. The corresponding partial width is given by
\begin{equation} \label{eq:width_chichi}
	\Gamma(\Phi_i \to \chi_j \chi_k) = \left( \frac{2}{1+\delta_{ij}} \right) \frac{g_{\Phi_i\chi_j\chi_k}^2}{16\pi} m_{\Phi_i} \left[ 1- \frac{\left(m_{\chi_j}+m_{\chi_k}\right)^2}{m_{\Phi_i}^2} \right]^{(1+\gamma)} \left[ 1- \frac{\left(m_{\chi_j}-m_{\chi_k}\right)^2}{m_{\Phi_i}^2} \right]^{(1-\gamma)} \;,
\end{equation}
where $\gamma=1/2$ ($\gamma=-1/2$) for a CP-even (CP-odd) $\Phi_i$. Here, we will present two scenarios: Either, we choose $g_{h \chi_1 \chi_1} = g_{a \chi_1 \chi_1} = 0$, which can also be interpreted as a ``pure'' 2HDM+S without additional states $\chi_i$, or, we choose $g_{h \chi_1 \chi_1} = g_{a \chi_1 \chi_1} = 2.5 \approx \sqrt{2\pi}$ close to the largest values allowed by perturbativity, such that ($h \to \chi_1 \chi_1$) and ($a \to \chi_1 \chi_1$) will be the dominant decay mode of $h$ and $a$.\footnote{The couplings of the remaining mass eigenstates, $h_{125}$, $H$, and $A$, to additional fermions $\chi_i$ are set to zero throughout this note.} Note that for the purposes of this note, we treat the $g_{\Phi_i \chi_j \chi_k}$ as free parameters, see e.g. Ref.~\cite{Baum:2017enm} for a discussion of such couplings in the context of consistent dark matter phenomenology. 

In the first two columns of Table~\ref{tab:BP_h125S} we present benchmark scenarios for ($H \to h_{125} h$) and ($A \to h_{125} a$) resonant production. Depending on the couplings $g_{\Phi_i\chi_1\chi_1}$, these scenarios result in the daughter Higgs bosons decaying either visibly (2$b$) or invisibly ($\cancel{\it{E}}_{T}$). We set all the free trilinear couplings between the interaction states, listed in Eq.~\eqref{eq:trilin}, to zero, cf. the discussion below Eq.~\eqref{eq:freecoup_combination}. The third column corresponds to a scenario where we choose non-zero values for the free trilinear couplings, in particular $g_{H^{\rm NSM} H^{\rm S} H^{\rm S}}$, $g_{H^{\rm NSM} A^{\rm S} A^{\rm S}}$, and $g_{H^{\rm S} A^{\rm NSM} A^{\rm S}}$, which are most relevant for ($H \to hh$), ($H \to aa$), and ($A \to ha$) decays, respectively. For this latter benchmark, we do not include the decays of $h$ and $a$ in the results presented here, thus, there is no need to fix the couplings of $h$ and $a$ to $\chi_1$.

We have chosen heavy Higgs bosons with masses of $700\,$GeV, such that they are heavy enough to have evaded detection so far, but are light enough to have sizable production cross section at the 13\,TeV LHC. The singlet components for both of them, $(S_H^{\rm S})^2$ and $(P_A^{\rm S})^2$, are fixed to 0.1, such that the gluon fusion production cross section is not unduly suppressed and decays into $Z$ are disfavored. For the low values of $\tan\beta$ considered here, the dominant decay mode for the heavy Higgs bosons out of the decays into pairs of SM particles would be to $t\bar{t}$, which is challenging to probe at the LHC~\cite{Dicus:1994bm,Barcelo:2010bm,Barger:2011pu,Bai:2014fkl,Jung:2015gta,Craig:2015jba,Gori:2016zto,Carena:2016npr}. The masses of the mostly singlet-like $a$ and $h$ are set to 200\,GeV, light enough for the heavy Higgs bosons to readily decay to them. The mass of the additional dark matter candidate $\chi_1$ is set to 50\,GeV. The trilinear couplings are not expected to be relevant unless very large, as can be seen from Eq.~\eqref{eq:freecoup_combination}, hence we set them to 0 for the first two benchmarks. 

\begin{table}[h!]
   \centering
   \begin{tabular}{c||c|c|c}
      \hline\hline
      $Z$-Phobic & Visible~(2$b$) & Invisible~($\cancel{\it{E}}_{T}$) & Double Singlet \\
      \hline\hline
      $m_H$ [GeV] & \multicolumn{3}{c}{$700$} \\
      $m_A$ [GeV] & \multicolumn{3}{c}{$700$} \\
      $m_h$ [GeV] & \multicolumn{3}{c}{$200$} \\
      $m_a$ [GeV] & \multicolumn{3}{c}{$200$} \\
      $m_\chi$ [GeV] & \multicolumn{3}{c}{$50$} \\
      $\tan \beta$ &   \multicolumn{3}{c}{$2.5$} \\
      \hline
      $(S_H^{\rm S})^2$ & \multicolumn{3}{c}{$0.1$} \\
      $(P_A^{\rm S})^2$ & \multicolumn{3}{c}{$0.1$} \\
      $g_{\Phi_i\chi_1\chi_1}$ & $0$ & $2.5$ & --\\
      $g_{H^{\rm NSM} H^{\rm S} H^{\rm S}}$ [GeV] & \multicolumn{2}{c|}{$0$} & 174 \\
      $g_{H^{\rm NSM} A^{\rm S} A^{\rm S}}$ [GeV] & \multicolumn{2}{c|}{$0$} & 174 \\
      $g_{A^{\rm NSM} H^{\rm NSM} H^{\rm S}}$ [GeV] & \multicolumn{2}{c|}{$0$} & 174 \\
      \hline
      $\sigma(ggH)$ [pb] & \multicolumn{3}{c}{$0.13$} \\
      $\sigma(ggA)$ [pb] & \multicolumn{3}{c}{$0.19$} \\
      \hline
      BR($H \to h_{125} h$) & \multicolumn{2}{c|}{$0.30$} & $0.29$\\
      BR($H \to hh$) & \multicolumn{2}{c|}{$0$} & $0.0094$ \\
      BR($H \to aa$) & \multicolumn{2}{c|}{$0$} & $0.0094$ \\
      BR($H \to Z a$) & \multicolumn{2}{c|}{$0.45$} & $0.44$ \\
      \hline
      BR($A \to h_{125} a$) & \multicolumn{2}{c|}{$0.28$} & $0.27$\\
      BR($A \to h a$) & \multicolumn{2}{c|}{$0$} & $0.018$\\
      BR($A \to Z h$) & \multicolumn{2}{c|}{$0.42$} & $0.41$ \\
      \hline
      BR($h \to \chi_1\chi_1$) & $0$ & $1.00$ & -- \\
      BR($h \to b \bar{b}$) & $0.94$ & $0.00$ & -- \\
      \hline
      BR($a \to \chi_1\chi_1$) & 0 & $1.00$ & -- \\
      BR($a \to b \bar{b}$) & $0.94$ & $0.00$ & -- \\
      \hline\hline
   \end{tabular}
   \caption{$Z$-Phobic benchmark scenarios. Perfect alignment is assumed, and all free trilinear couplings not listed above, cf. Eq.~\eqref{eq:trilin}, are set to 0.}
   \label{tab:BP_h125S}
\end{table}

\subsubsection{$Z$-Phobic: $h_{125}$+Visible ($H\to h_{125} a$ and $A\to h_{125} h$)}
\label{sec:h125S_bb}

\begin{figure}[hp]
   \begin{centering}
      \includegraphics[width = 2.5in]{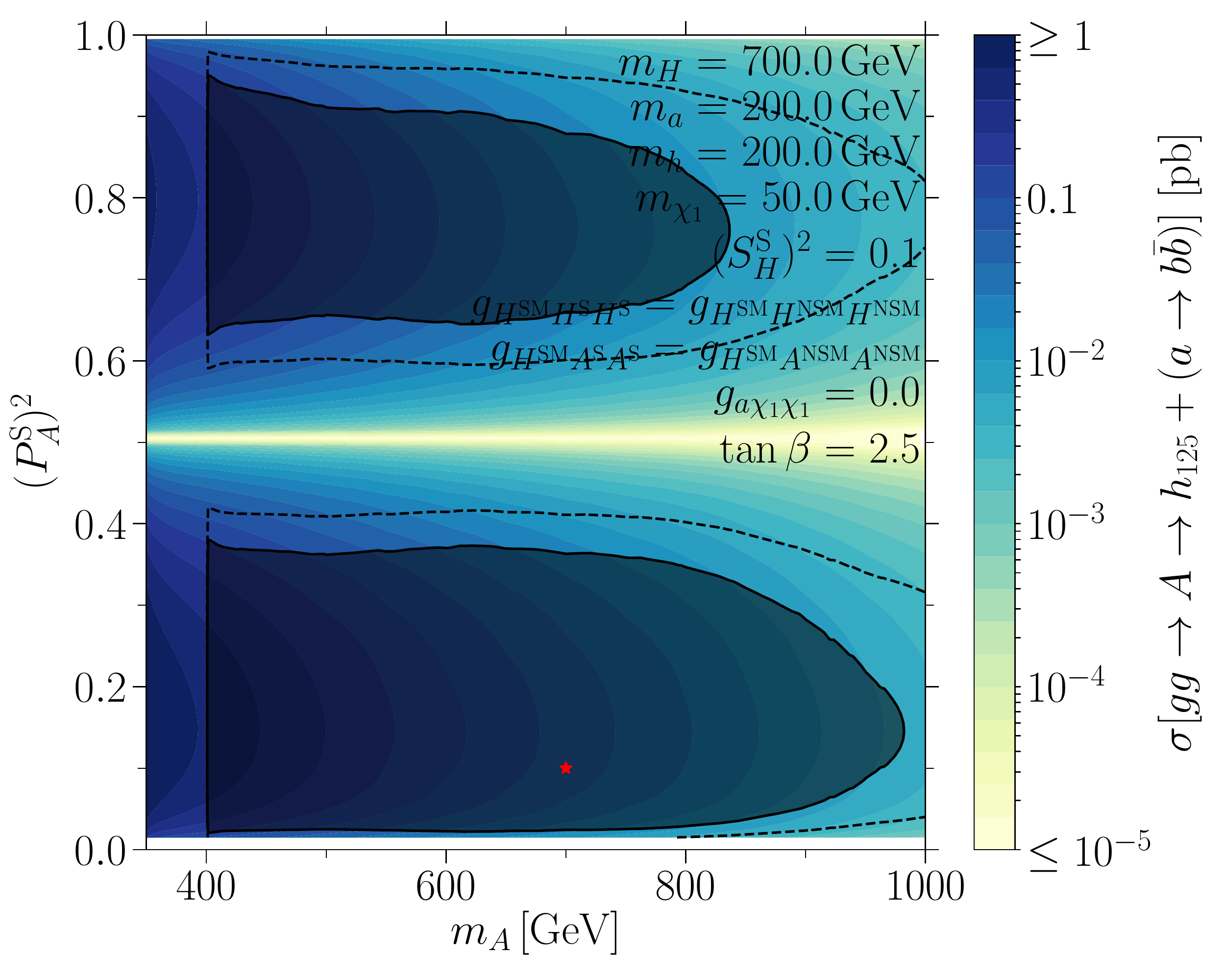}
      \hspace{.5in}
      \includegraphics[width = 2.5in]{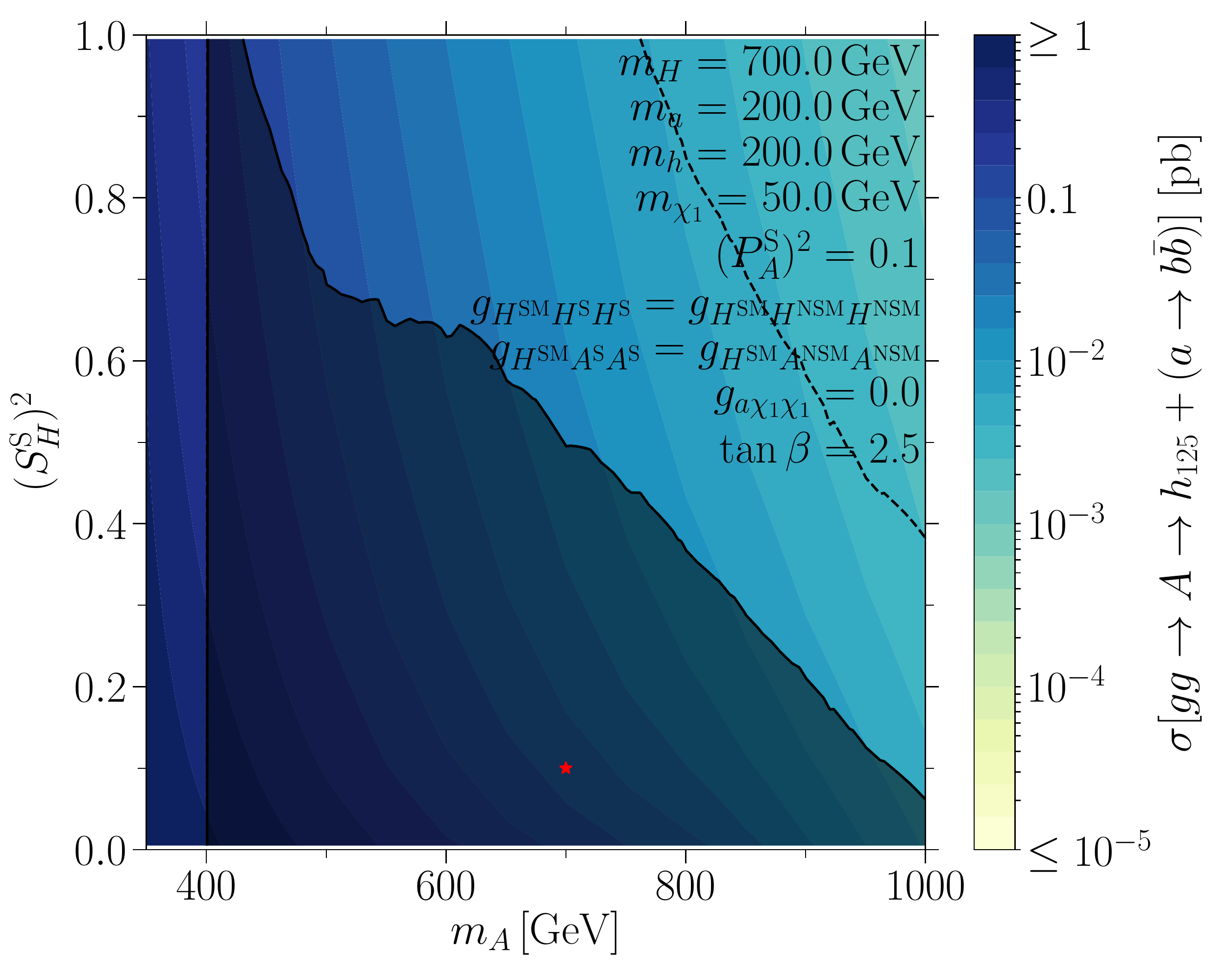}

      \includegraphics[width = 2.5in]{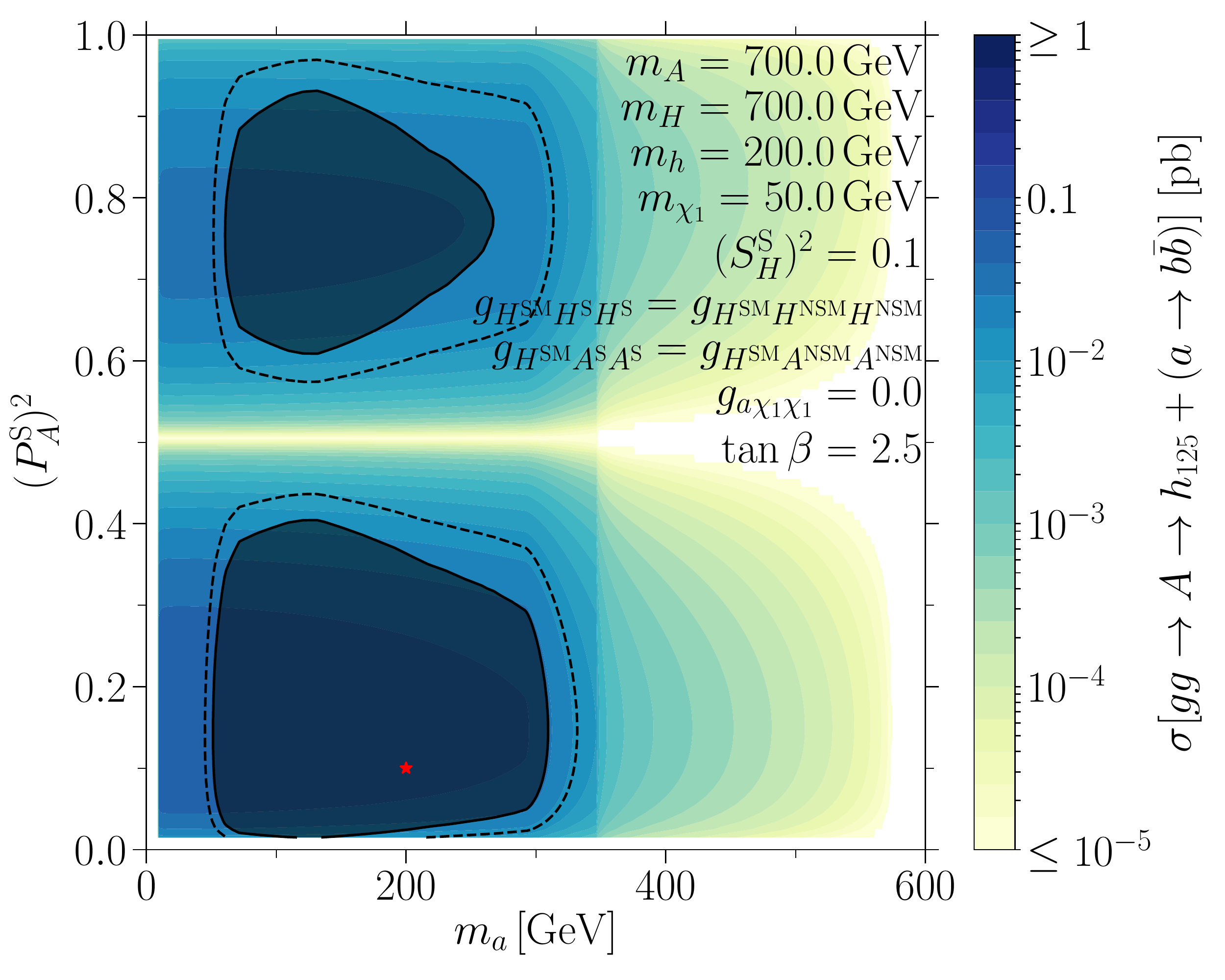}
      \hspace{.5in}
      \includegraphics[width = 2.5in]{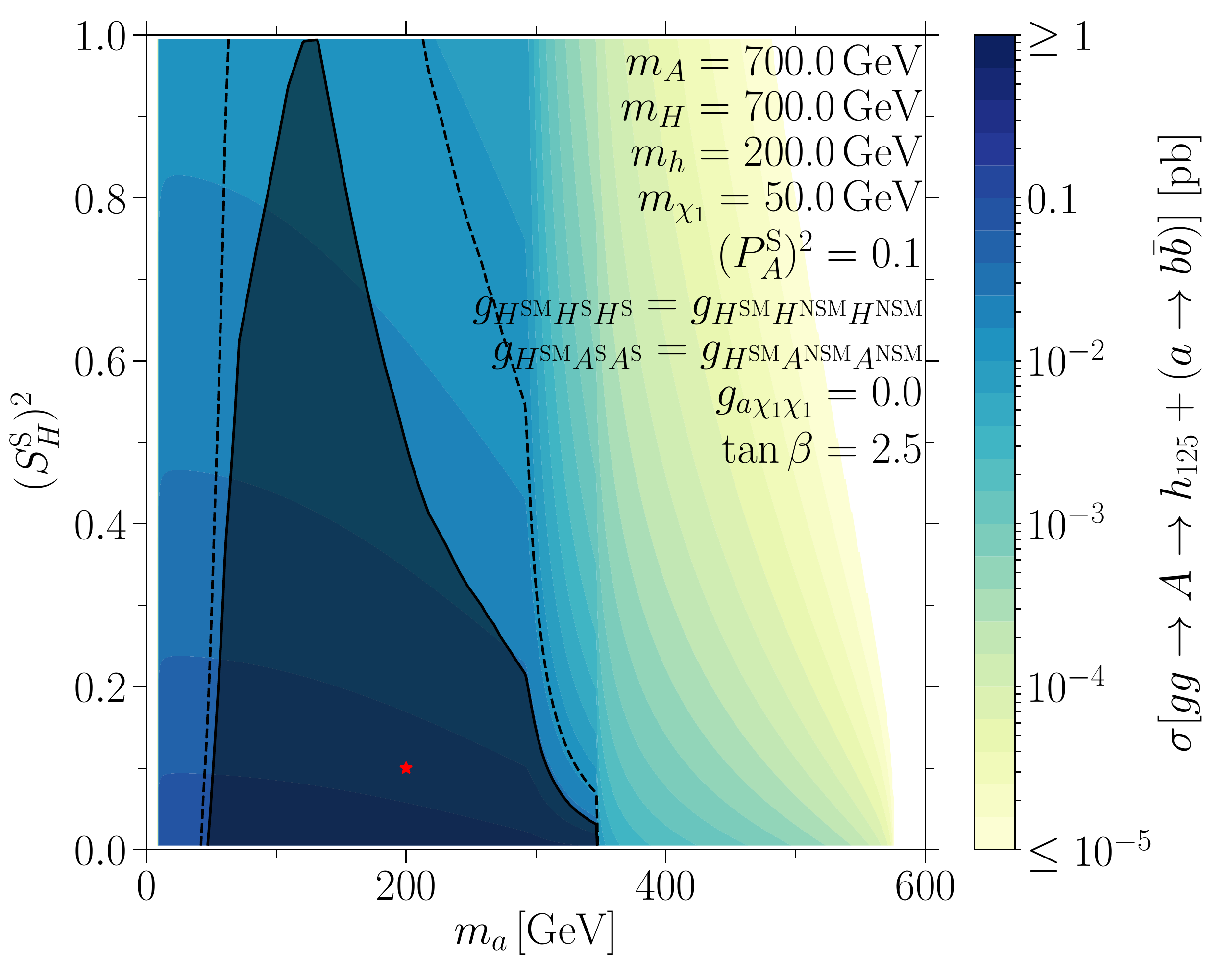}

      \includegraphics[width = 2.5in]{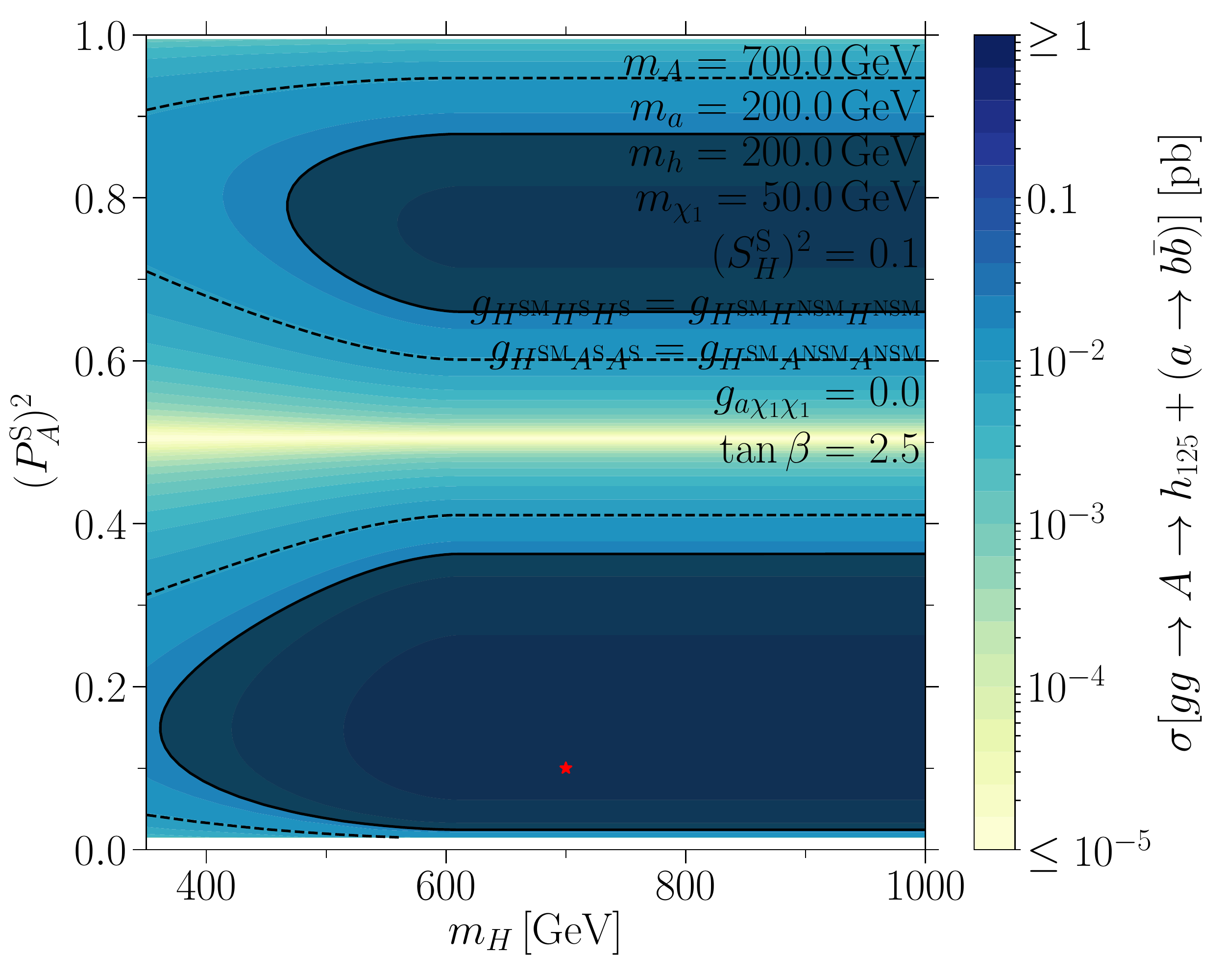}
      \hspace{.5in}
      \includegraphics[width = 2.5in]{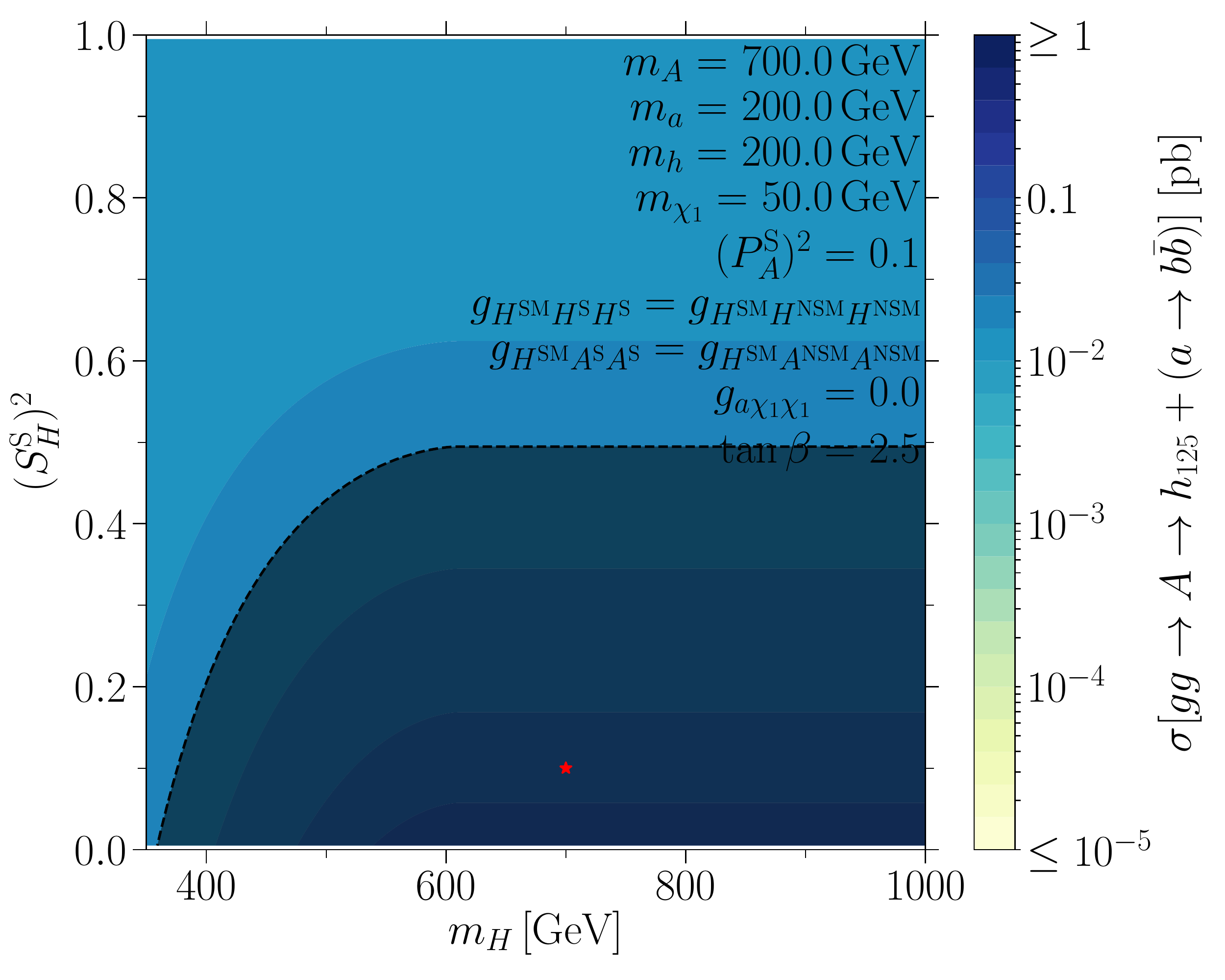}

      \includegraphics[width = 2.5in]{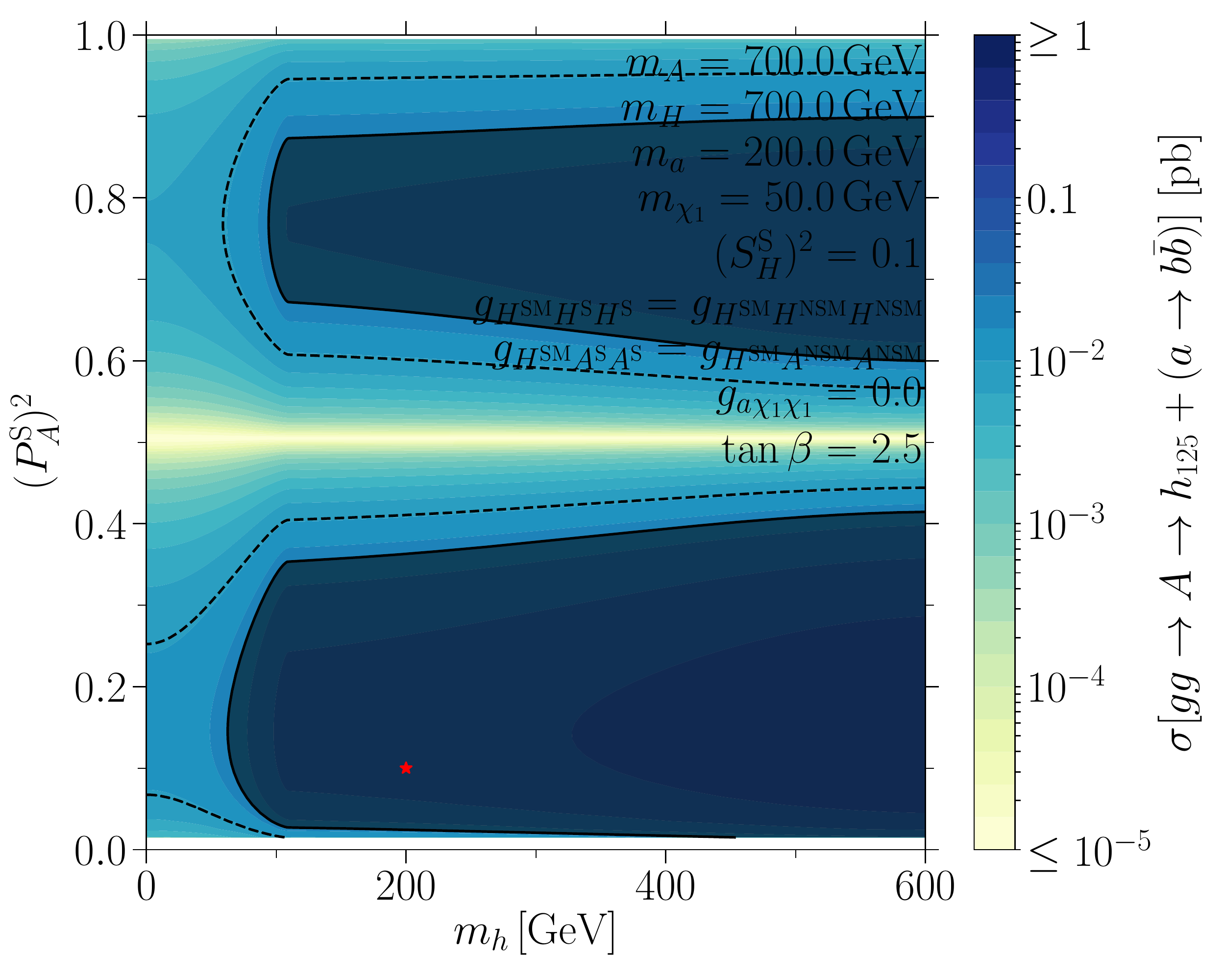}
      \hspace{.5in}
      \includegraphics[width = 2.5in]{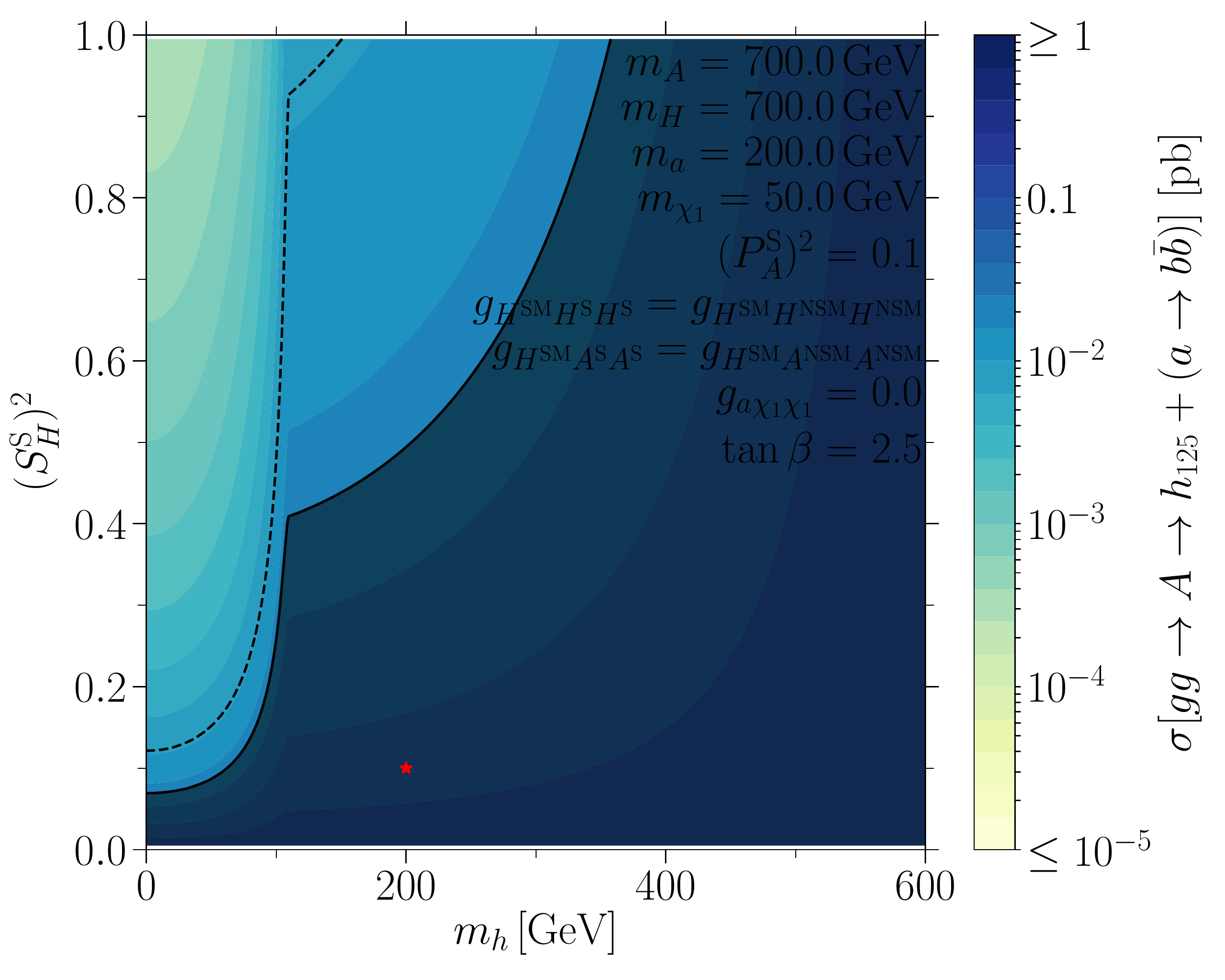}

      \caption{Cross sections and projected LHC sensitivities with 300\,fb$^{-1}$ of data for ($gg \to A \to h_{125} a\to 4b$) in the planes of the relevant masses and mixing angles. The color scale and contours denote cross-sections as labeled by the color bars. The dark shaded regions denote the region where the cross section is larger than the projected sensitivity of the LHC. The dashed black lines denote cross sections a factor of two smaller. The red stars indicate the benchmark point from the first column in Table~\ref{tab:BP_h125S}.}
      \label{fig:Ah125as_Vis}
   \end{centering}
\end{figure}

\begin{figure}[hp]
   \begin{centering}
      \includegraphics[width = 2.5in]{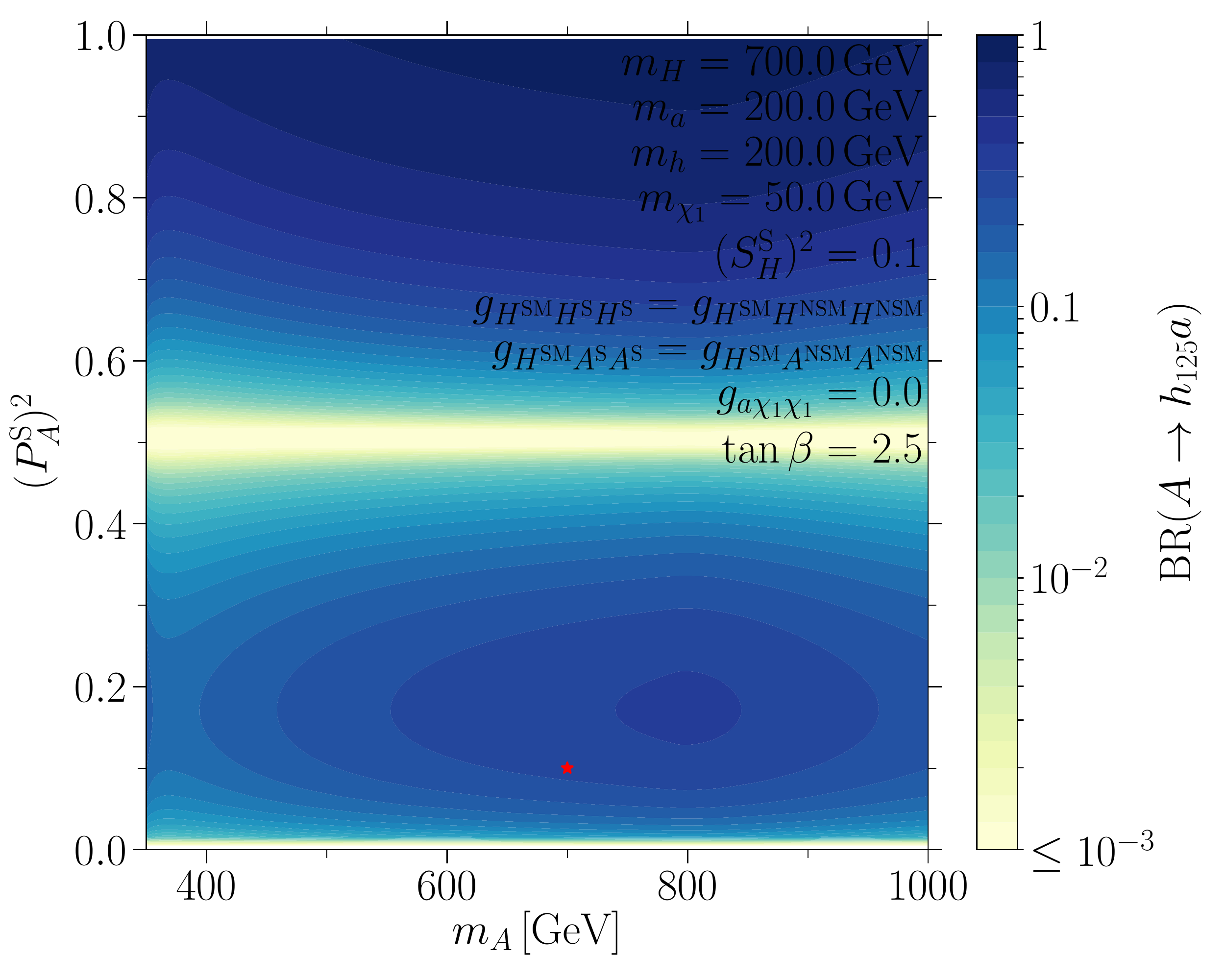}
      \hspace{.5in}
      \includegraphics[width = 2.5in]{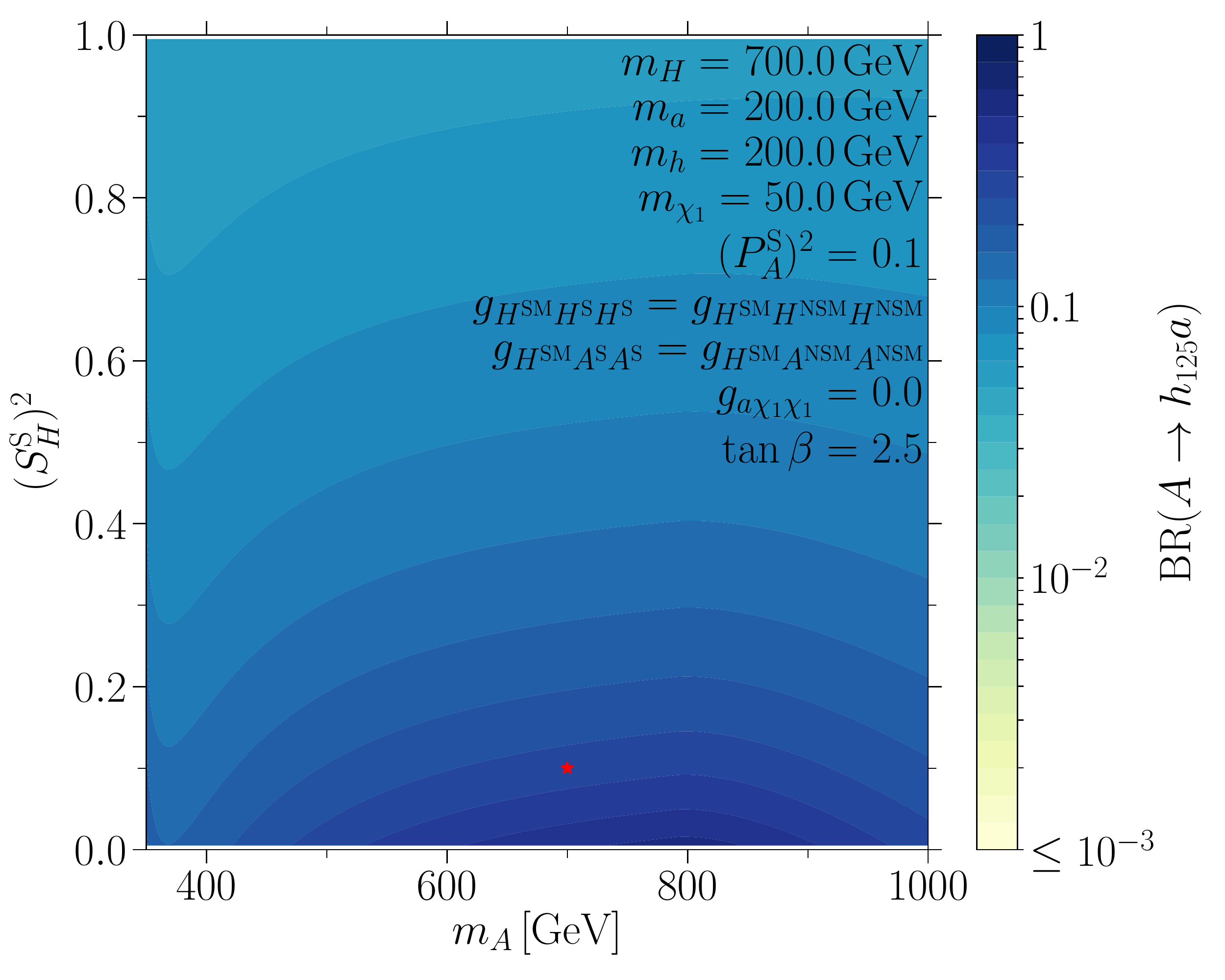}

      \includegraphics[width = 2.5in]{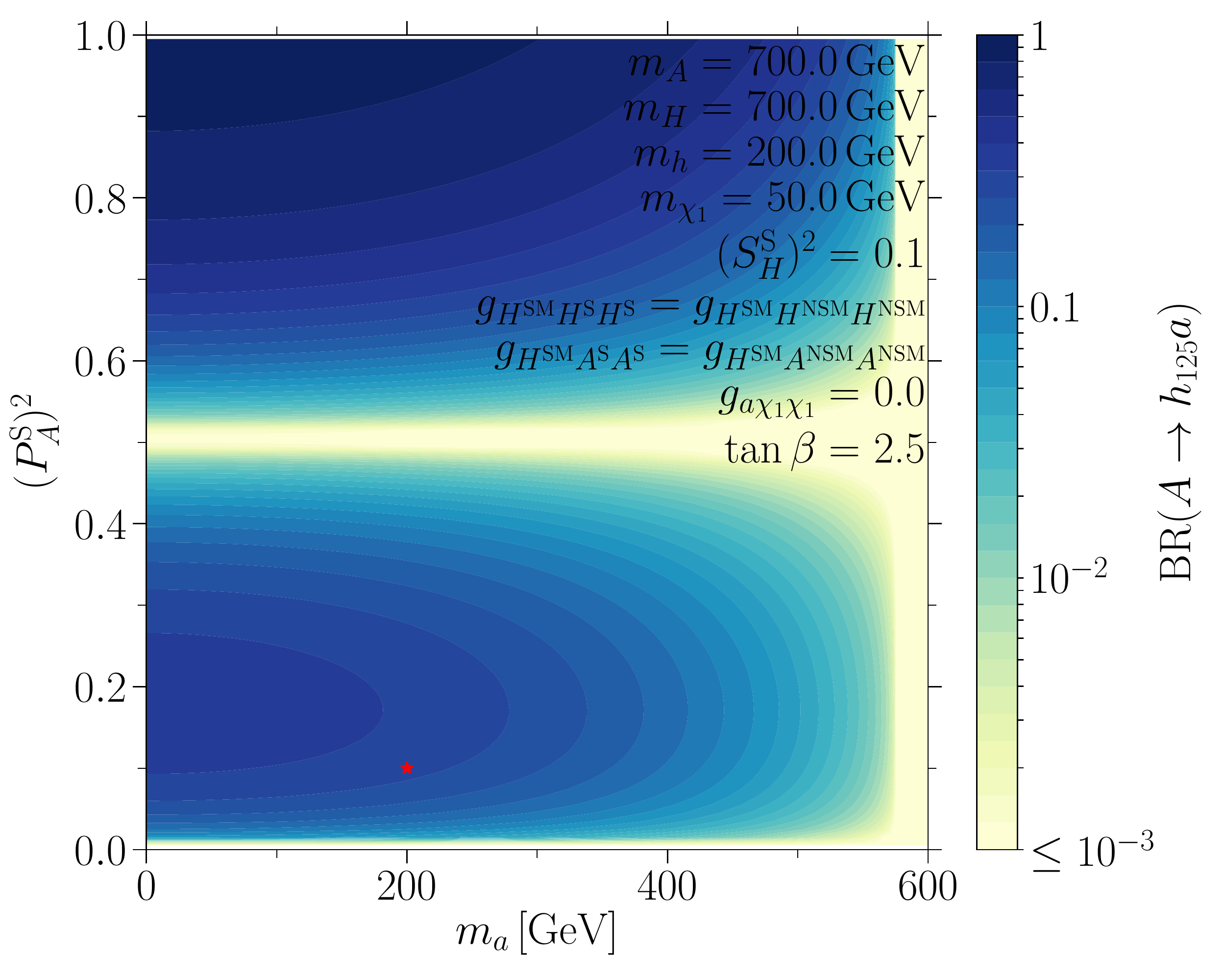}
      \hspace{.5in}
      \includegraphics[width = 2.5in]{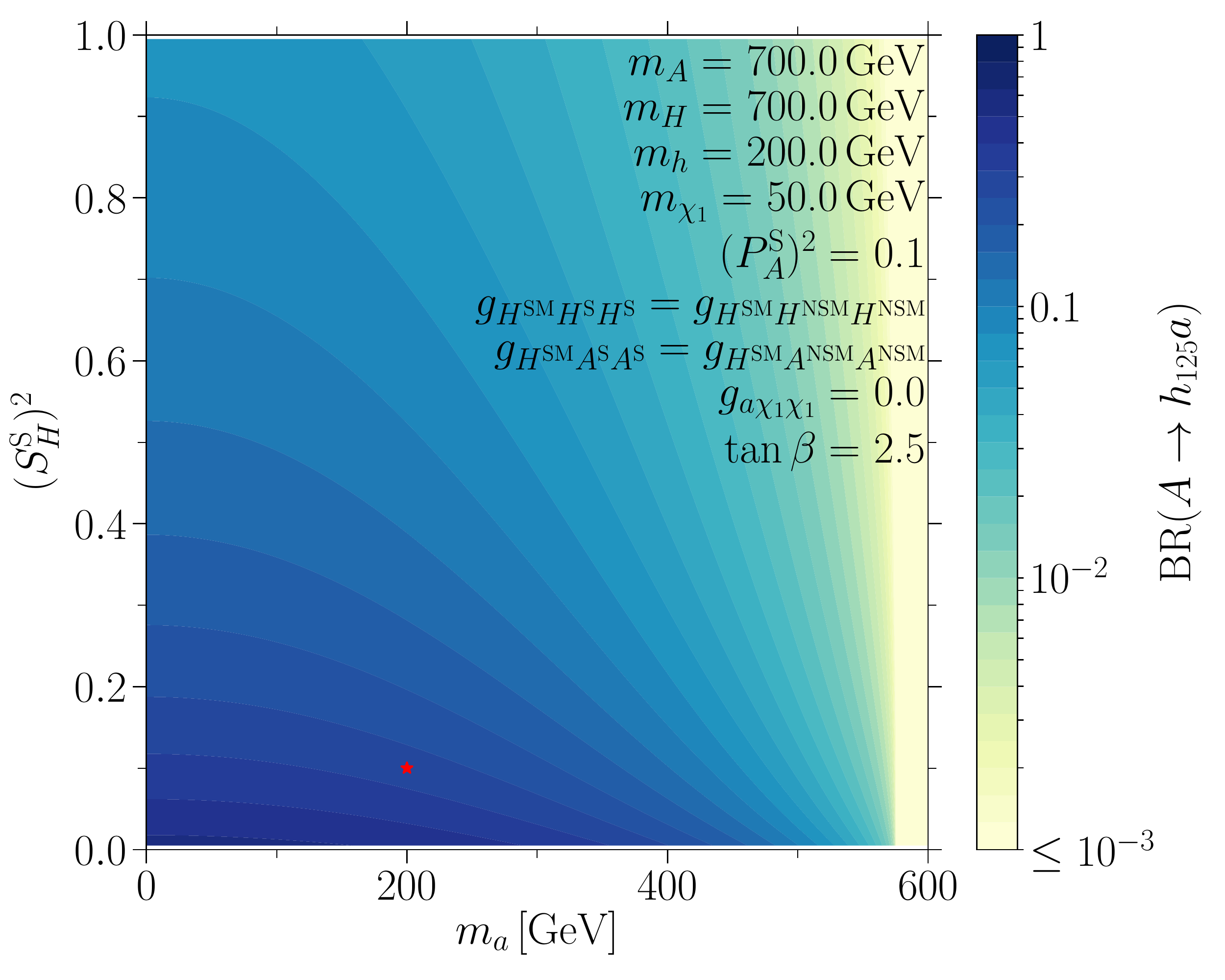}

      \includegraphics[width = 2.5in]{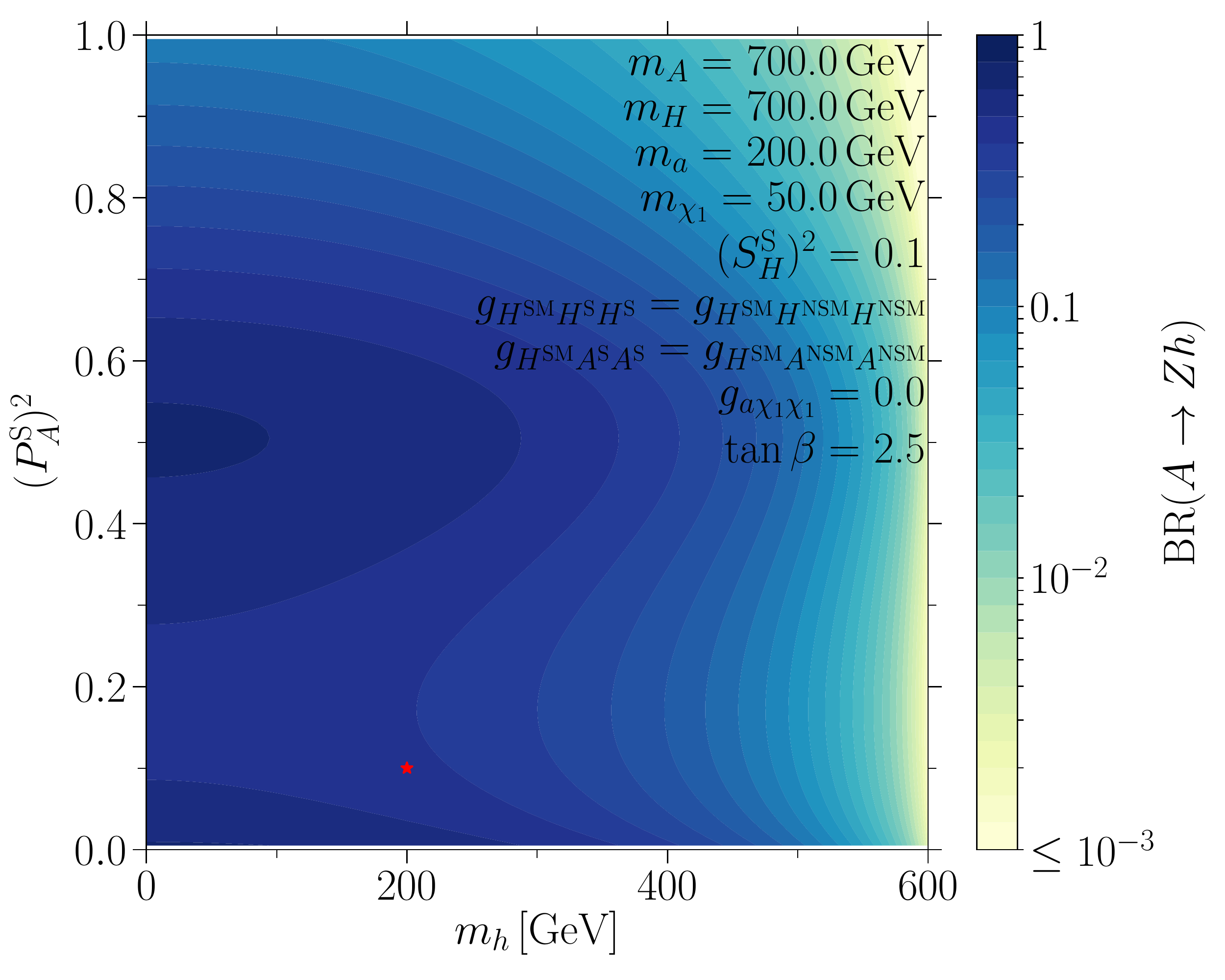}
      \hspace{.5in}
      \includegraphics[width = 2.5in]{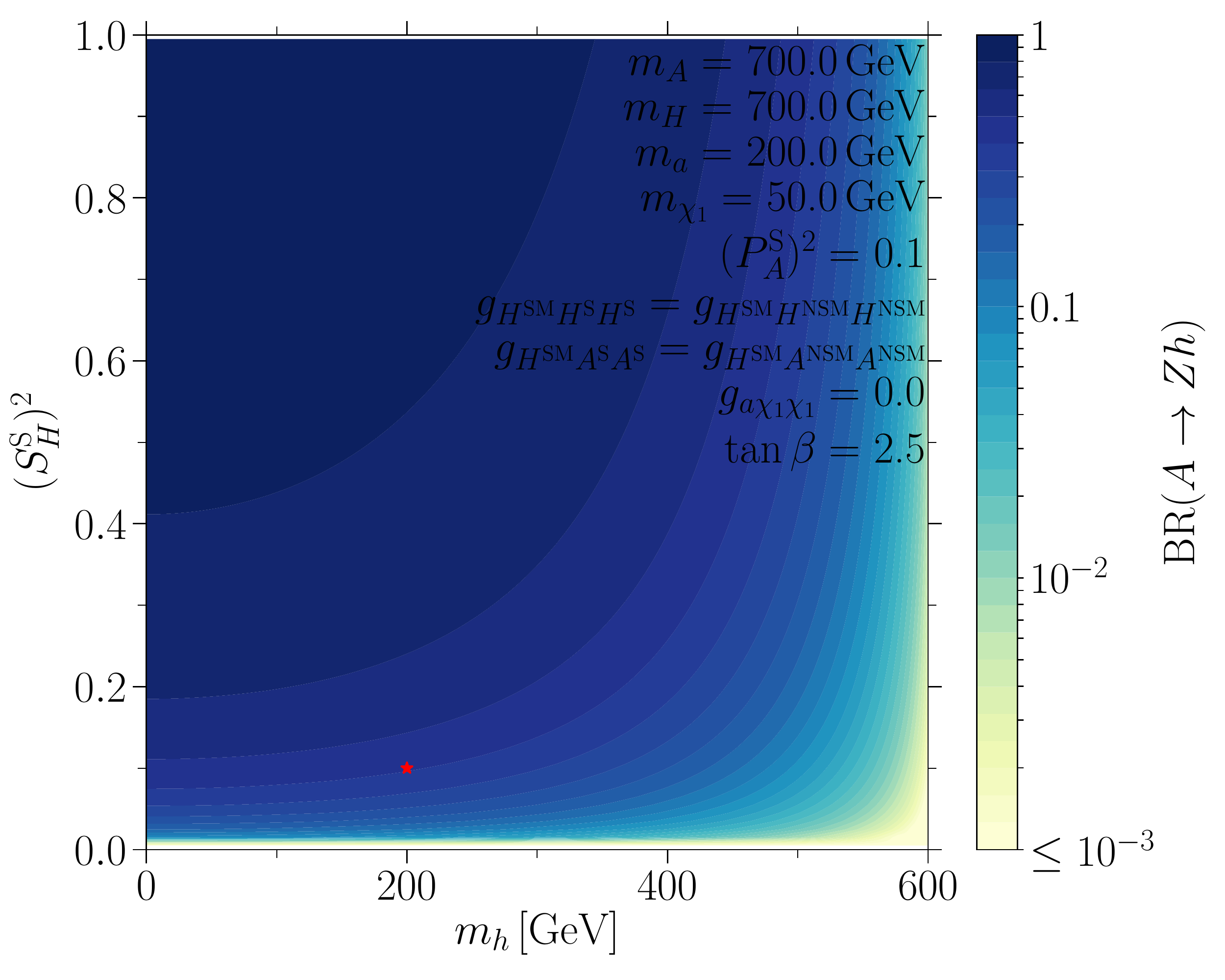}

      \includegraphics[width = 2.5in]{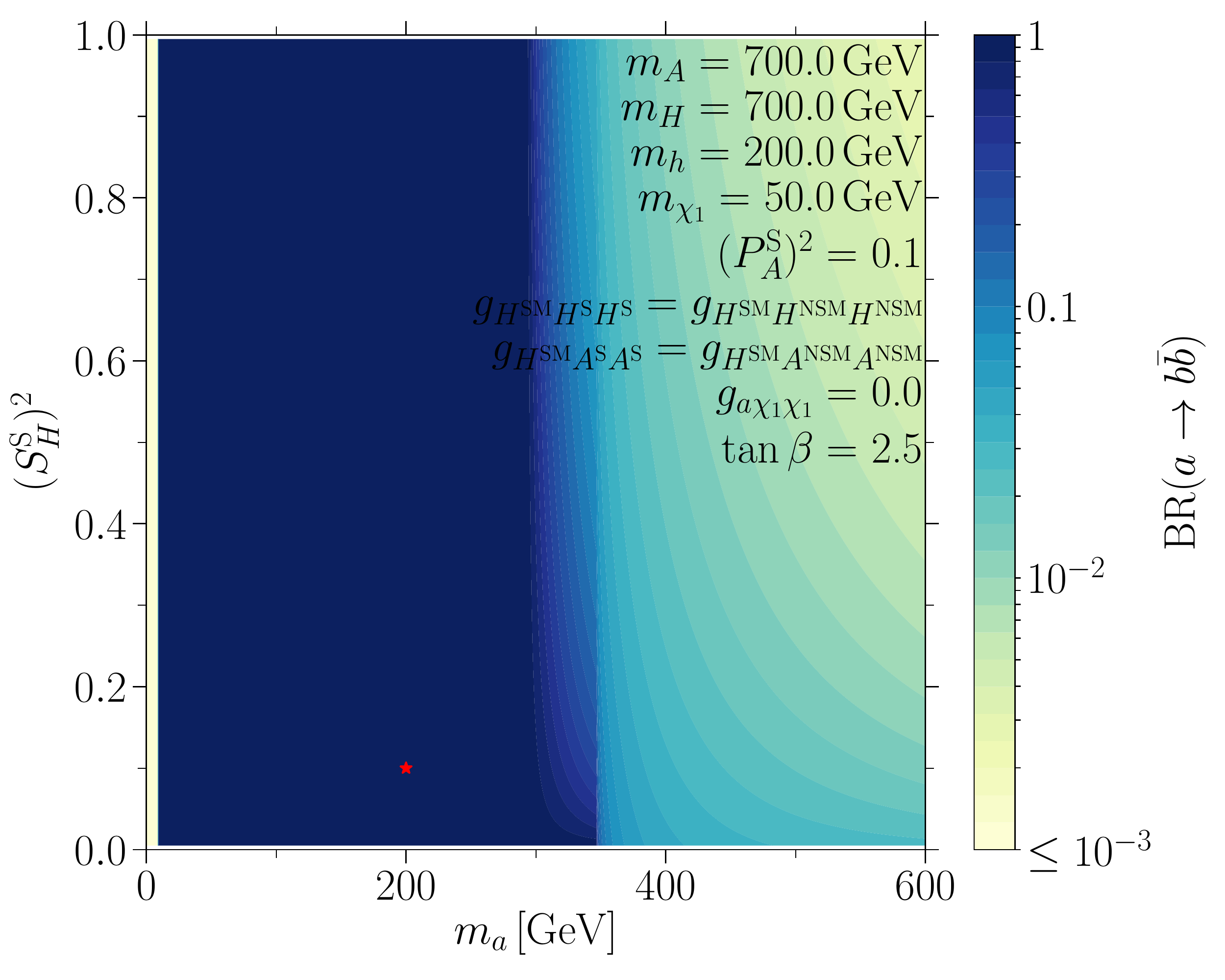}
      \hspace{.5in}
      \includegraphics[width = 2.5in]{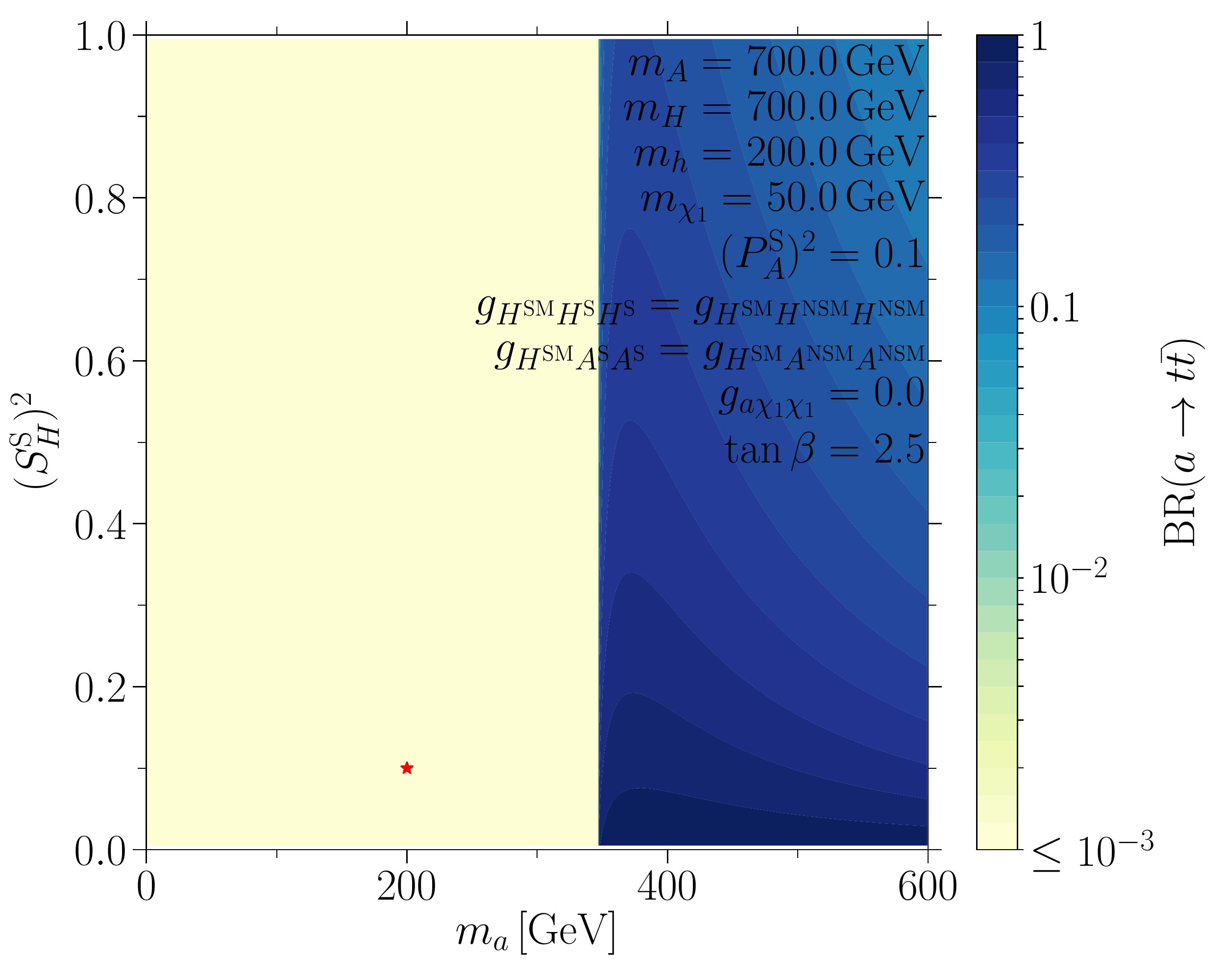}

      \caption{Most relevant branching ratios for $(gg \to A \to h_{125} a \to 4b)$ searches in planes of the relevant masses and mixing angles. The color scale denotes cross-sections as labeled by the color bars. The red stars indicate the benchmark point from the first column in Table~\ref{tab:BP_h125S}.}
      \label{fig:Ah125as_BRs}
   \end{centering}
\end{figure}

\begin{figure}[hp]
   \begin{centering}
      \includegraphics[width = 2.5in]{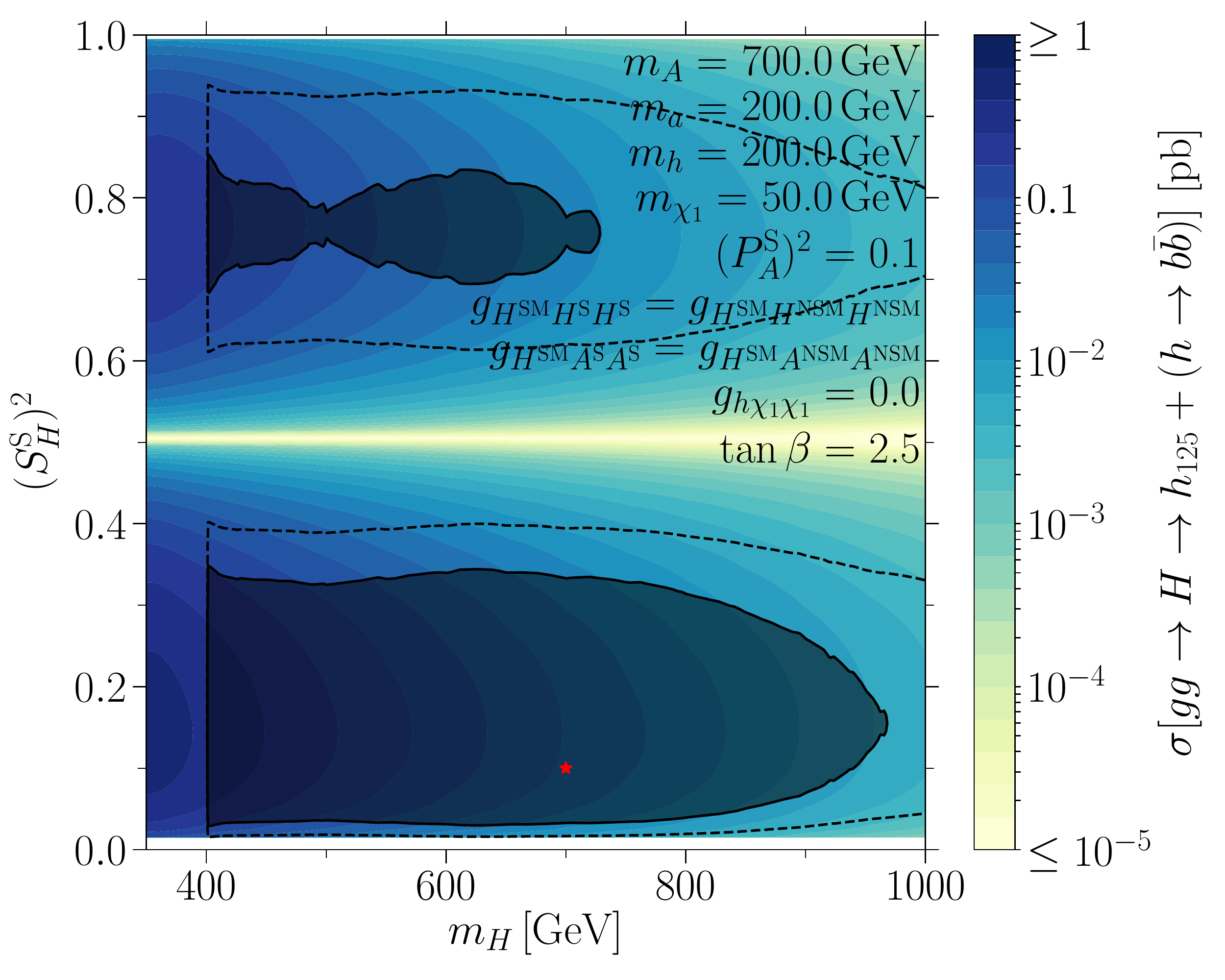}
      \hspace{.5in}
      \includegraphics[width = 2.5in]{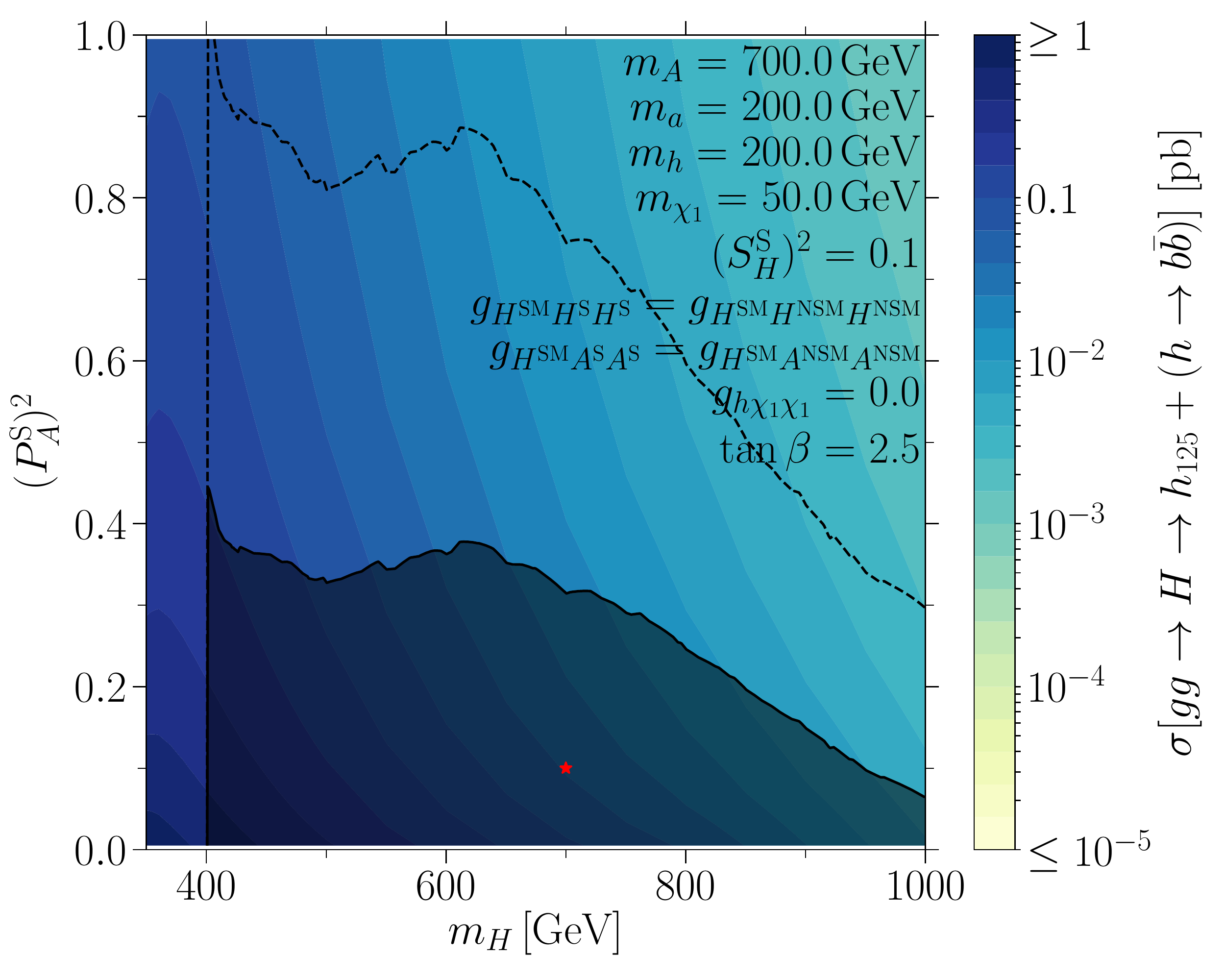}

      \includegraphics[width = 2.5in]{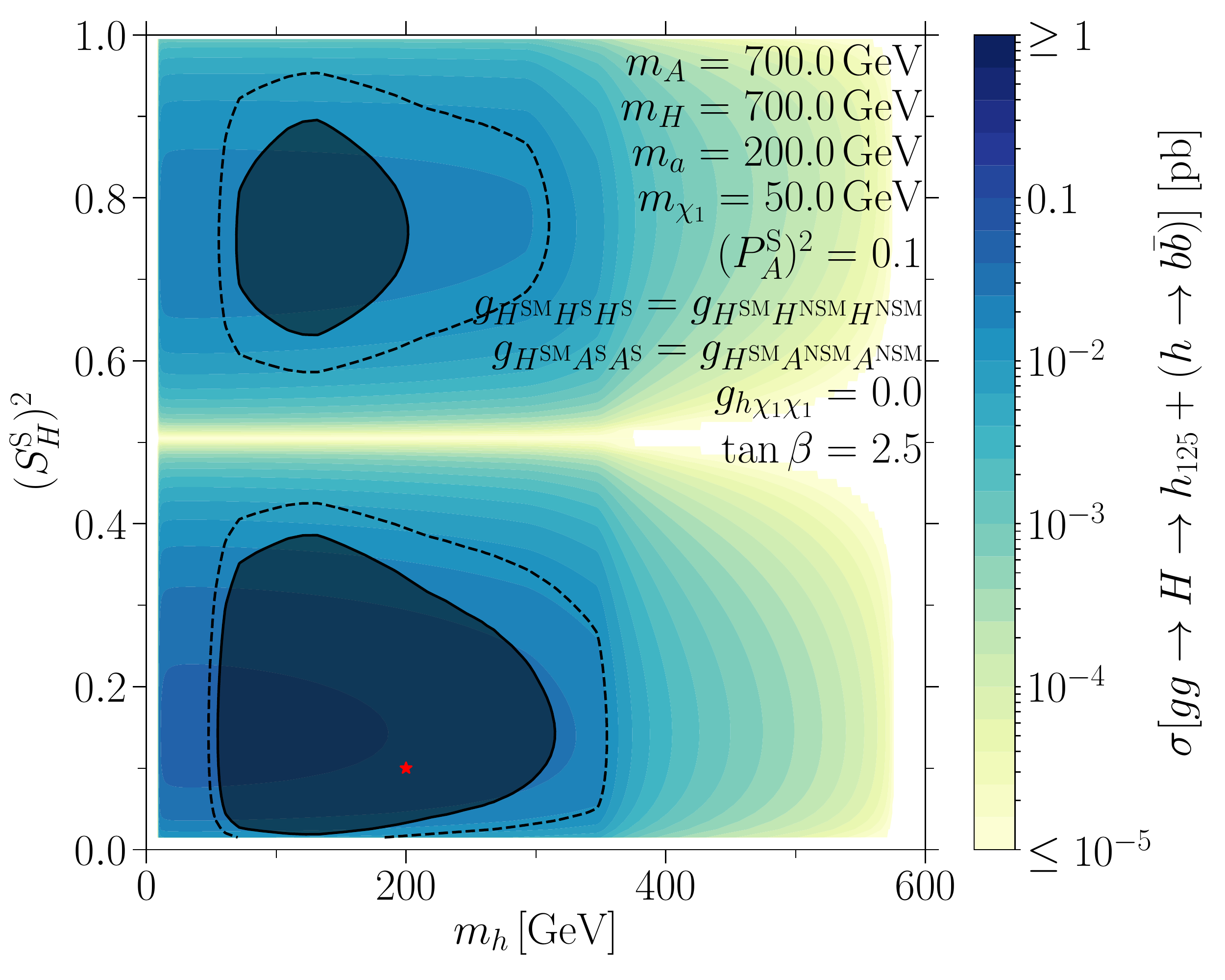}
      \hspace{.5in}
      \includegraphics[width = 2.5in]{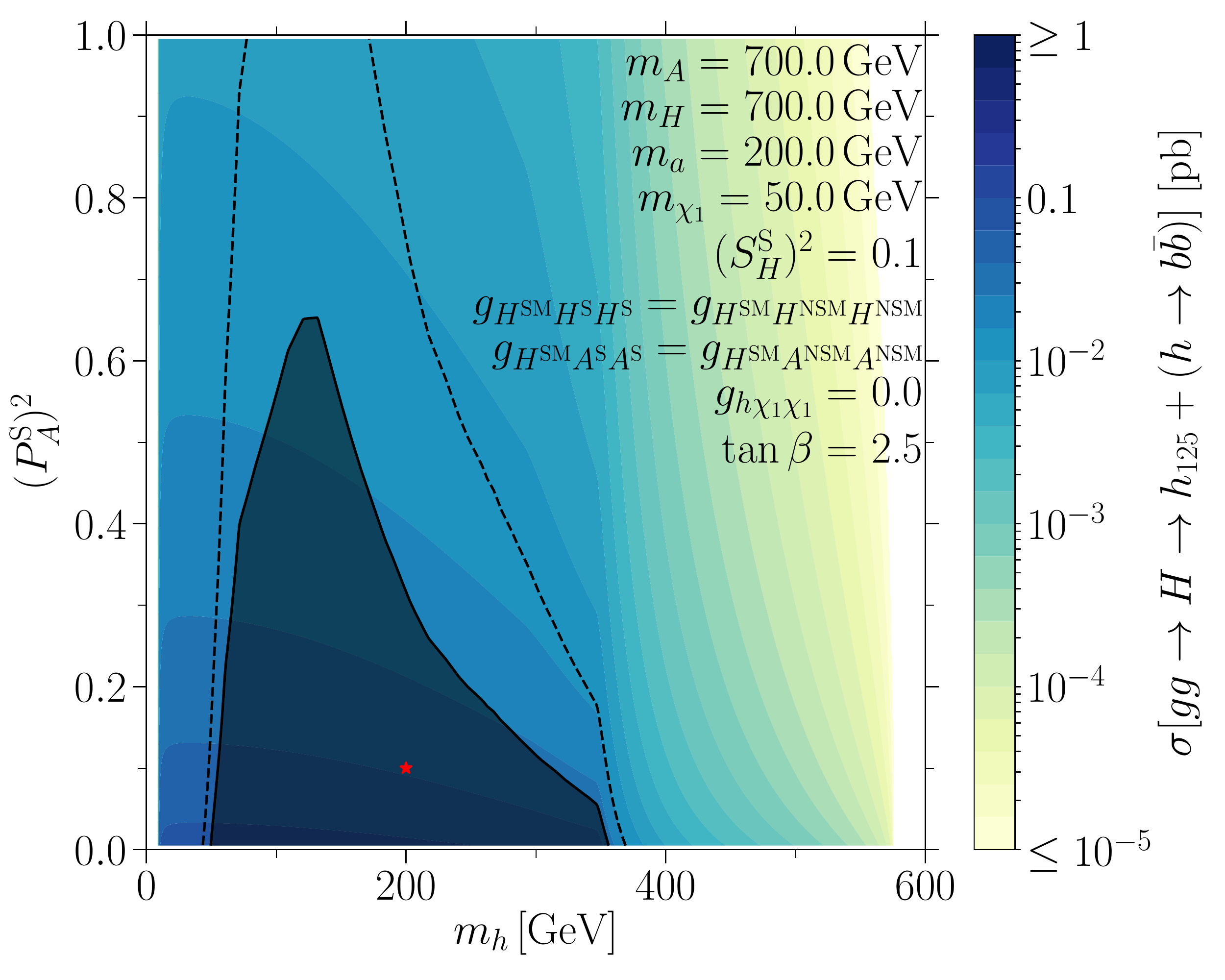}

      \includegraphics[width = 2.5in]{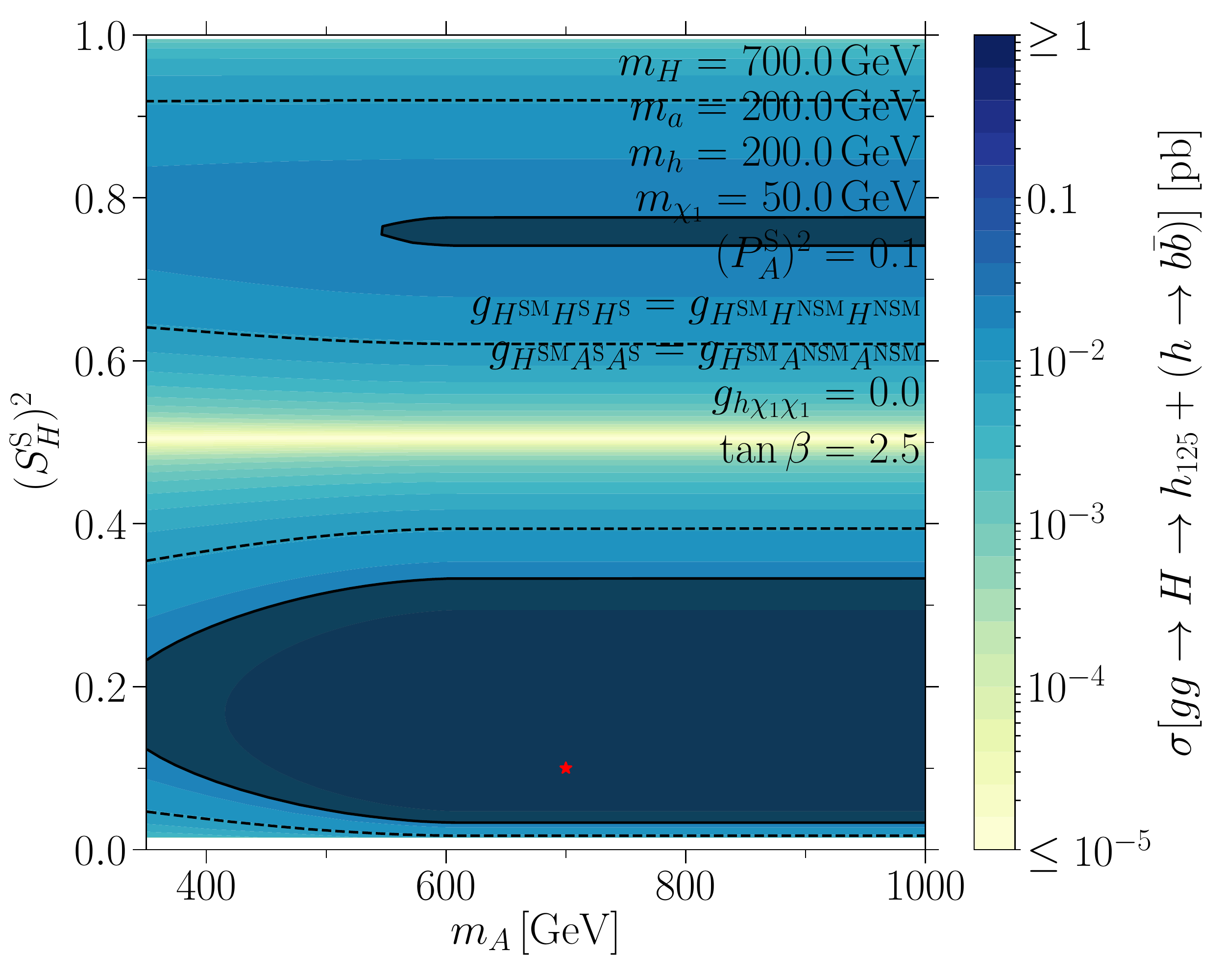}
      \hspace{.5in}
      \includegraphics[width = 2.5in]{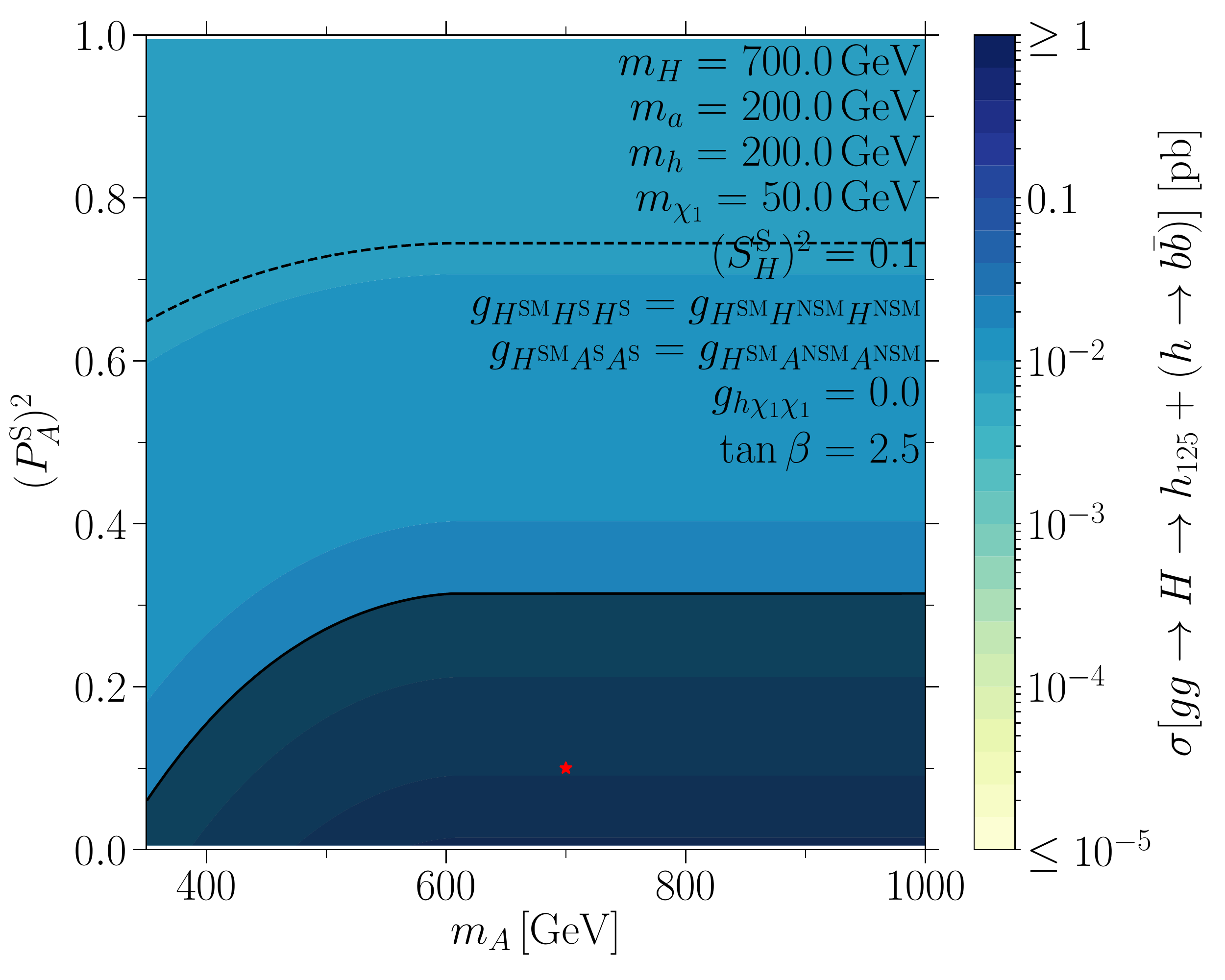}

      \includegraphics[width = 2.5in]{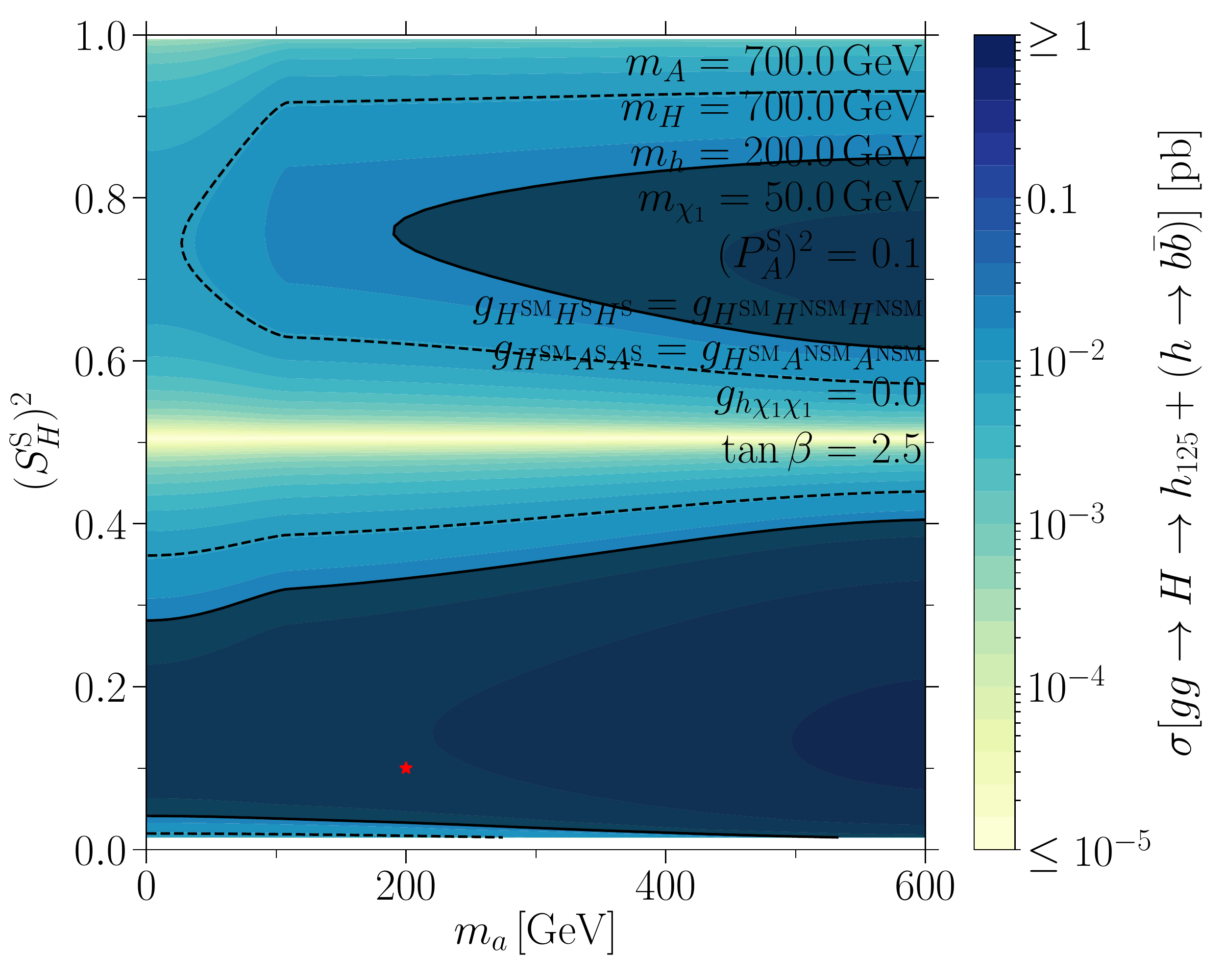}
      \hspace{.5in}
      \includegraphics[width = 2.5in]{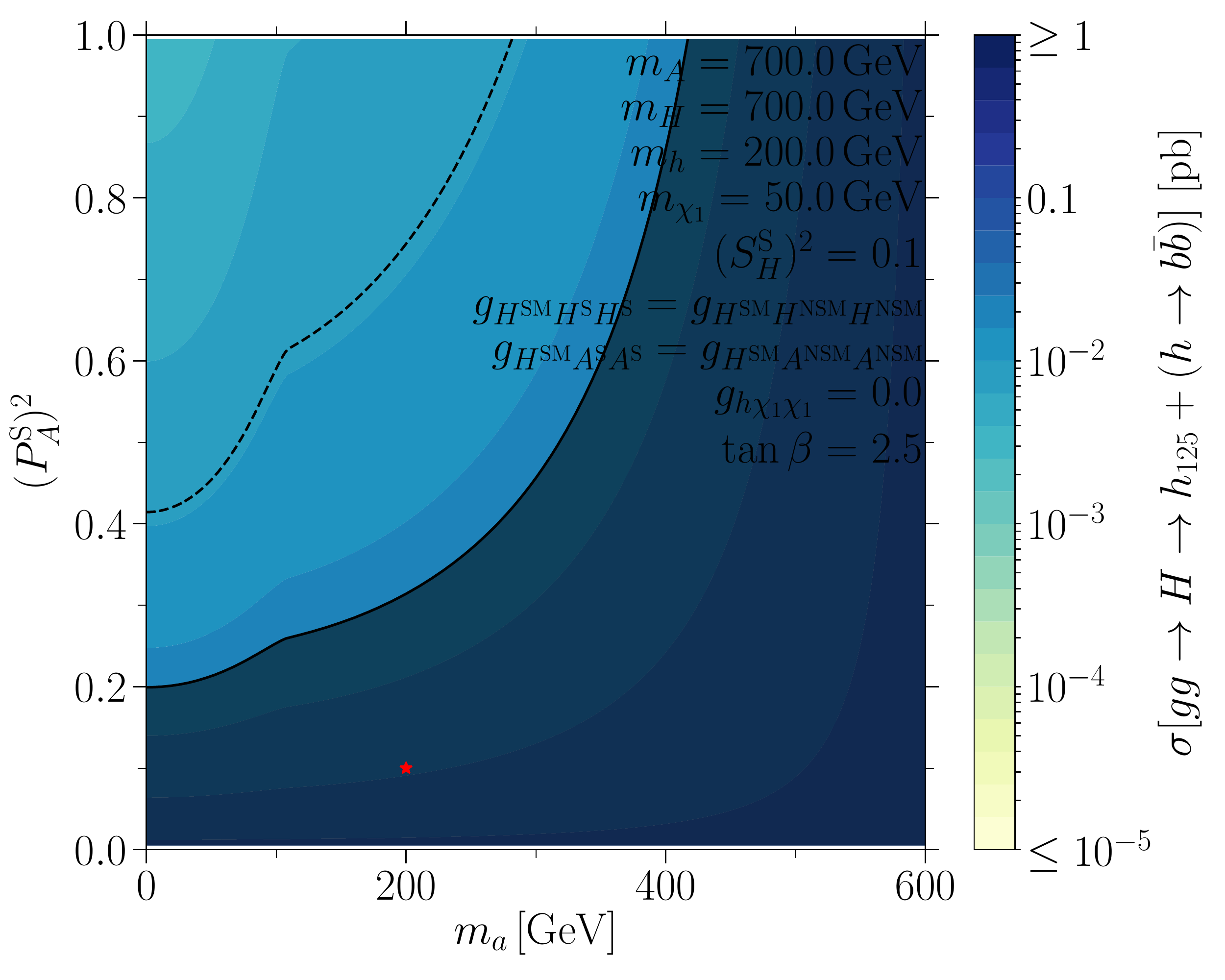}
      \caption{Cross sections and projected LHC sensitivities with 300\,fb$^{-1}$ of data for ($gg \to H \to h_{125} h\to 4b$) in the planes of the relevant masses and mixing angles. The color scale and contours denote cross-sections as labeled by the color bars. The dark shaded regions denote the region where the cross section is larger than the projected sensitivity of the LHC. The dashed black lines denote cross sections a factor of two smaller. The red stars indicate the benchmark point from the first column in Table~\ref{tab:BP_h125S}.}
\label{fig:Hh125hs_Vis}
\end{centering}
\end{figure}

\begin{figure}[hp]
   \begin{centering}
      \includegraphics[width = 2.5in]{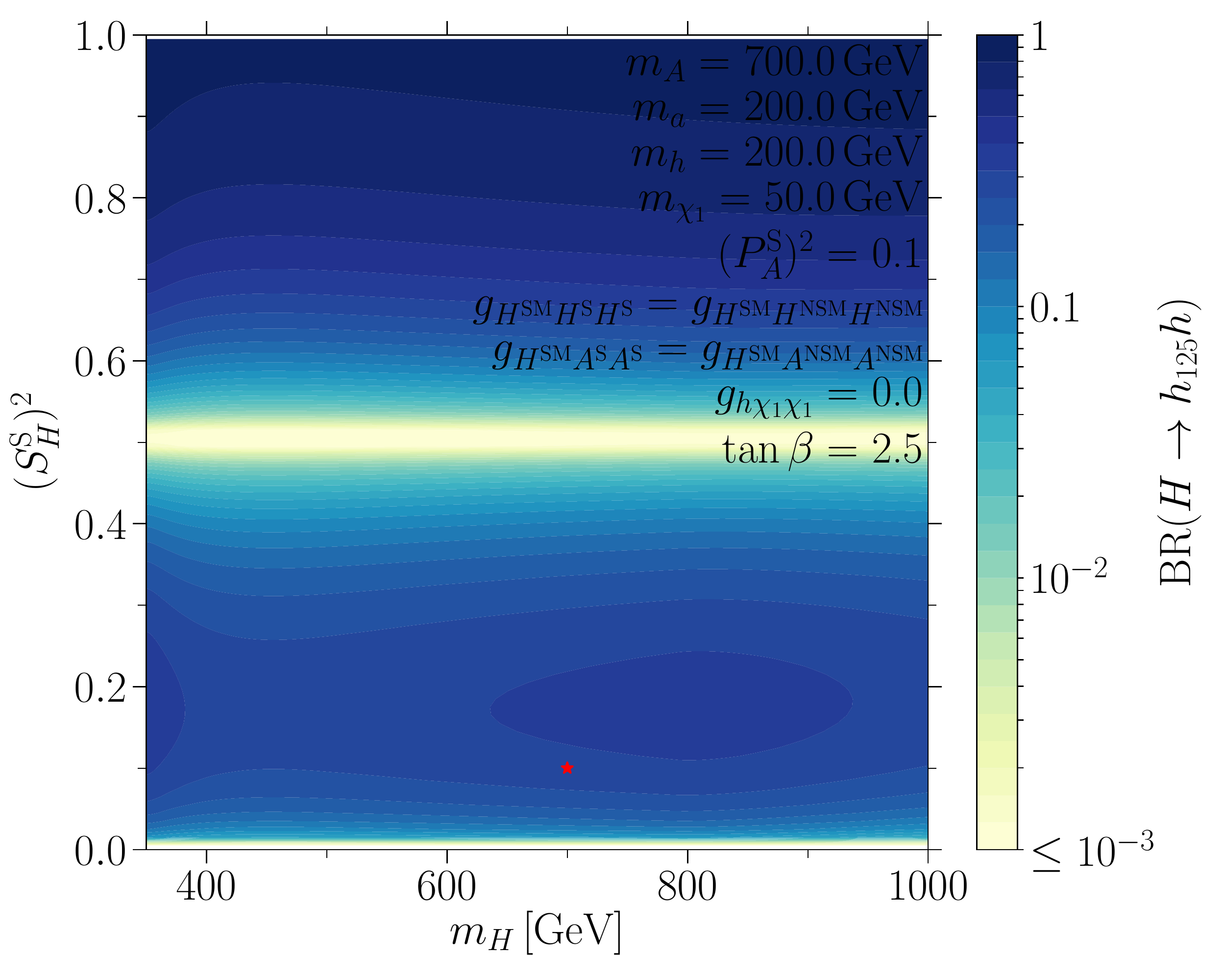}
      \hspace{.5in}
      \includegraphics[width = 2.5in]{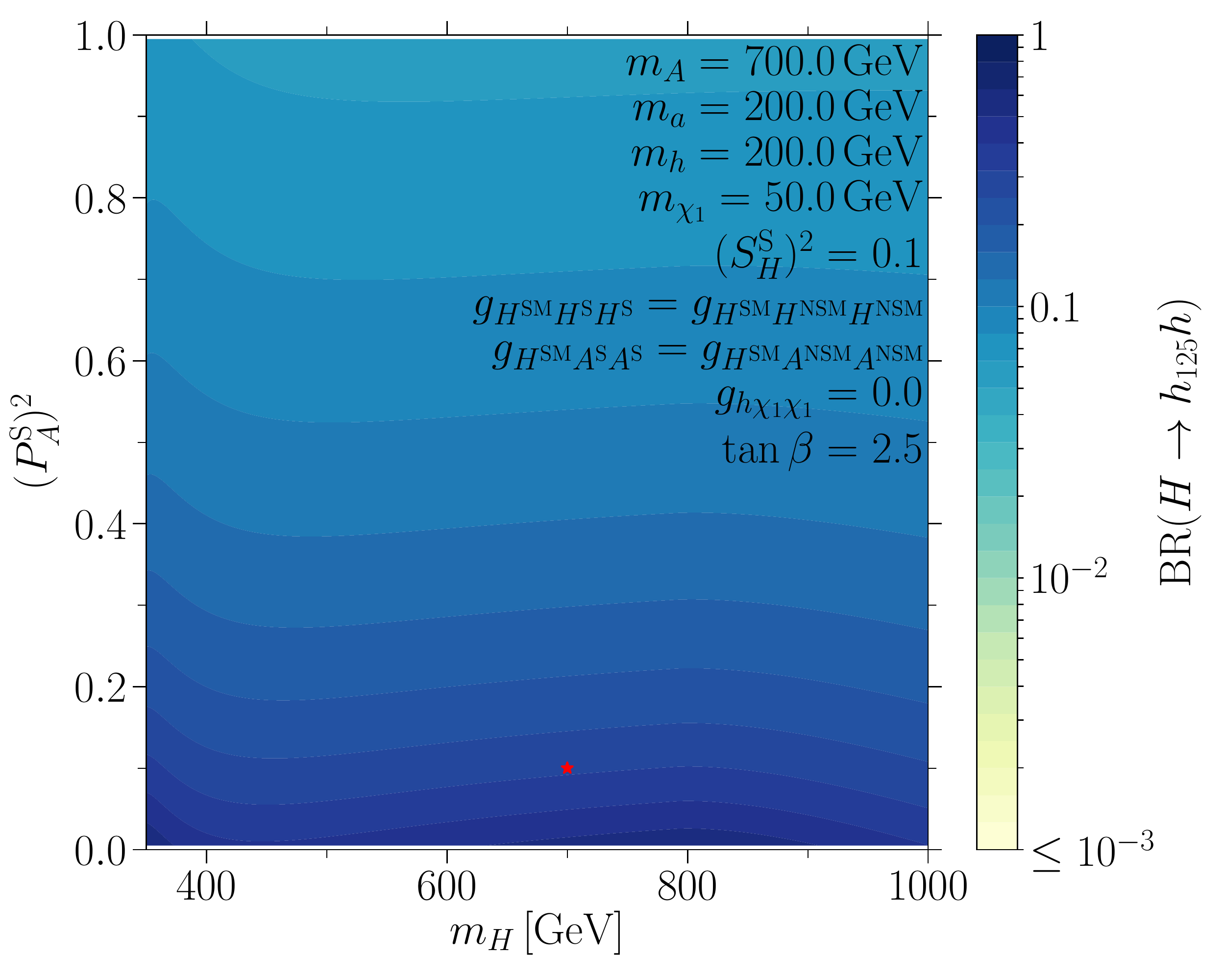}

      \includegraphics[width = 2.5in]{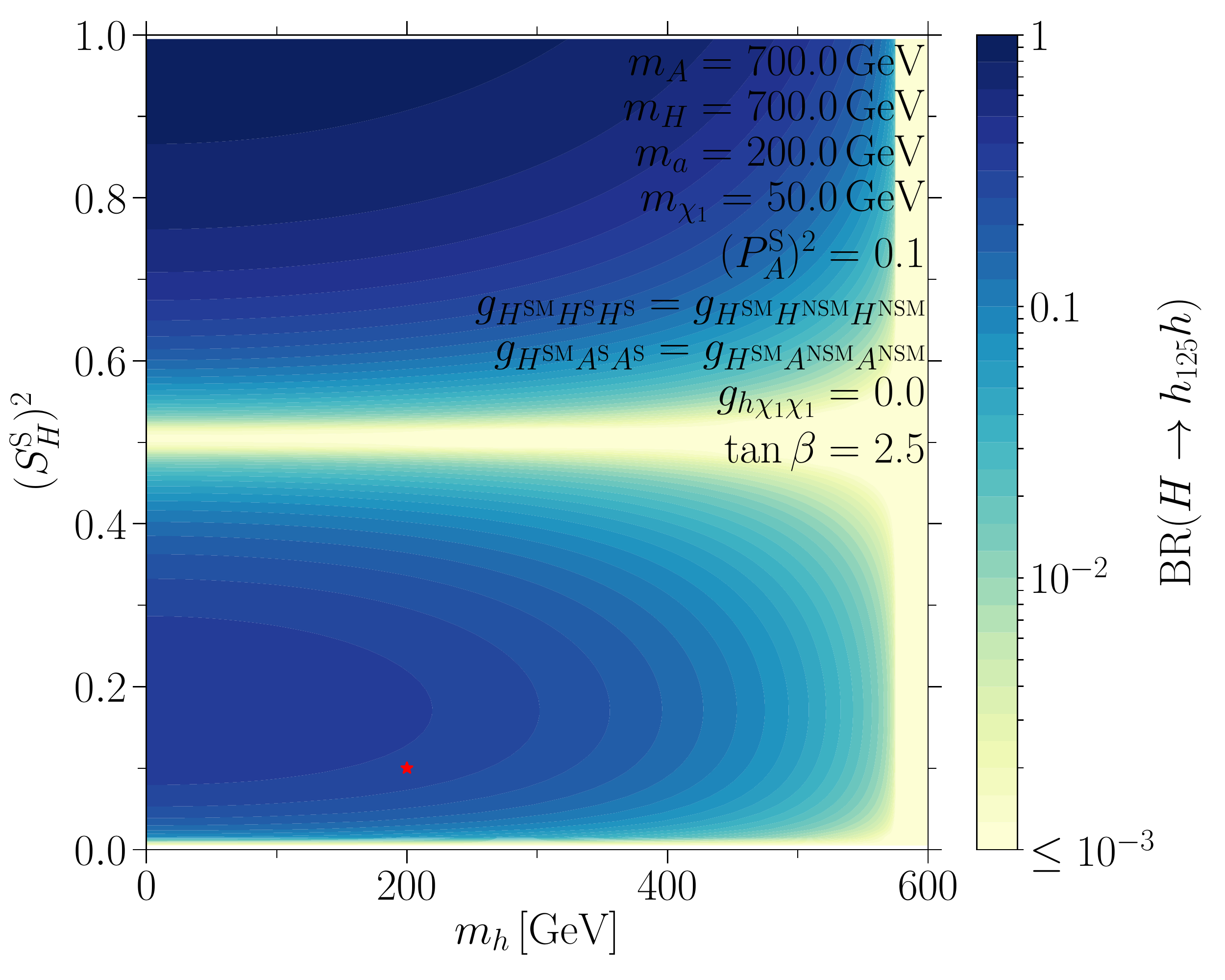}
      \hspace{.5in}
      \includegraphics[width = 2.5in]{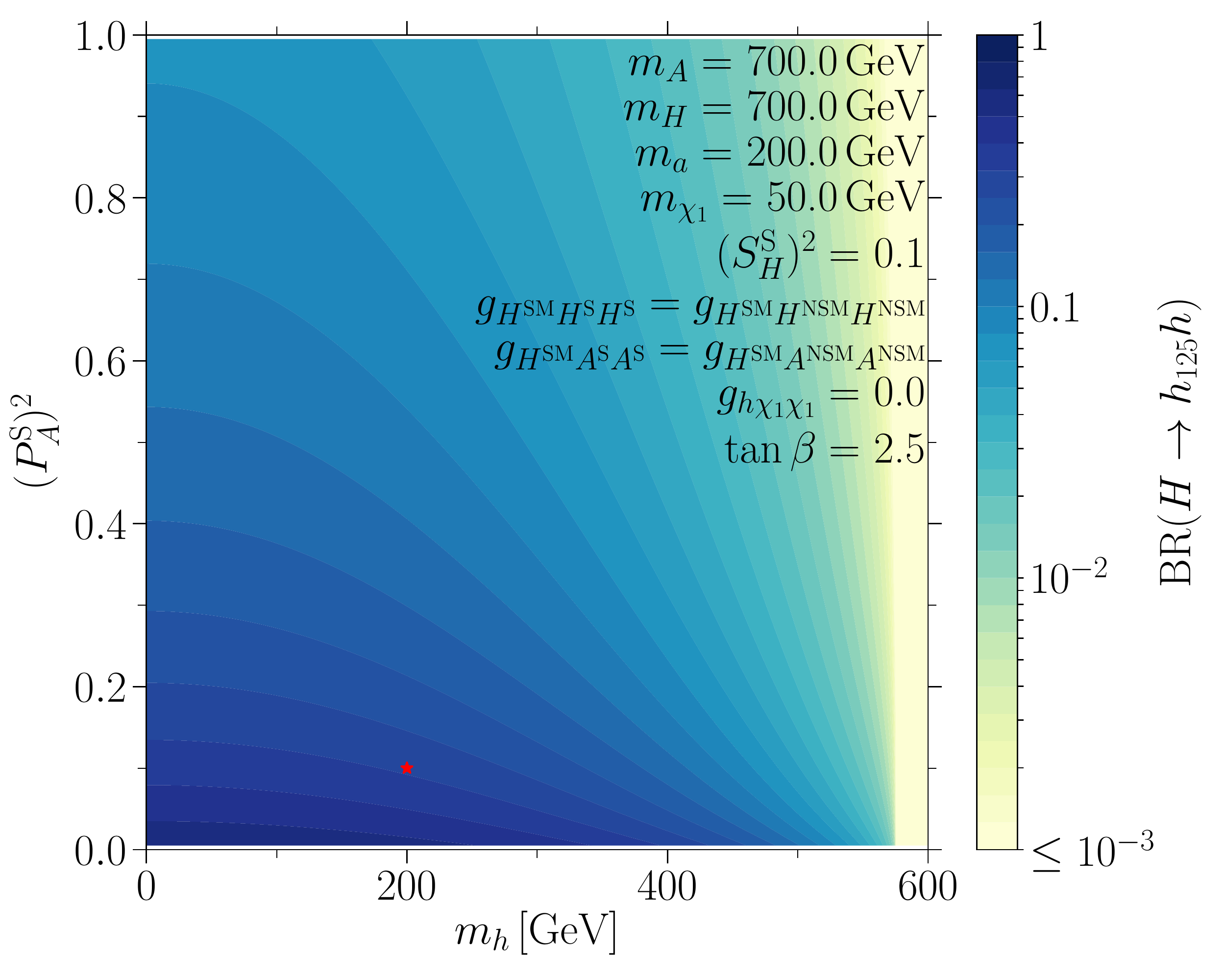}

      \includegraphics[width = 2.5in]{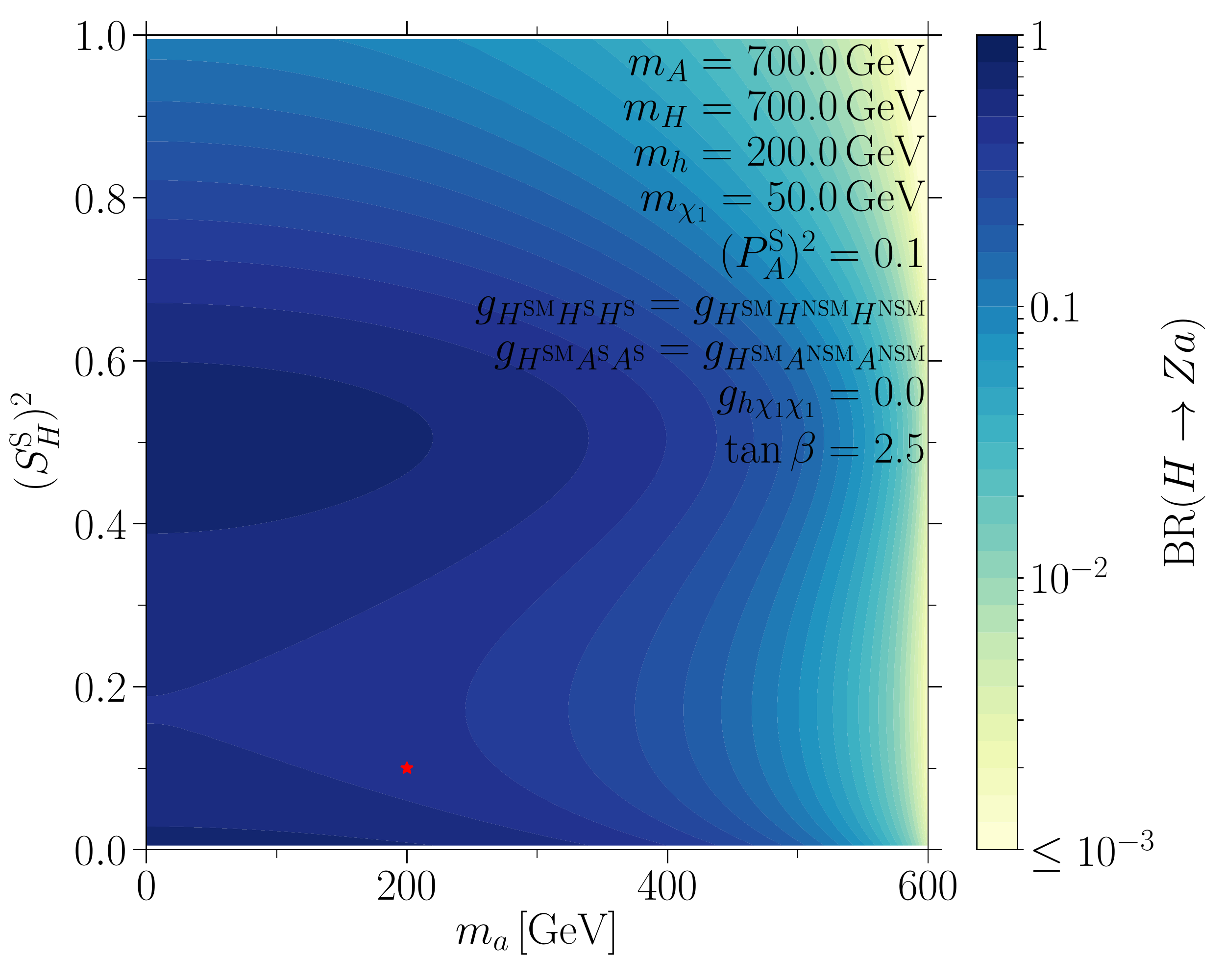}
      \hspace{.5in}
      \includegraphics[width = 2.5in]{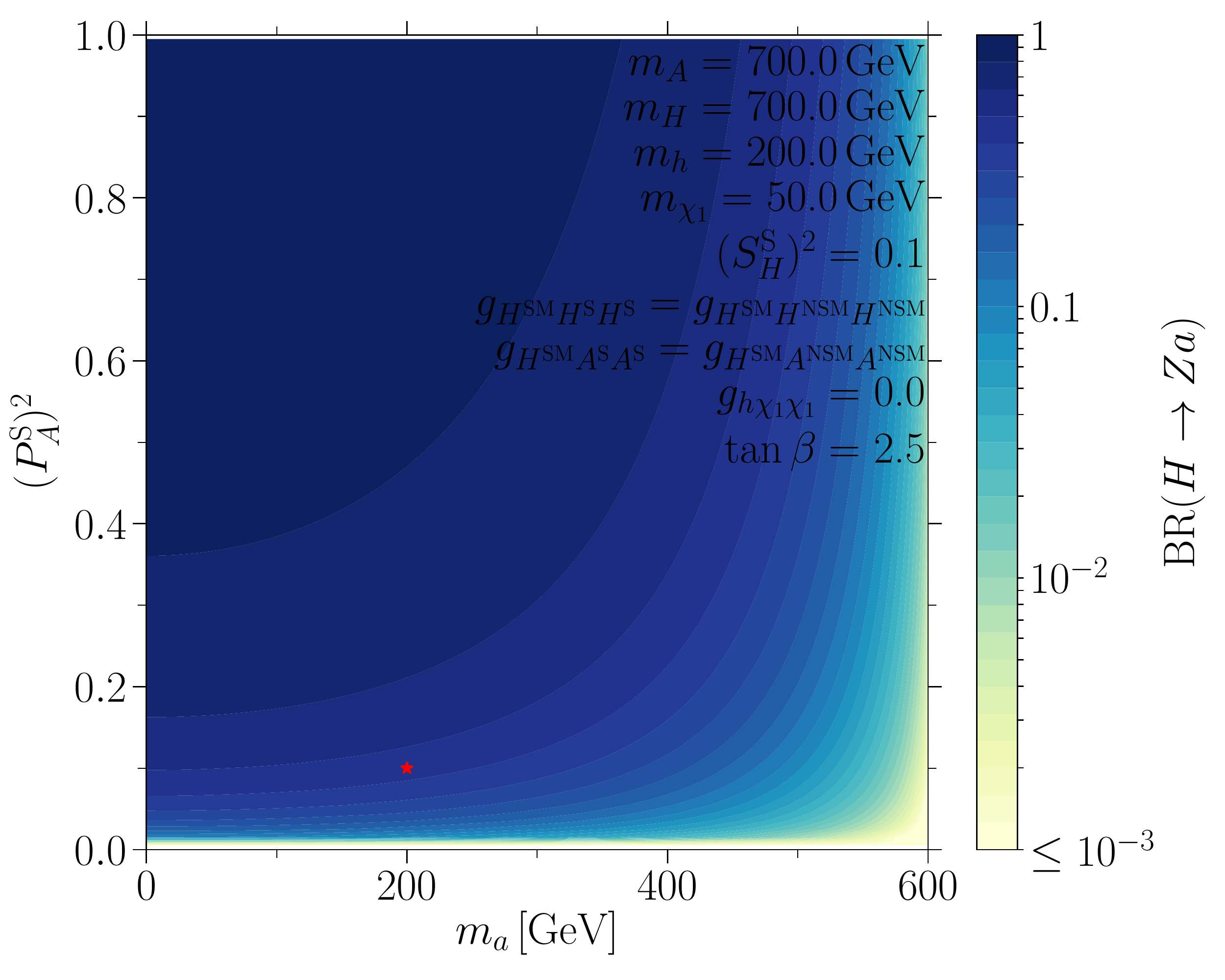}

      \includegraphics[width = 2.5in]{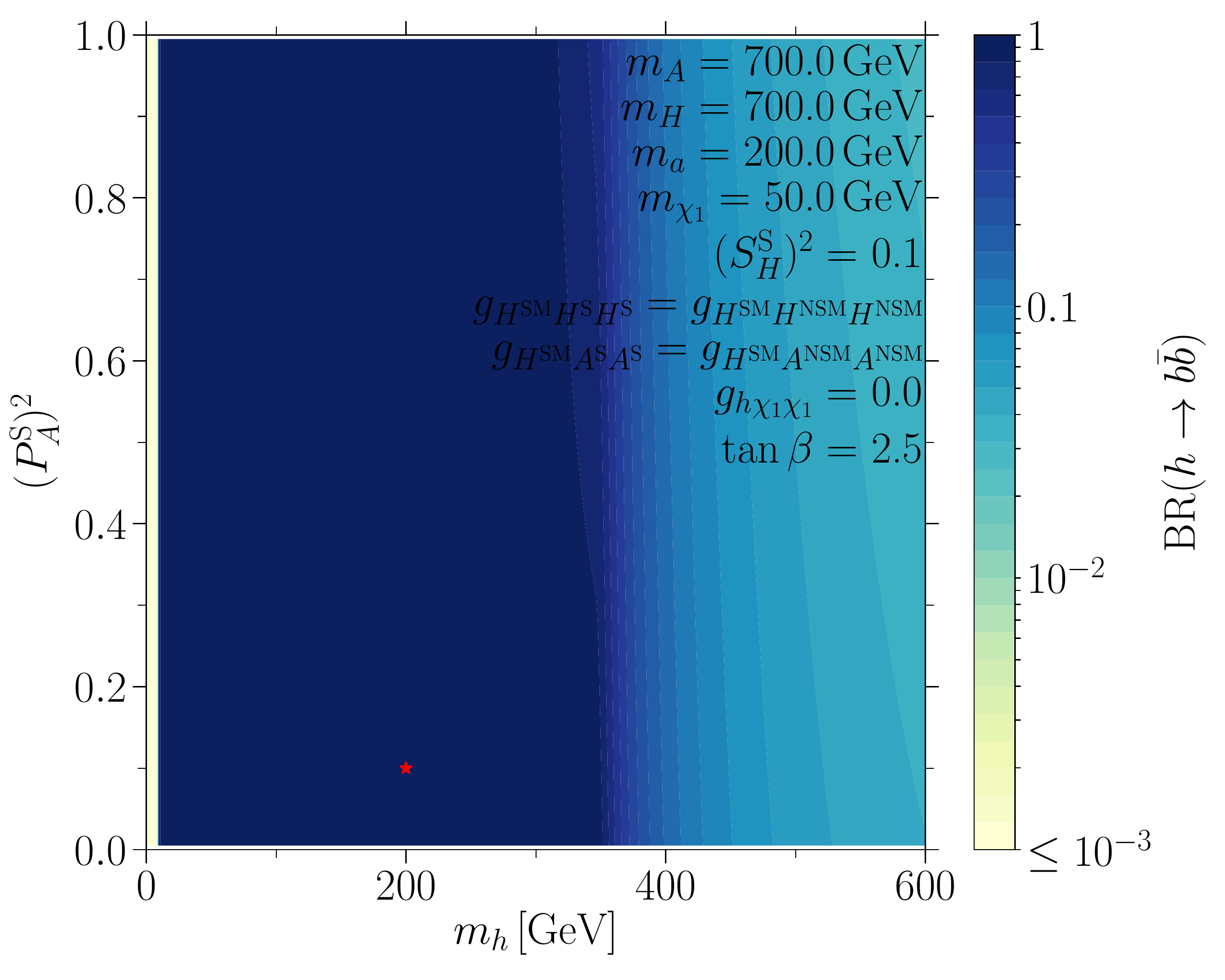}
      \hspace{.5in}
      \includegraphics[width = 2.5in]{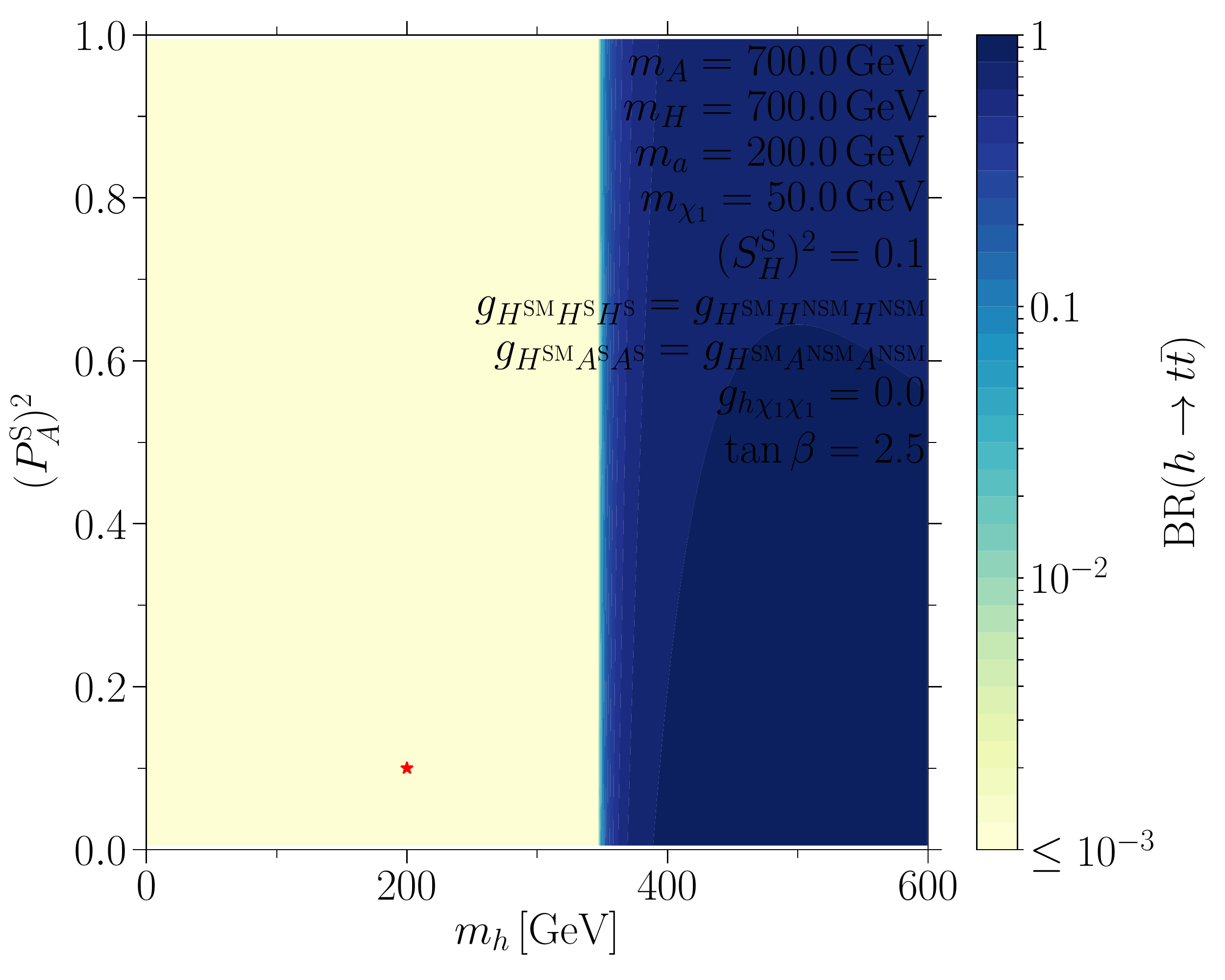}
      \caption{Most relevant branching ratios for $(gg \to H \to h_{125} h \to 4b)$ searches in planes of the relevant masses and mixing angles. The color scale denotes cross-sections as labeled by the color bars. The red stars indicate the benchmark point from the first column in Table~\ref{tab:BP_h125S}.}
\label{fig:Hh125hs_BRs}
\end{centering}
\end{figure}

In Figs.~\ref{fig:Ah125as_Vis} and \ref{fig:Hh125hs_Vis} we show the variation in the final state cross-section when changing the values of the parameters listed in Table~\ref{tab:BP_h125S}. Here and in the all the plots, the parameters held fixed to benchmark values are portrayed in each plot label. The coupling of $h/a$ to possible dark matter particles $\chi_1$ is set to 0, hence, this benchmark scenario can also be interpreted as a ``pure'' 2HDM+S without the addition of the state $\chi_1$. The contours and color shading show the cross-section, and the gray shaded region denotes the region where the LHC is expected to probe this scenario with 300\,fb$^{-1}$ of data in the 4$b$ channel; we take the projected sensitivity from Ref.~\cite{Ellwanger:2017skc}. The dashed black lines denote cross-sections which are a factor of two smaller than the projected reach. The sensitivity in the $2b2\gamma$ channel is expected to be similar~\cite{Ellwanger:2017skc}. Note that the cross sections shown in the figures do not include $h_{125}$ branching ratios, which are taken to be the SM values, hence sensitivities for different $h_{125}$ decay modes can be easily compared. The red stars denote the benchmark point defined in Table~\ref{tab:BP_h125S}. Note that we treat the sensitivities independently for $H$ and $A$. 

In Fig.~\ref{fig:Ah125as_Vis} we show the variations in ($gg \to A\to h_{125} a$) and in Fig.~\ref{fig:Ah125as_BRs} the most relevant branching ratios for this channel. The most relevant quantities for this cross-section are the pseudoscalar mixing angle and masses shown in the two top left panels. The observed strong suppression for maximally mixed $A^{\rm NSM}-A^{\rm S}$ originates from the suppression of the relevant trilinear coupling in that region, cf. Eq.~\eqref{eq:freecoup_combination}. The main effect of varying $m_A$ is in the gluon fusion production cross section, which results in an overall scaling of the cross sections shown. We further note that the LHC loses sensitivity as $a$ approaches the $t\bar{t}$ threshold. Above the top threshold and for low values of $\tan\beta$, $a$ will decay preferentially to top quark pairs, cf. Fig.~\ref{fig:Ah125as_BRs}. It is interesting to note that in such a case, the final state signature will be $t\bar{t}h_{125}$, and the cross section may be of the same order as the $t\bar{t}h_{125}$ associated production cross section in the SM. To the best of our knowledge, no projections exist for this final state. However, it seems possible to get stringent constraints in this case by demanding that the top pairs reconstruct to $m_a$, combined with the demand that $m_a+m_{h_{125}}=m_A$.

The dependence on the scalar mixing angles and masses is due to the correlated decays of $(A\to Z h/H$), cf. Fig.~\ref{fig:Ah125as_BRs}. Since the coupling $(A\to Z h/H$) is proportional to the NSM component of $h/H$, cf. Eq.~\eqref{eq:width_hh}, the observed decrease in sensitivity for $(A\to h_{125} a$) with increasing NSM component of the scalars can be easily understood. Similarly, as the $h/H$ become heavier, $Z h/H$ channels are not kinematically accessible, and hence the sensitivity for $(A\to h_{125} a$) increases. Therefore, the dark regions shown in the bottom four panels and top two right panels of Fig.~\ref{fig:Ah125as_Vis} should be understood to denote regions mainly driven by where the $(A\to Z h/H$) decay is sufficiently suppressed so as to not impact the sensitivity to the $(A\to h_{125} a$) channel.

Fig.~\ref{fig:Hh125hs_Vis} shows the variations in ($gg \to H\to h_{125} h$) and Fig.~\ref{fig:Hh125hs_BRs} the most relevant braching ratios. The same qualitative behavior for the cross-sections and LHC sensitivities is observed as in Fig.~\ref{fig:Ah125as_Vis}, with the obvious interchange of the masses and mixing angles for the scalars and pseudoscalars. The overall reduction in LHC sensitivities/cross sections can be understood from the fact that even for $m_A=m_H$, the gluon fusion production cross section for the scalar $H$ can be almost a factor of 2 smaller than the production cross section for the pseudoscalar $A$. 

As can be seen from these plots, the LHC will be sensitive to large regions of the parameter space in this scenario, even with only 300\,fb$^{-1}$ of data. However, we stress that whereas more data will certainly lead to the exploration of a larger region of allowed parameters, to optimize coverage, complimentary and correlated search channels such as $(A\to Z h)$ and $(H\to Za)$ should be utilized as well. 

\FloatBarrier

\subsubsection{$Z$-Phobic: $h_{125}$+Invisible ($H\to h_{125} a$ and $A\to h_{125} h$)}
\label{sec:h125S_xx}

\begin{figure}[hp]
   \begin{centering}
      \includegraphics[width=2.5in]{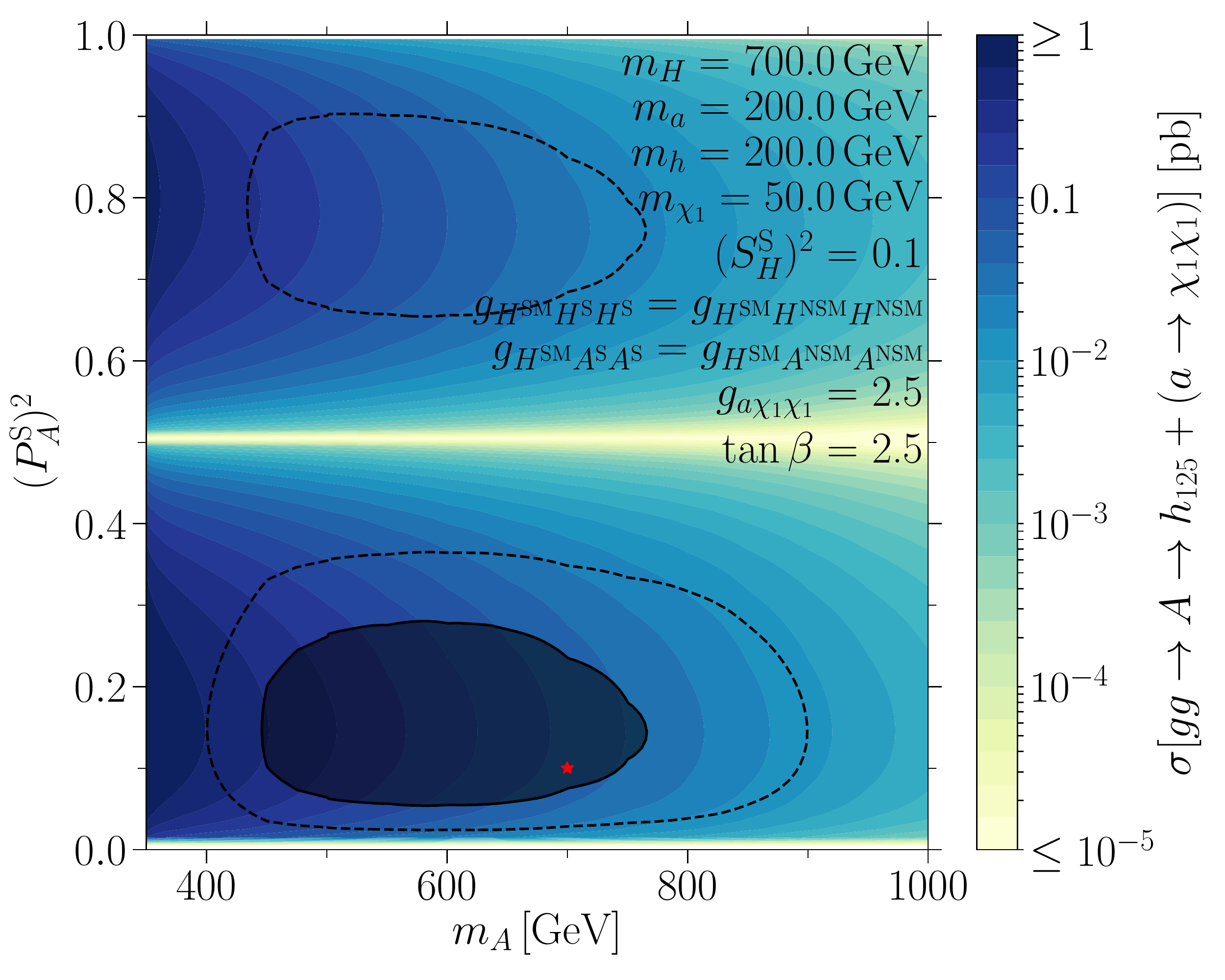}
      \hspace{.5in}
      \includegraphics[width=2.5in]{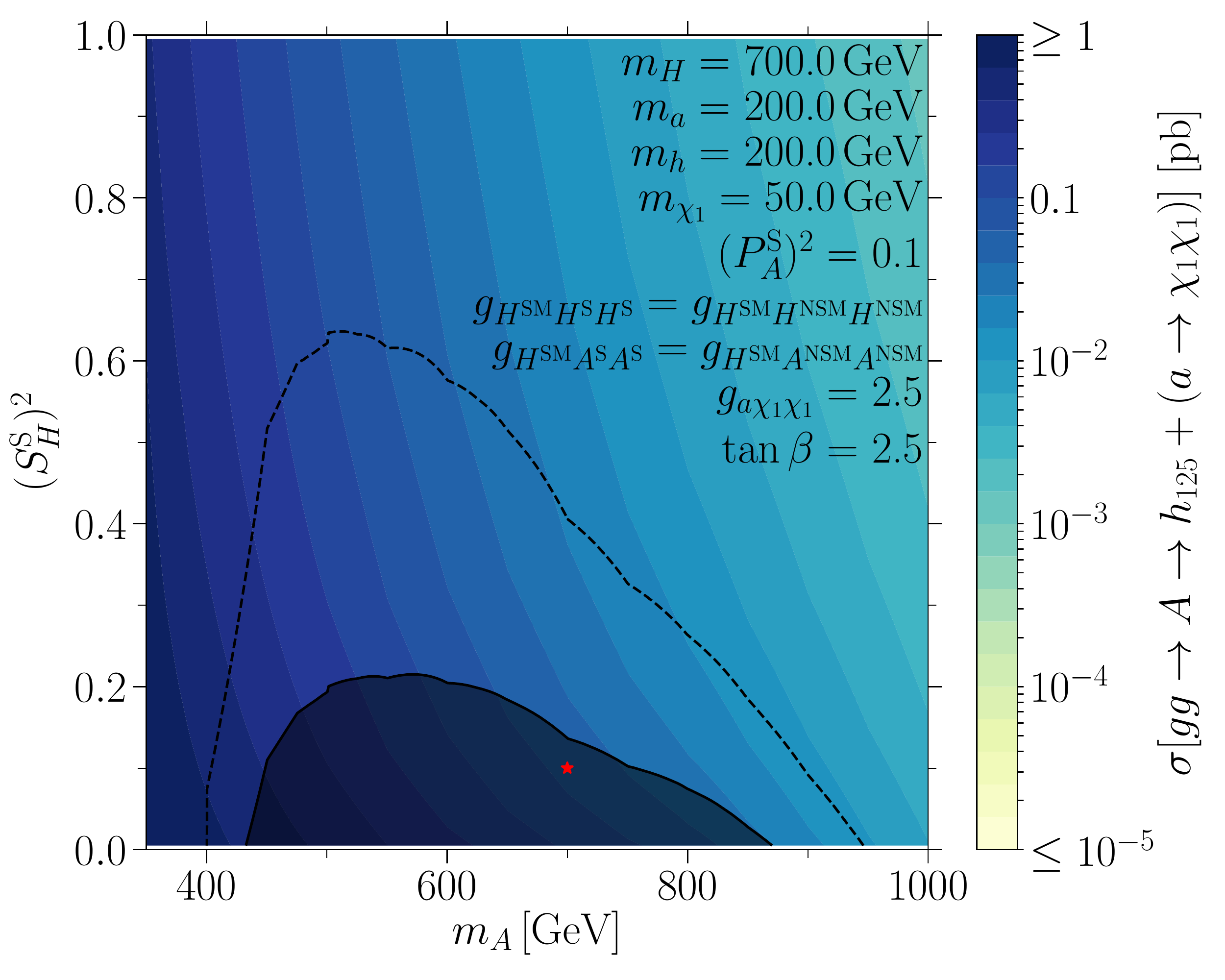}

      \includegraphics[width=2.5in]{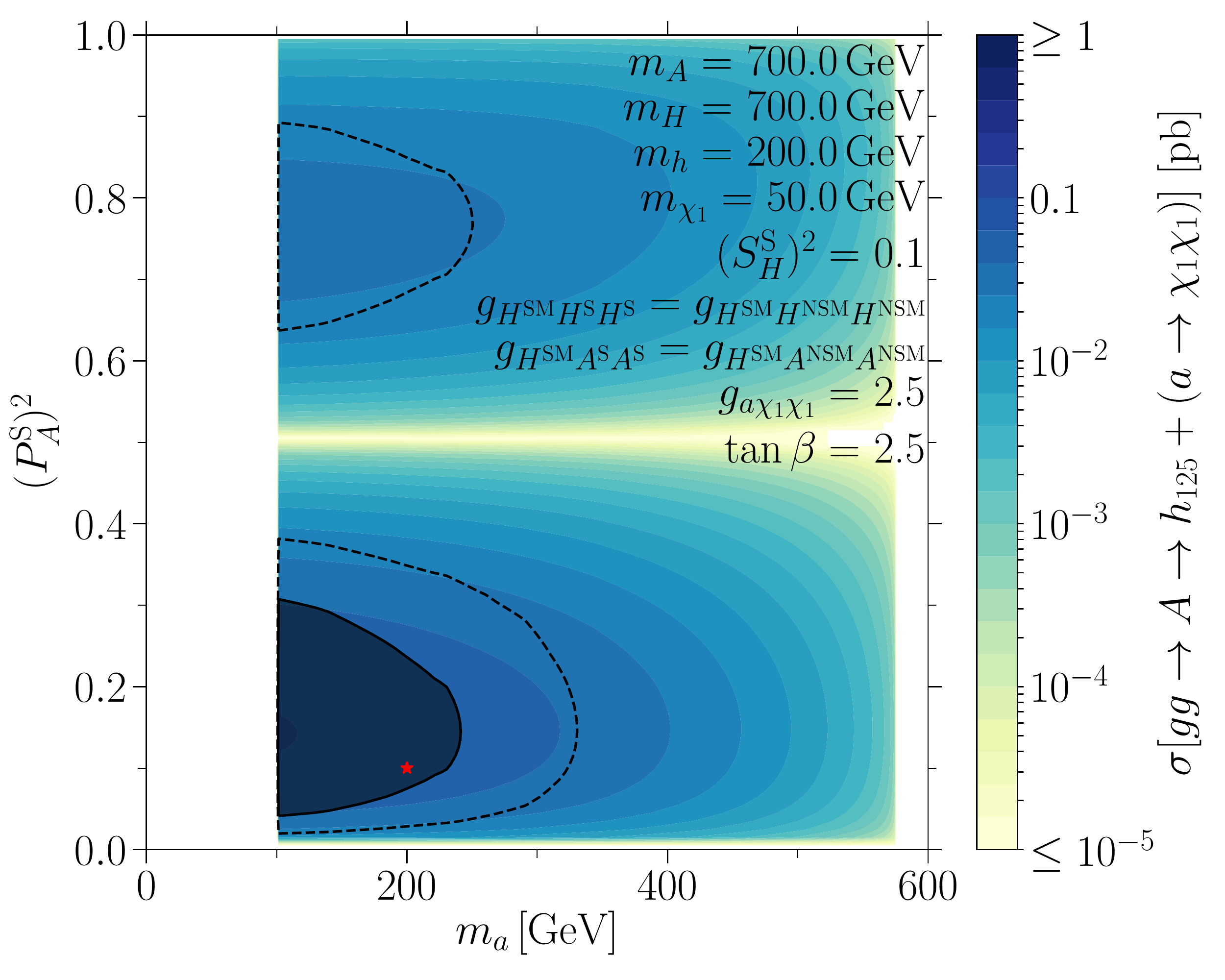}
      \hspace{.5in}
      \includegraphics[width=2.5in]{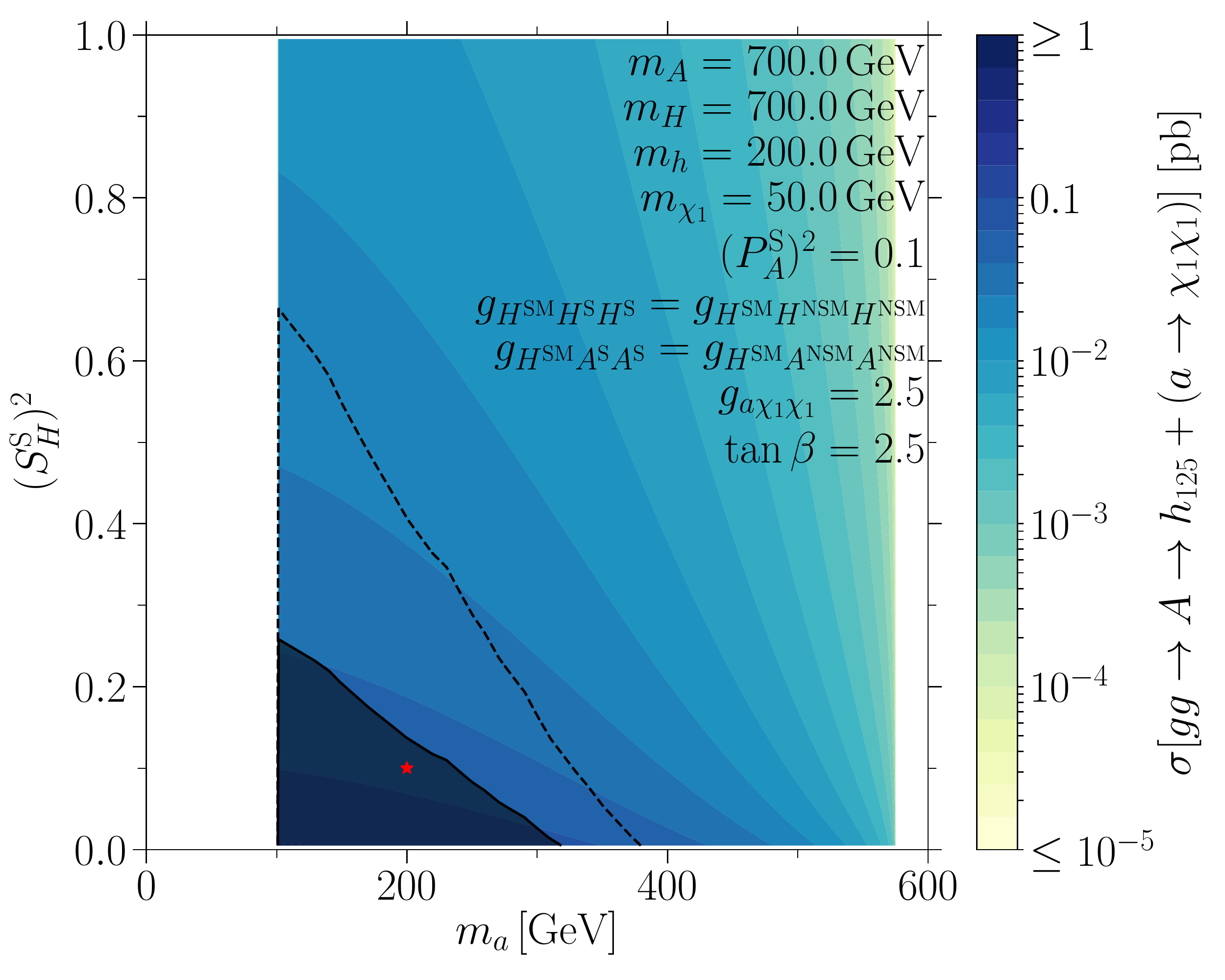}

      \includegraphics[width=2.5in]{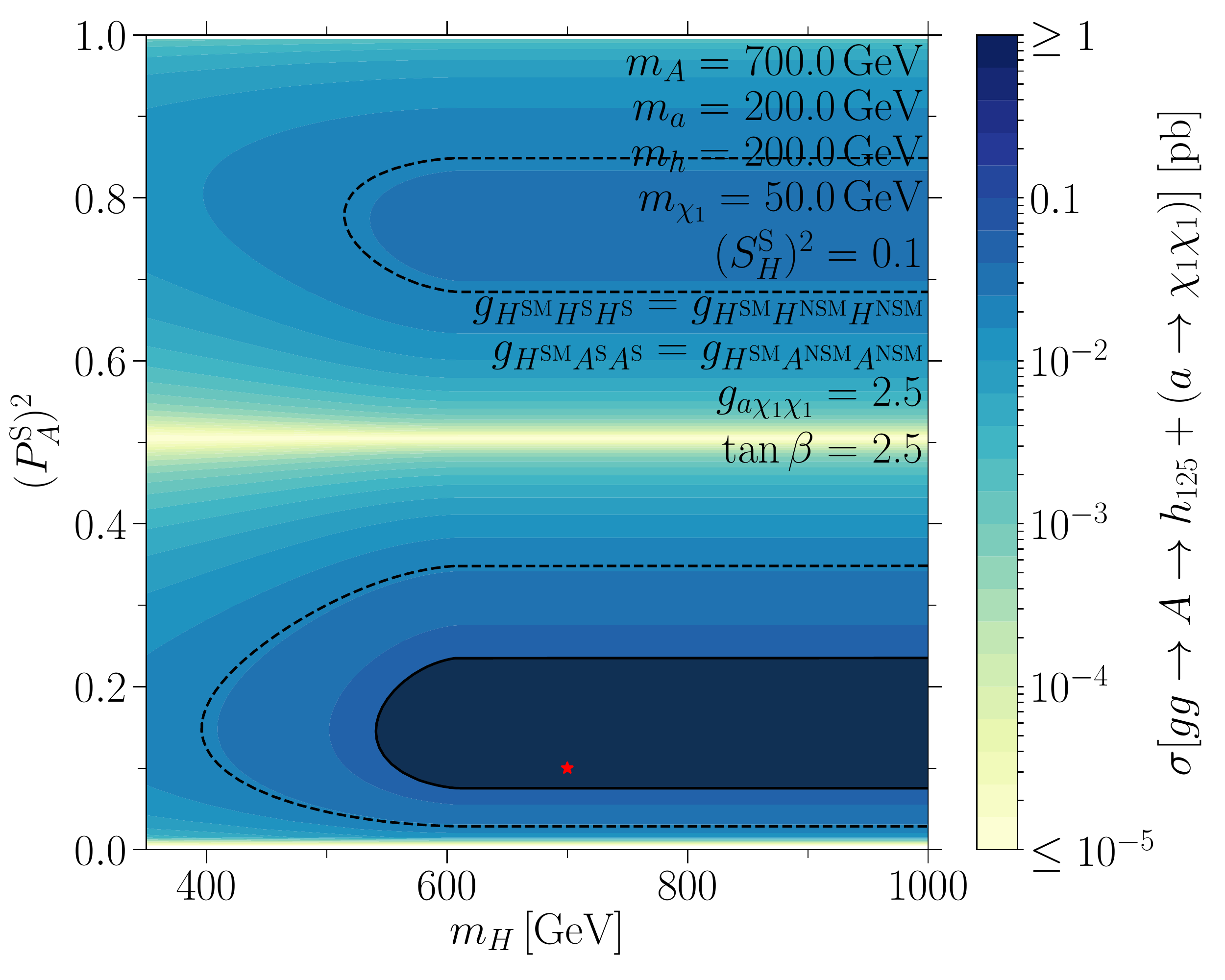}
      \hspace{.5in}
      \includegraphics[width=2.5in]{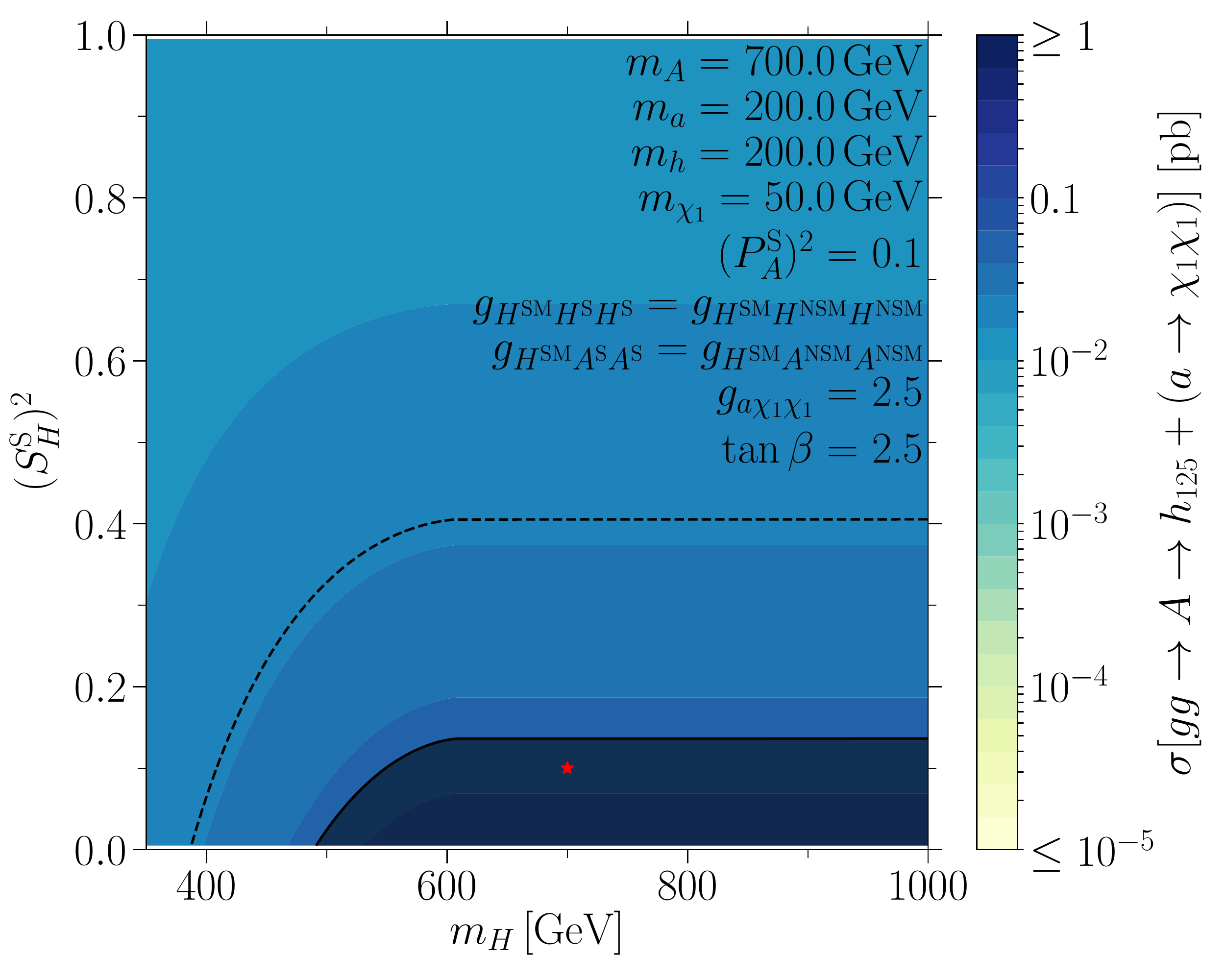}

      \includegraphics[width=2.5in]{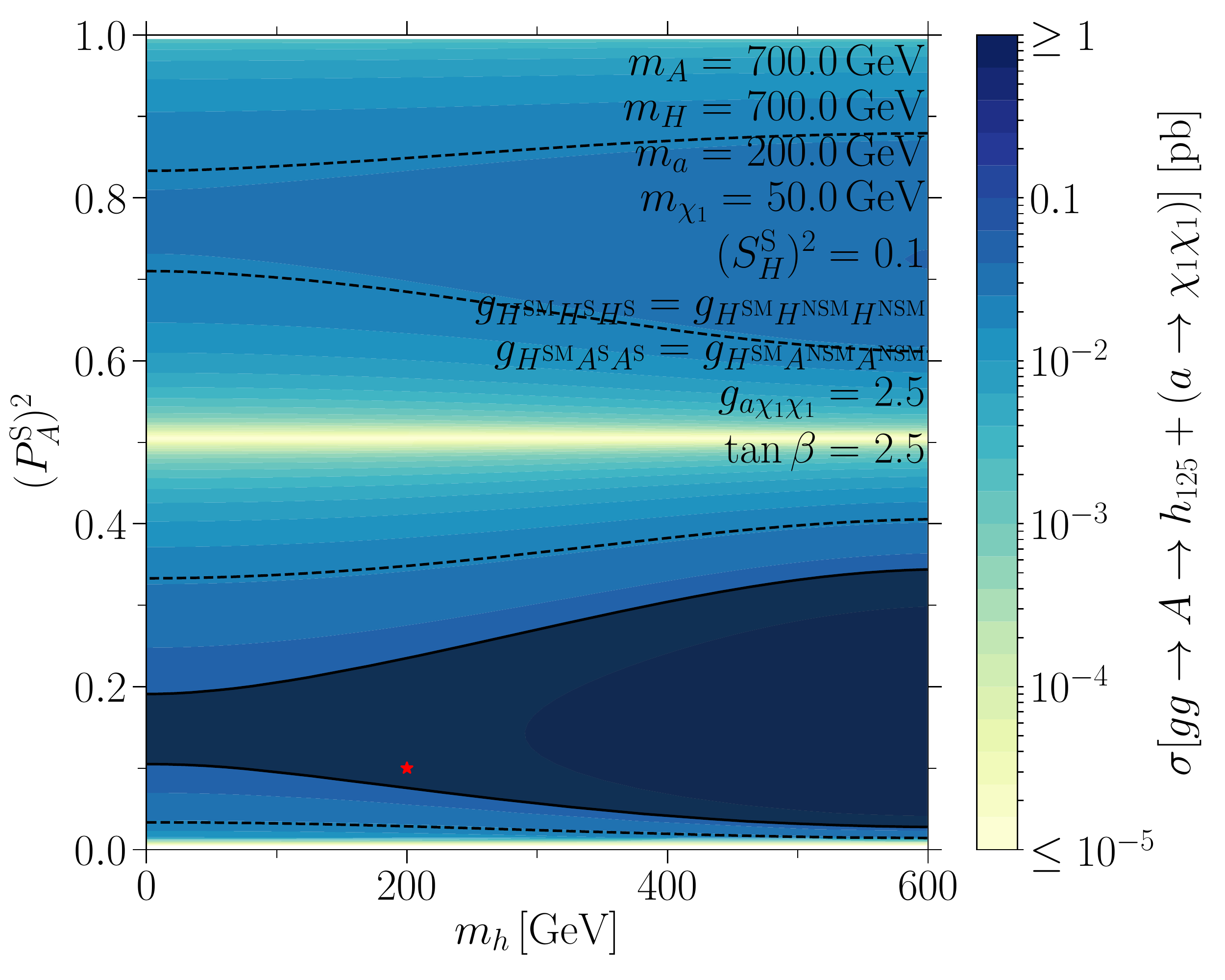}
      \hspace{.5in}
      \includegraphics[width=2.5in]{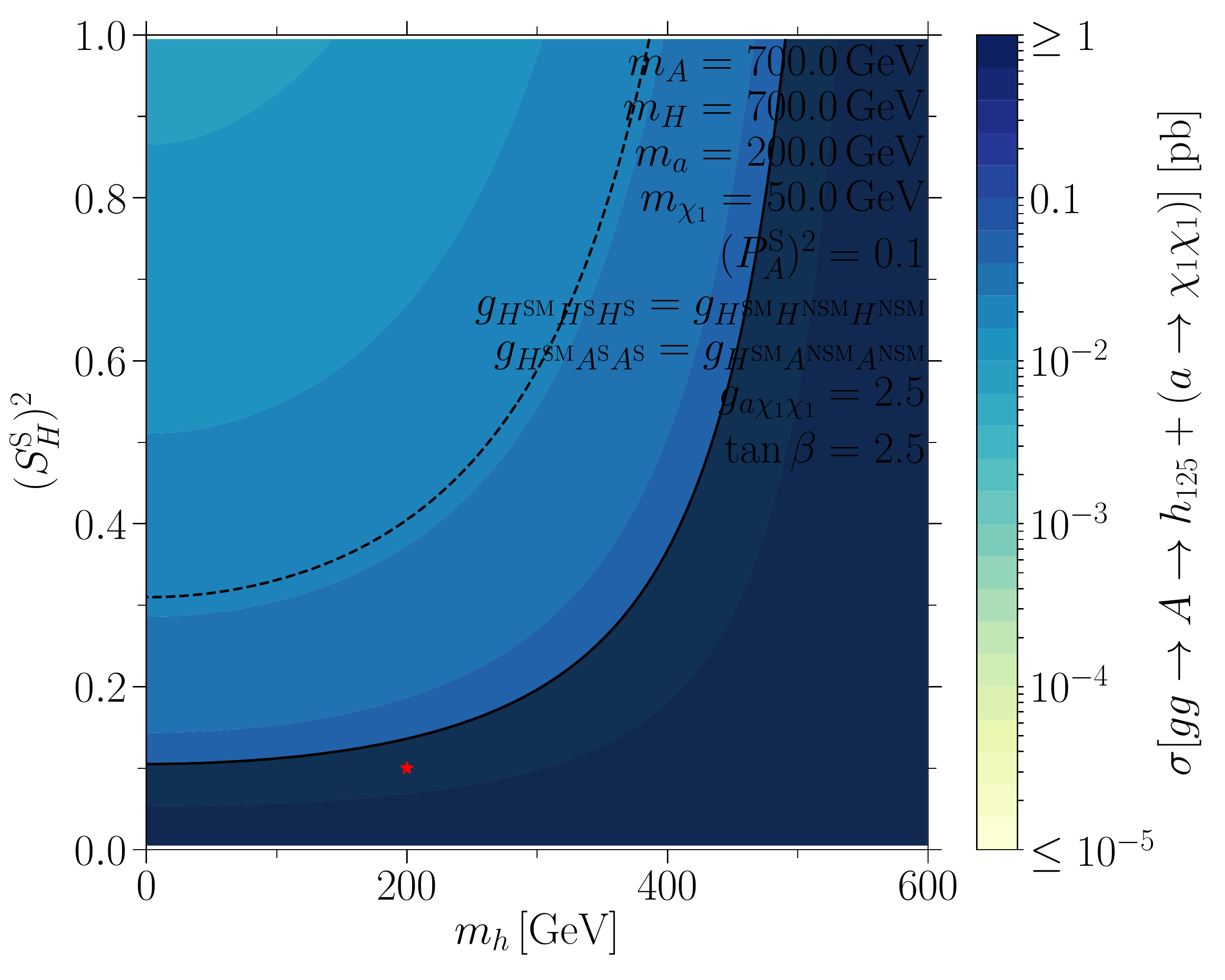}

      \caption{Cross sections and projected LHC sensitivities with 300\,fb$^{-1}$ of data for ($gg \to A \to h_{125} a\to 2b +\cancel{\it{E}}_{T} $) in the planes of the relevant masses and mixing angles. The color scale and contours denote cross-sections as labeled by the color bars. The dark shaded regions denote regions where the cross sections exceeds the projected LHC sensitivity. The dashed black lines denote cross sections a factor of two smaller. The red stars indicate the benchmark point from the second column in Table~\ref{tab:BP_h125S}. }
      \label{fig:Ah125as_Inv}
   \end{centering}
\end{figure}

\begin{figure}[hp]
   \begin{centering}
      \includegraphics[width = 2.5in]{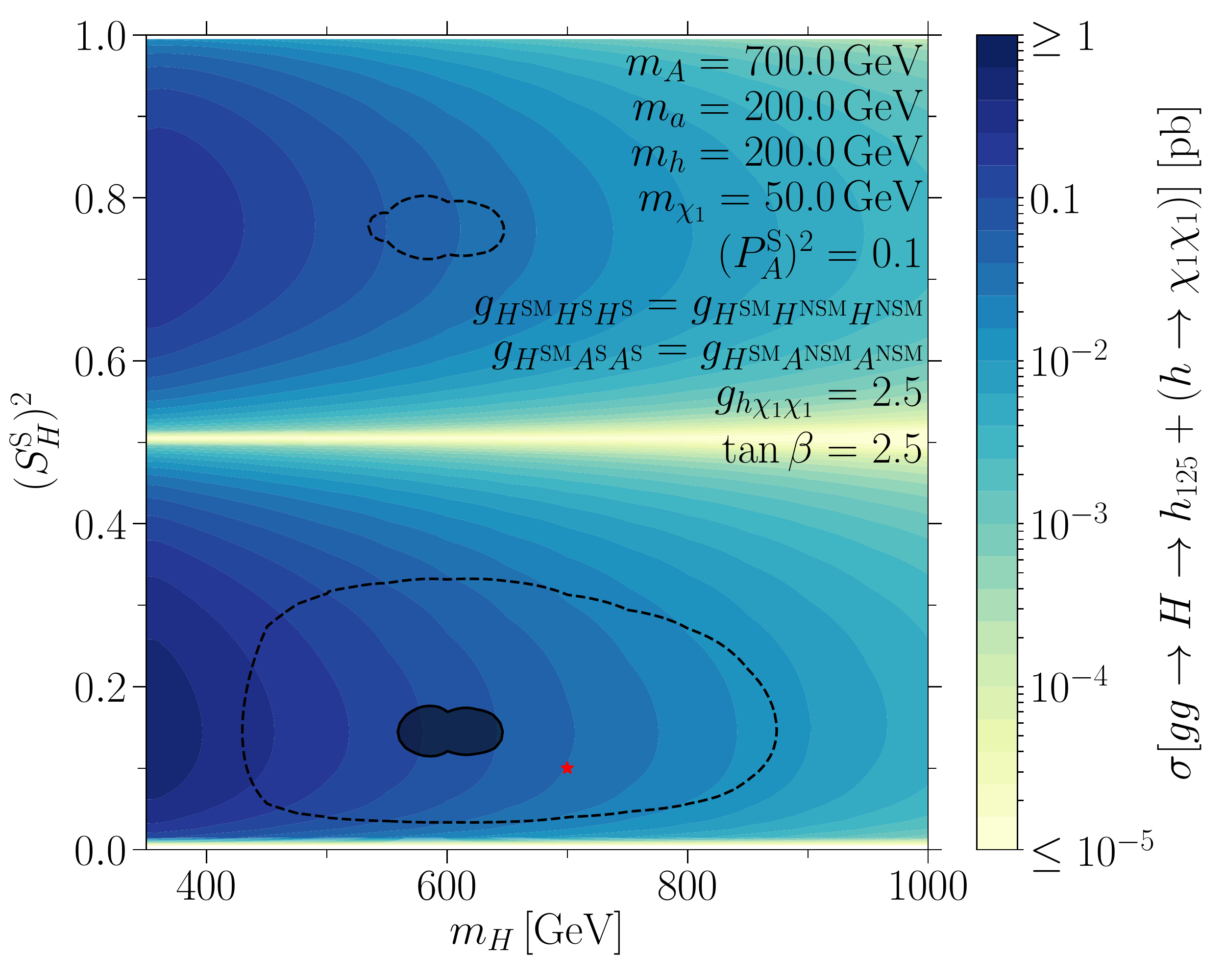}
      \hspace{.5in}
      \includegraphics[width = 2.5in]{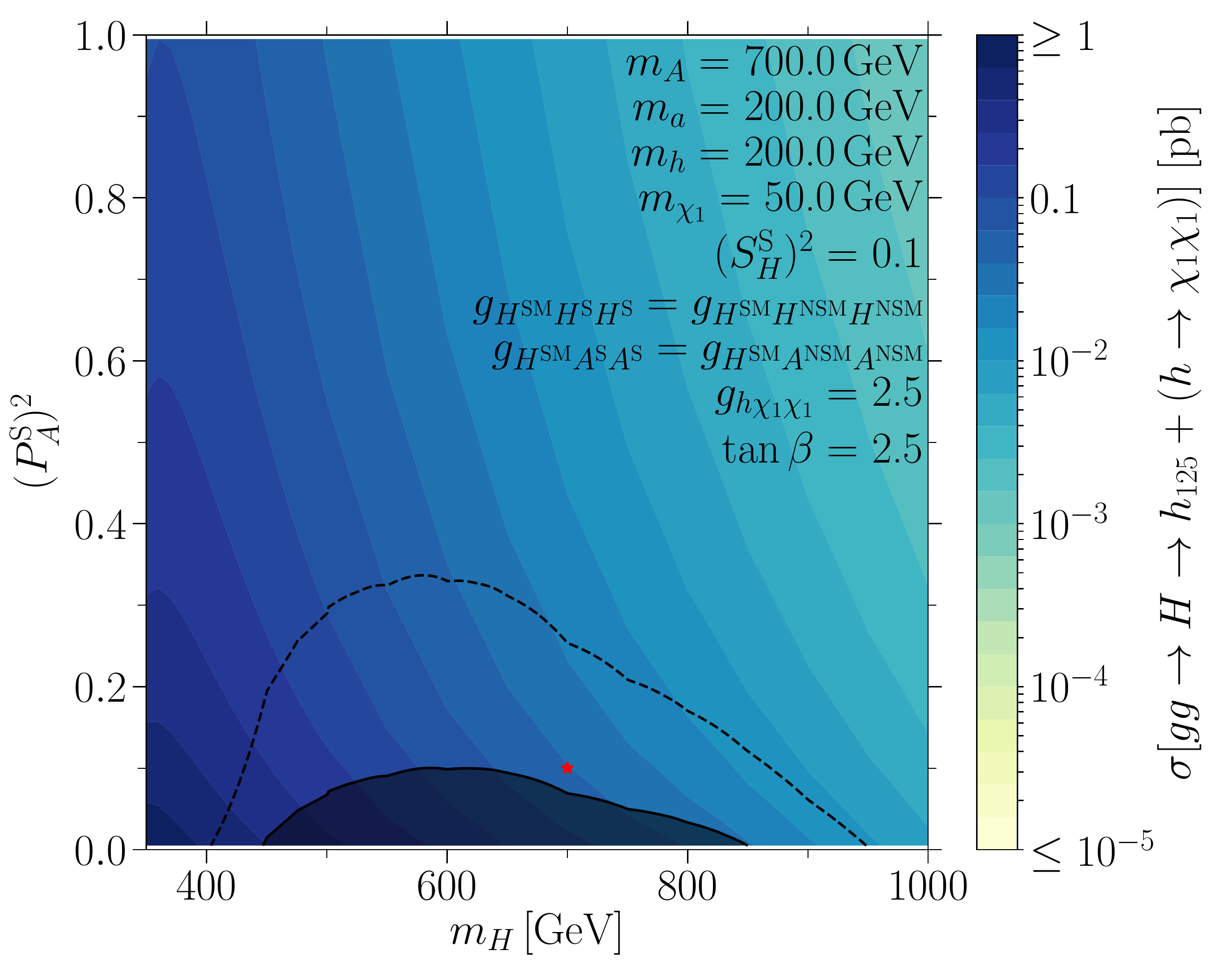}

      \includegraphics[width = 2.5in]{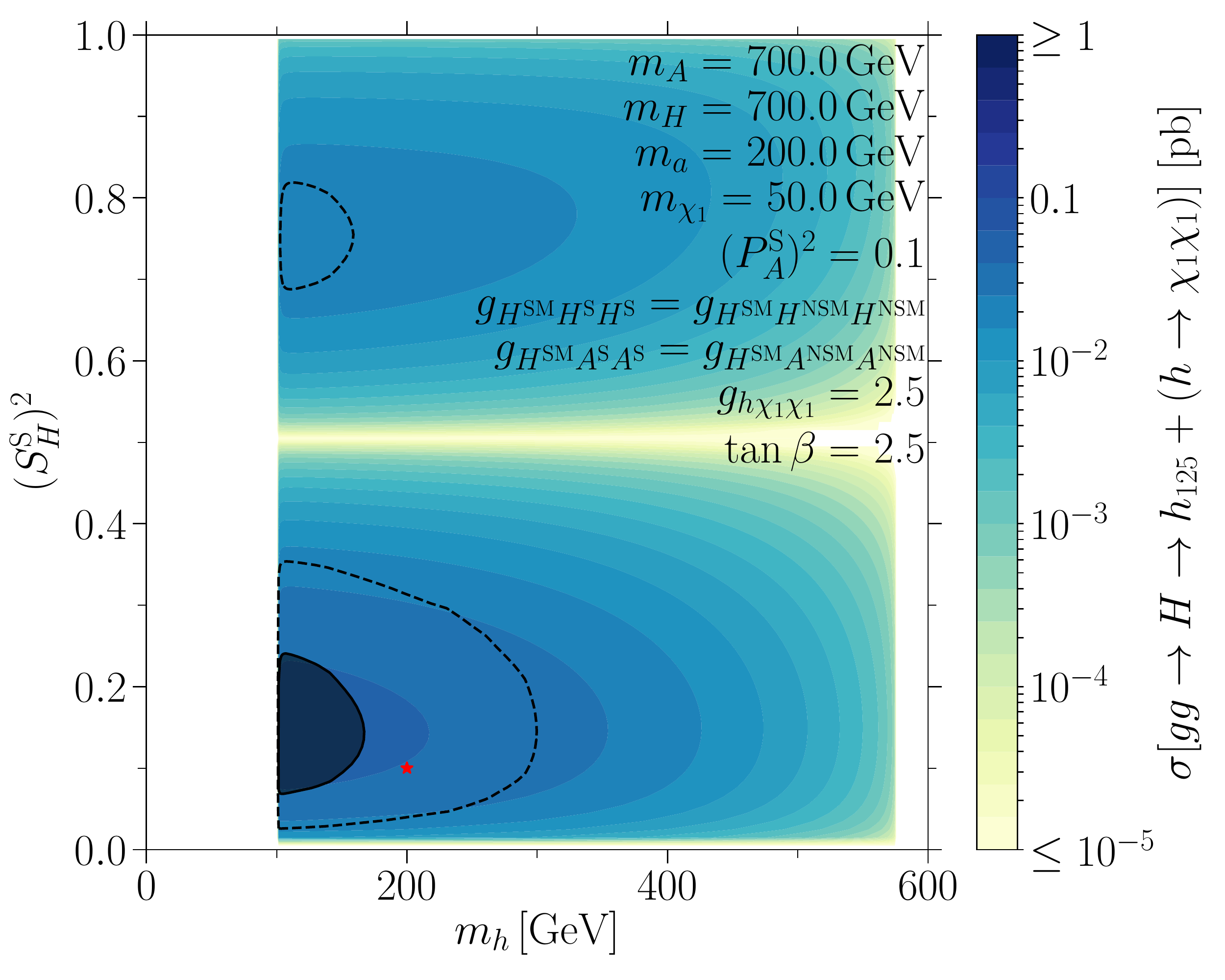}
      \hspace{.5in}
      \includegraphics[width = 2.5in]{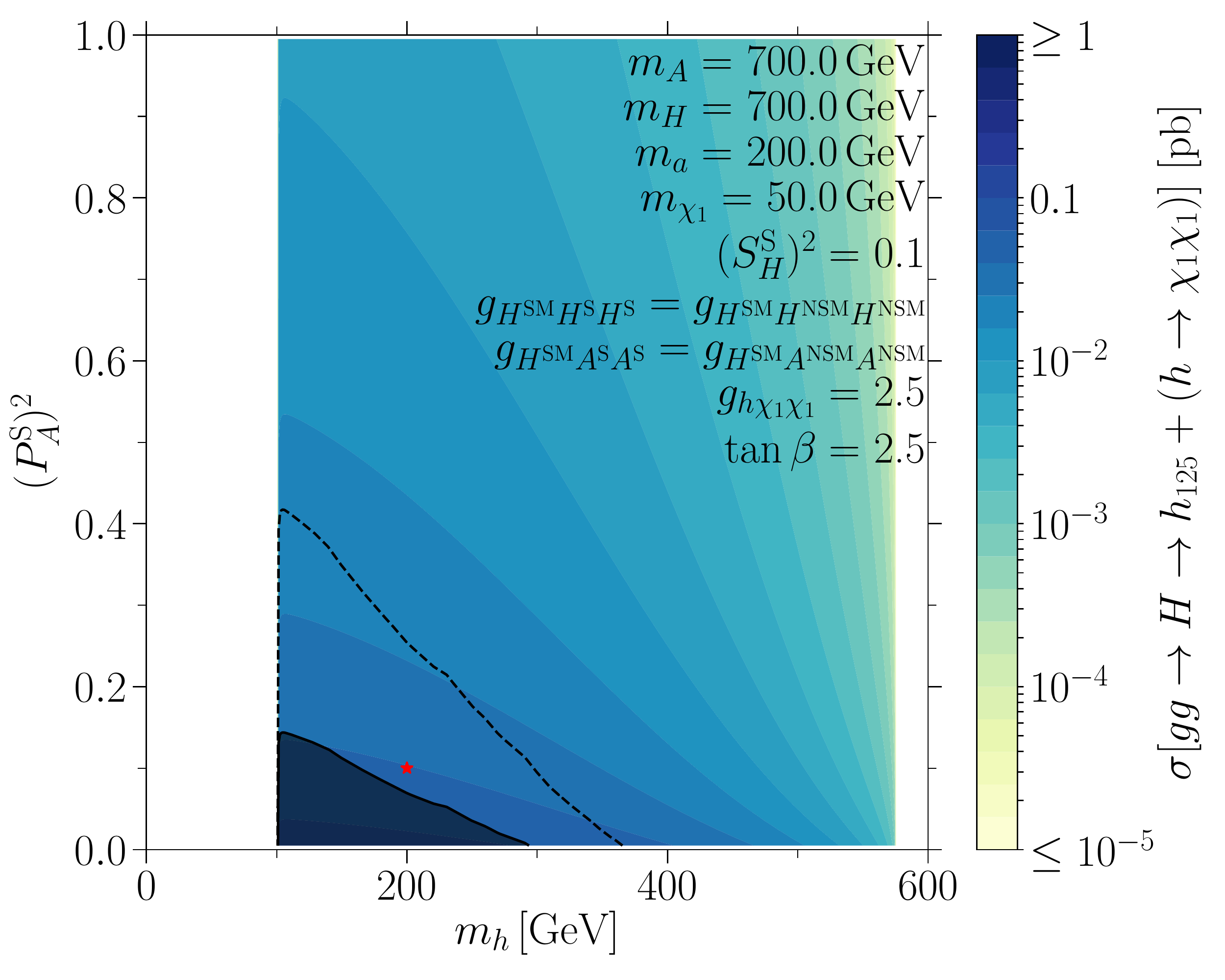}

      \includegraphics[width = 2.5in]{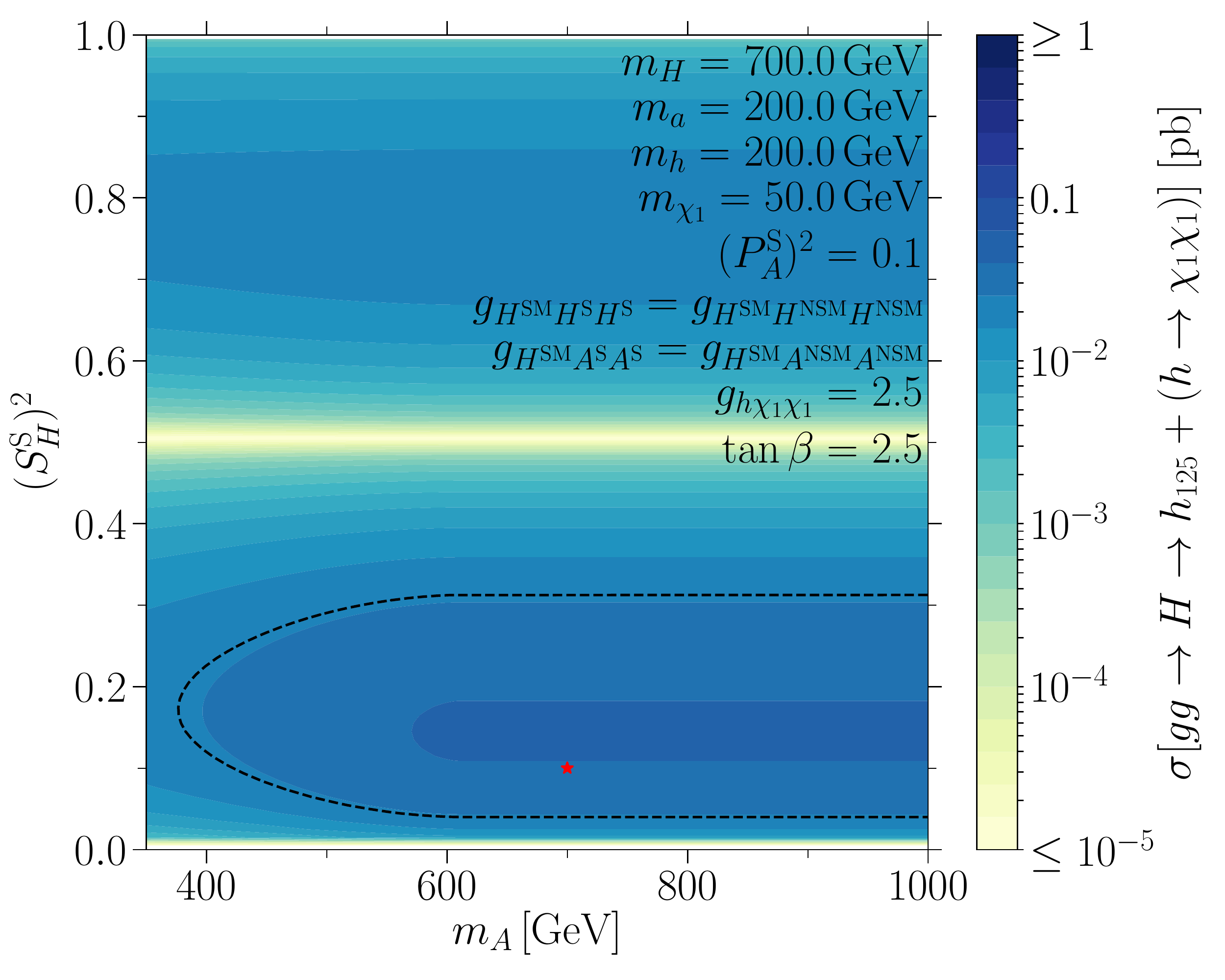}
      \hspace{.5in}
      \includegraphics[width = 2.5in]{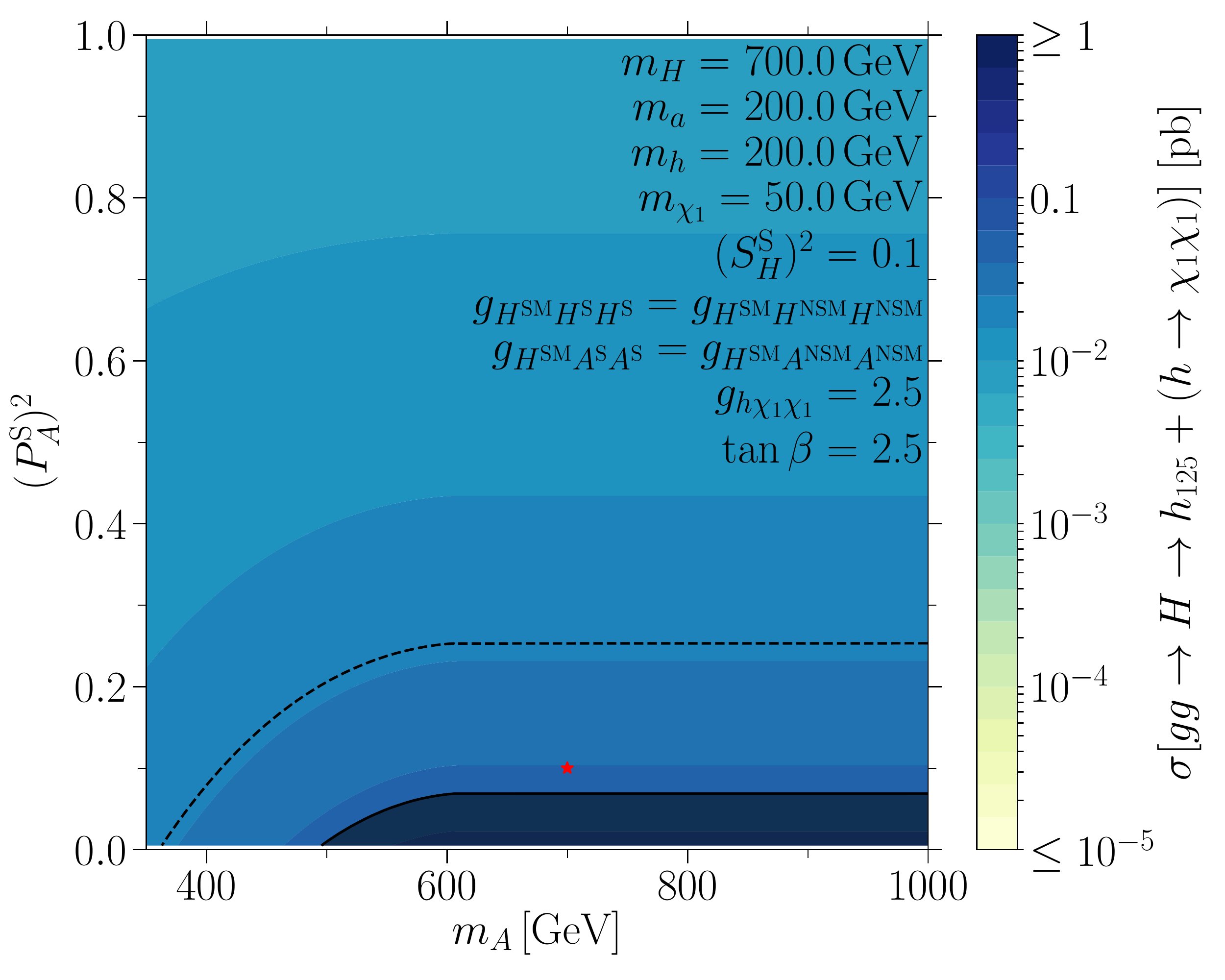}

      \includegraphics[width = 2.5in]{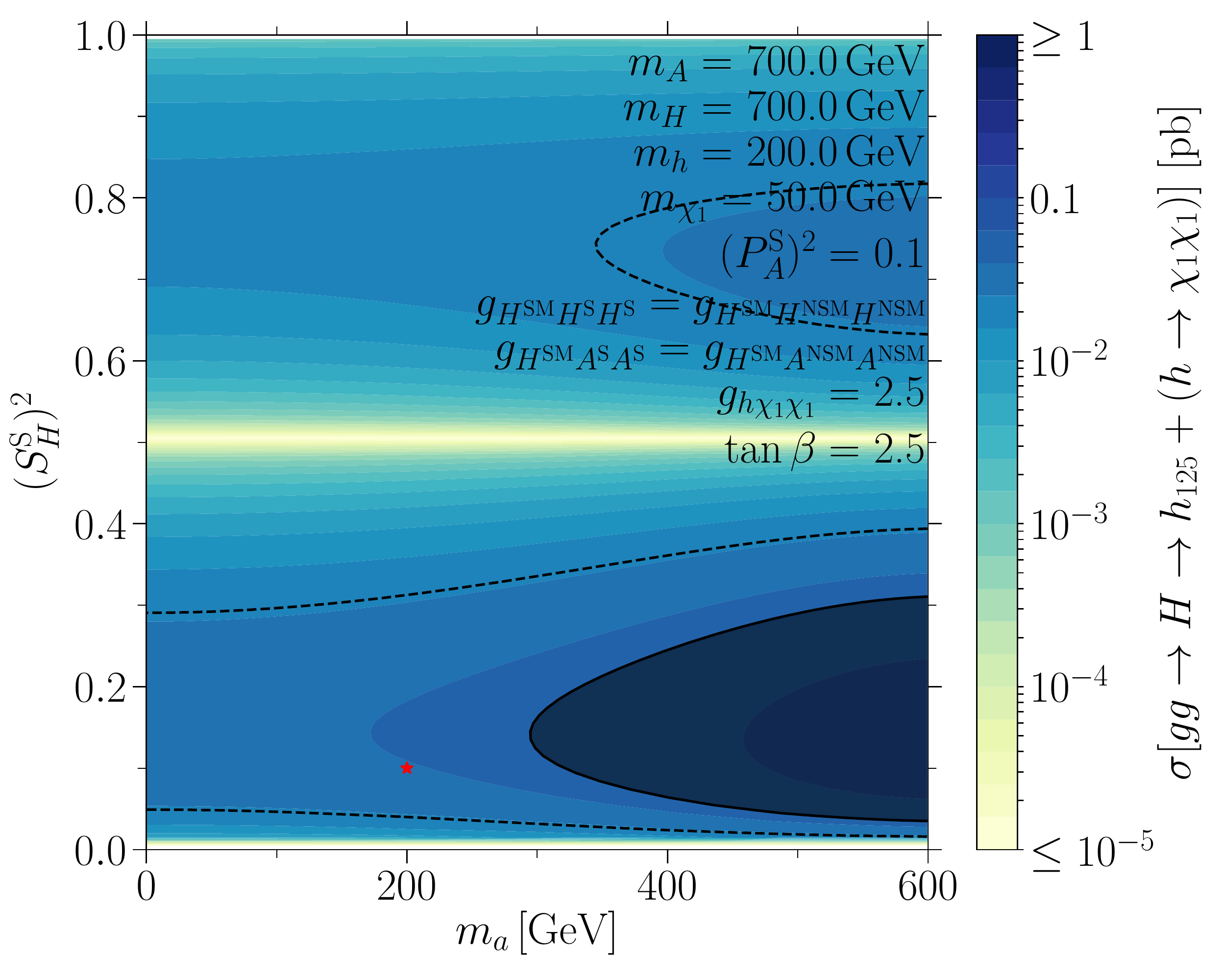}
      \hspace{.5in}
      \includegraphics[width = 2.5in]{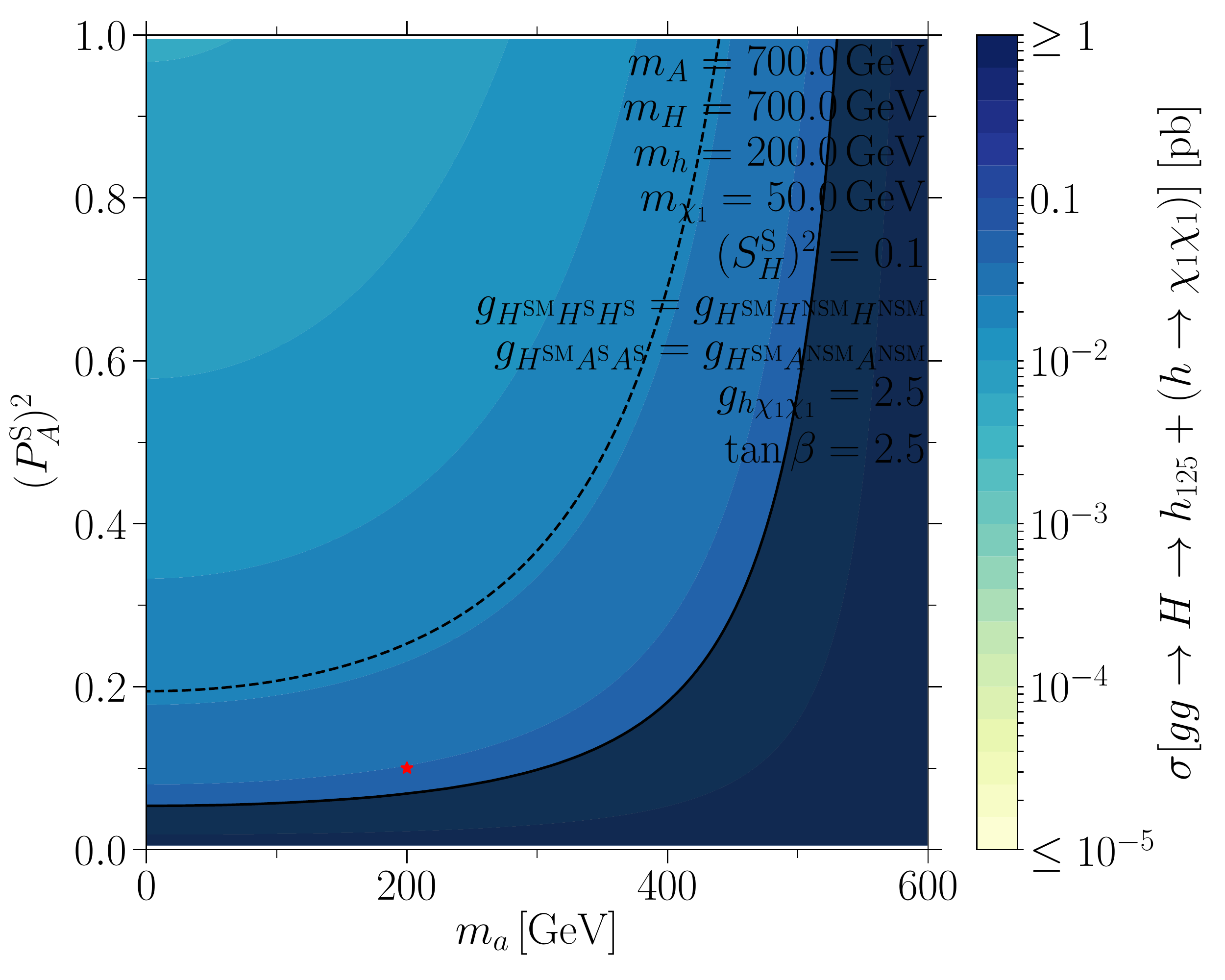}

      \caption{Cross sections and projected LHC sensitivities with 300\,fb$^{-1}$ of data for ($gg \to H \to h_{125} h\to 2b +\cancel{\it{E}}_{T} $) in the planes of the relevant masses and mixing angles. The color scale and contours denote cross-sections as labeled by the color bars. The dark shaded regions denote regions where the cross sections exceeds the projected LHC sensitivity. The dashed black lines denote cross sections a factor of two smaller. The red stars indicate the benchmark point from the second column in Table~\ref{tab:BP_h125S}.}
      \label{fig:Hh125hs_Inv}
   \end{centering}
\end{figure}

In Figs.~\ref{fig:Ah125as_Inv} and \ref{fig:Hh125hs_Inv} the variation in the final state cross-sections and the related LHC sensitivities, taken from Ref.~\cite{Baum:2017gbj}, are portrayed for the same channels as in the previous section, but with a large values of $g_{h\chi_1\chi_1} = g_{a\chi_1\chi_1} = 2.5$, such that BR$(h/a\to \chi_1\chi_1)\approx1$. The cross-sections shown and their behavior are very similar to what was shown/discussed in the Sec.~\ref{sec:h125S_bb}. However, somewhat smaller regions of parameter space are with in the projected sensitivity of the LHC. The cut-off for $m_{a/h}<100\,$GeV is due to fact that we fixed $m_{\chi_1}=50\,$GeV, hence, there decays of $a/h$ into pairs of $\chi_1$ are kinematically forbidden. 

While the LHC sensitivity is reduced compared to the Visible scenario presented in the previous section, upcoming runs of the LHC could still probe sizable regions of this scenario's parameter space.

\FloatBarrier

\subsubsection{$Z$-Phobic: Double Singlet Scenario ($H \to hh$, $H \to aa$, and $A \to ha$)} \label{sec:SS}

\begin{figure}[hp]
   \begin{centering}
      \includegraphics[width = 2.5in]{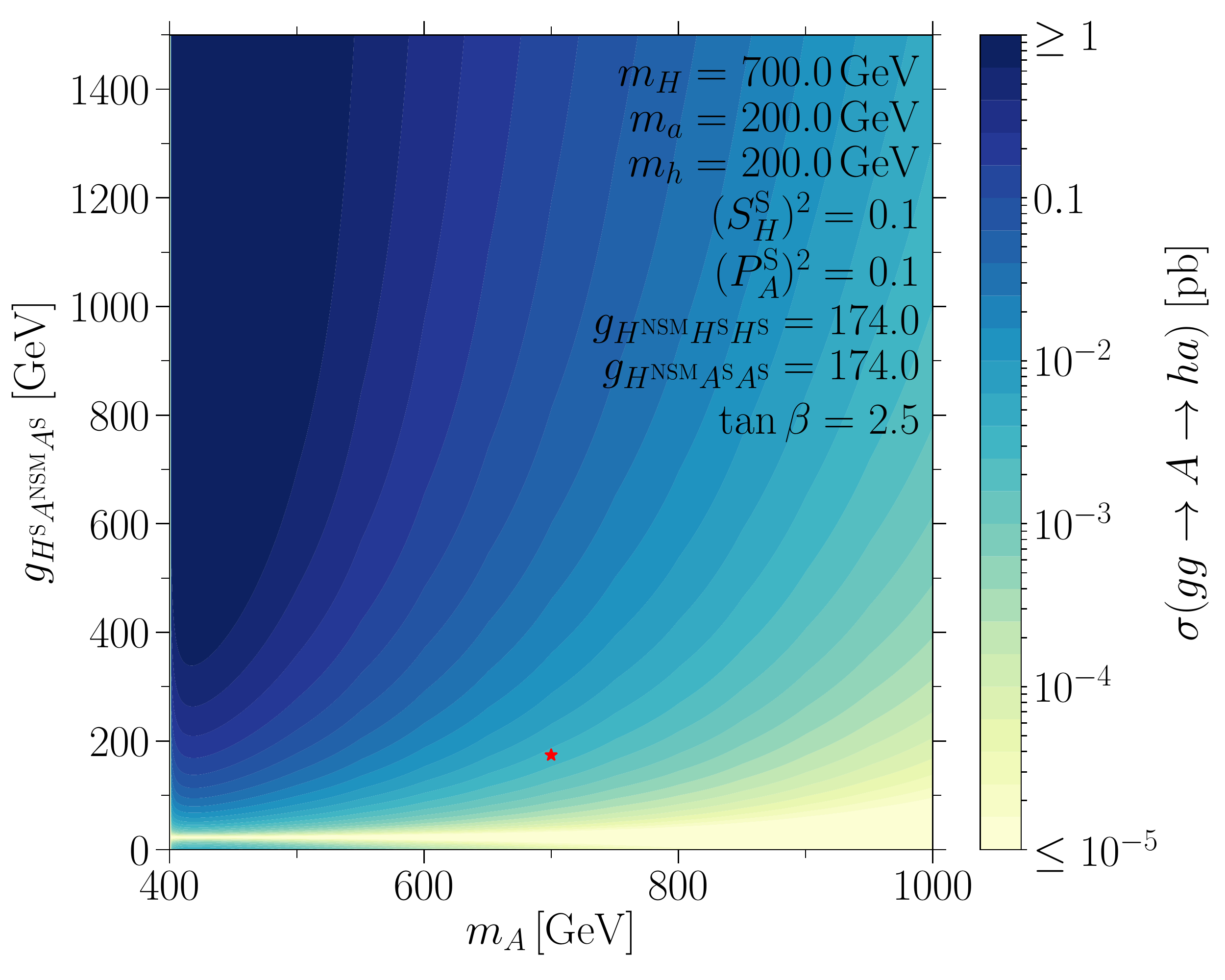}
      \hspace{.5in}
      \includegraphics[width = 2.5in]{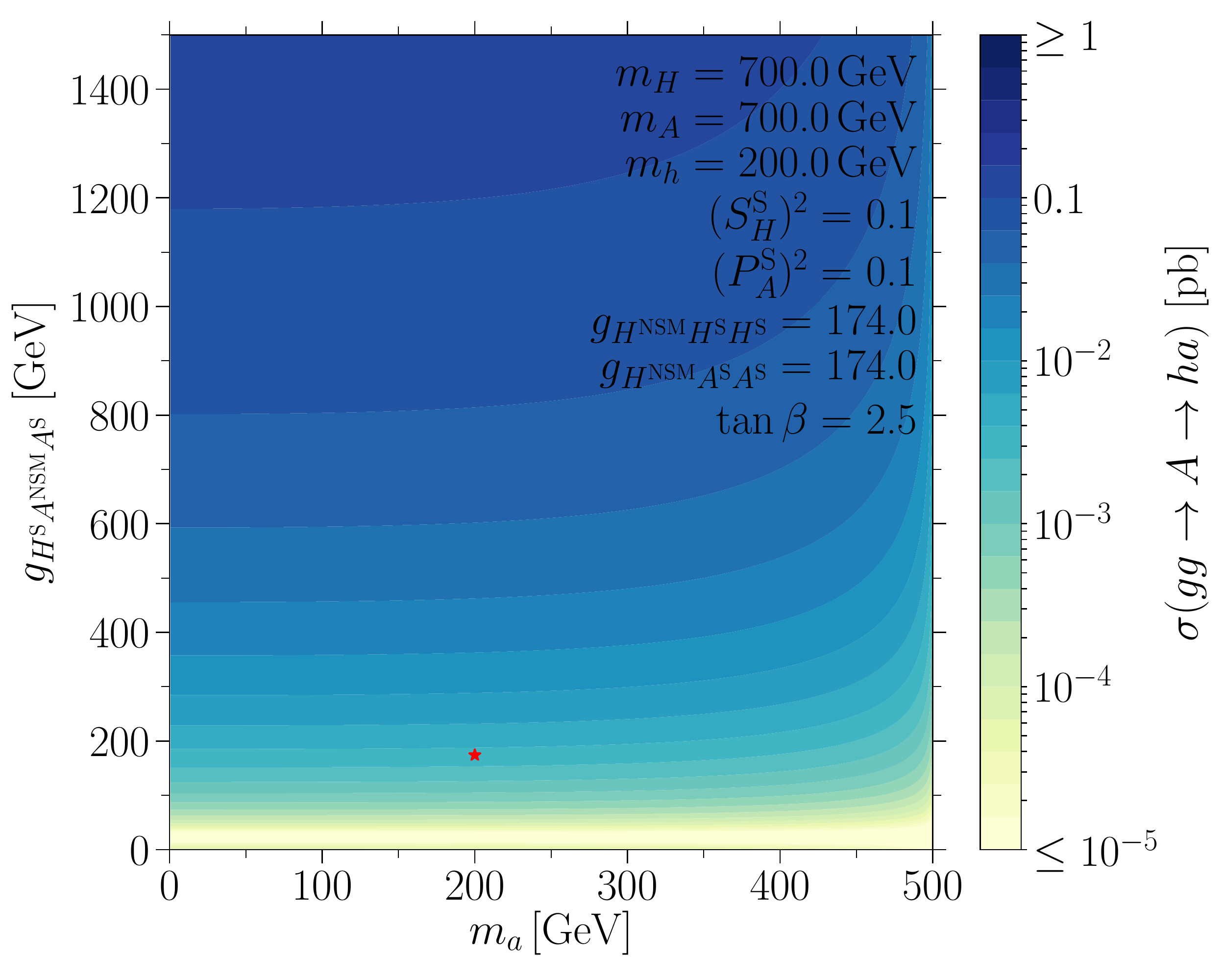}

      \includegraphics[width = 2.5in]{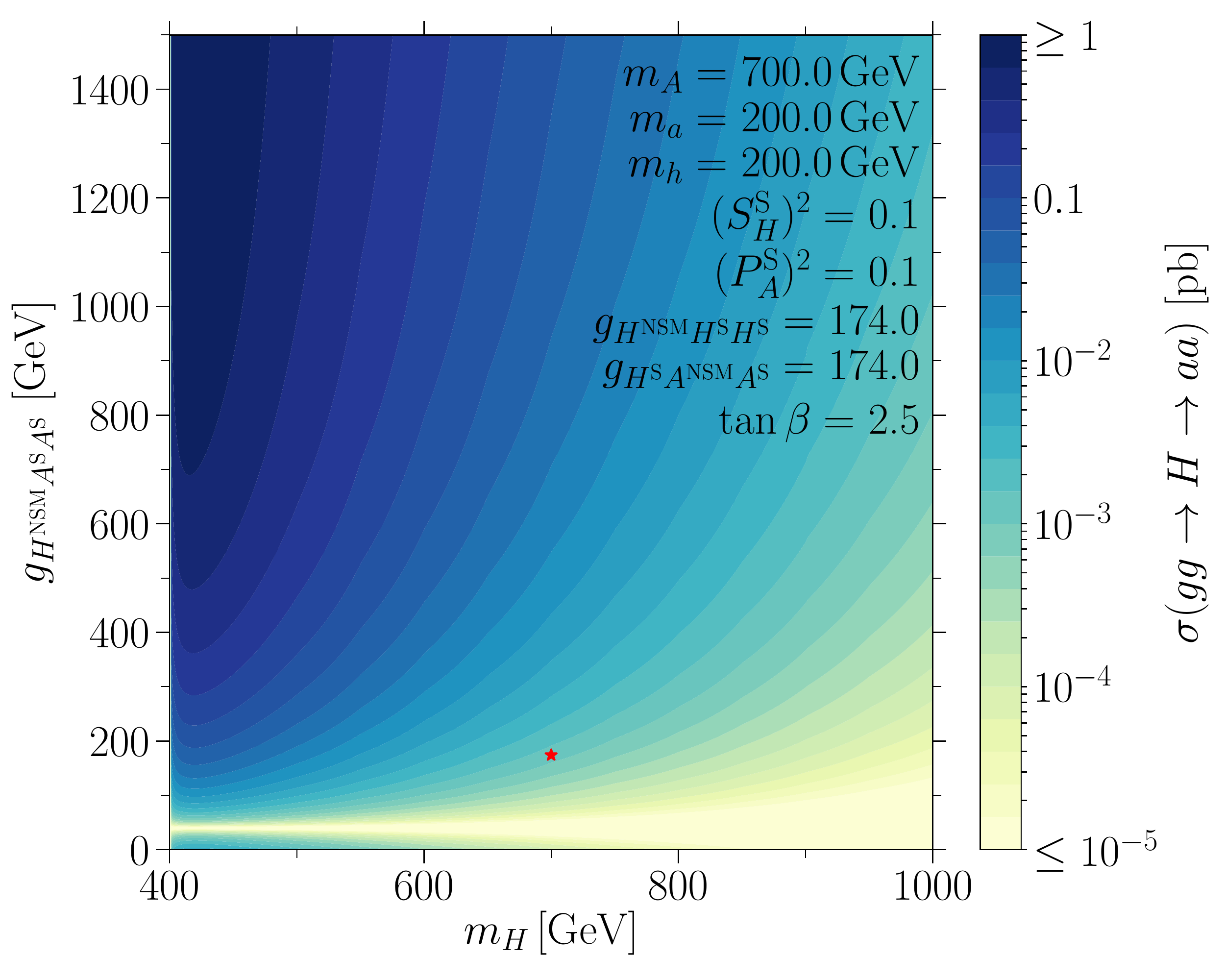}
      \hspace{.5in}
      \includegraphics[width = 2.5in]{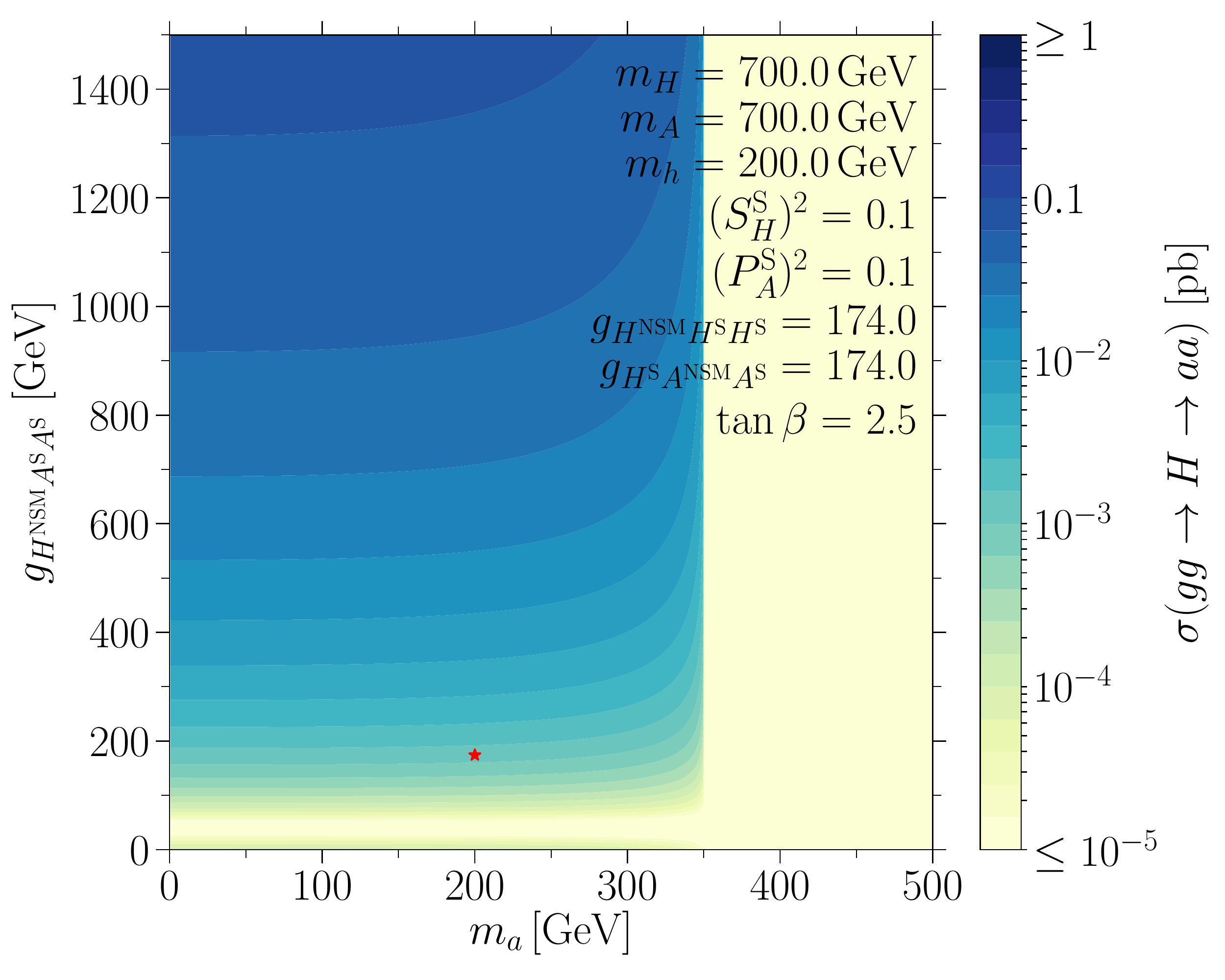}

      \includegraphics[width = 2.5in]{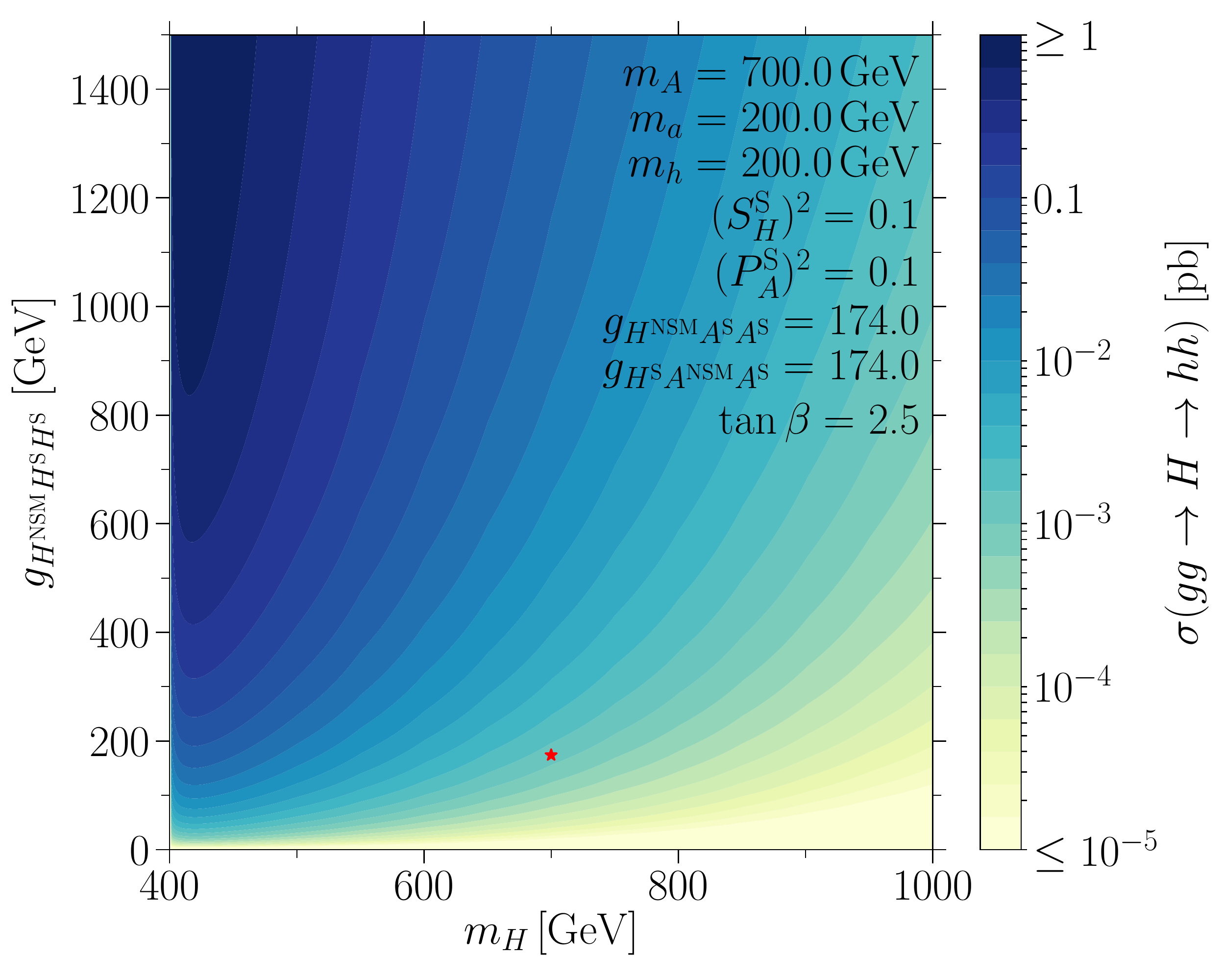}
      \hspace{.5in}
      \includegraphics[width = 2.5in]{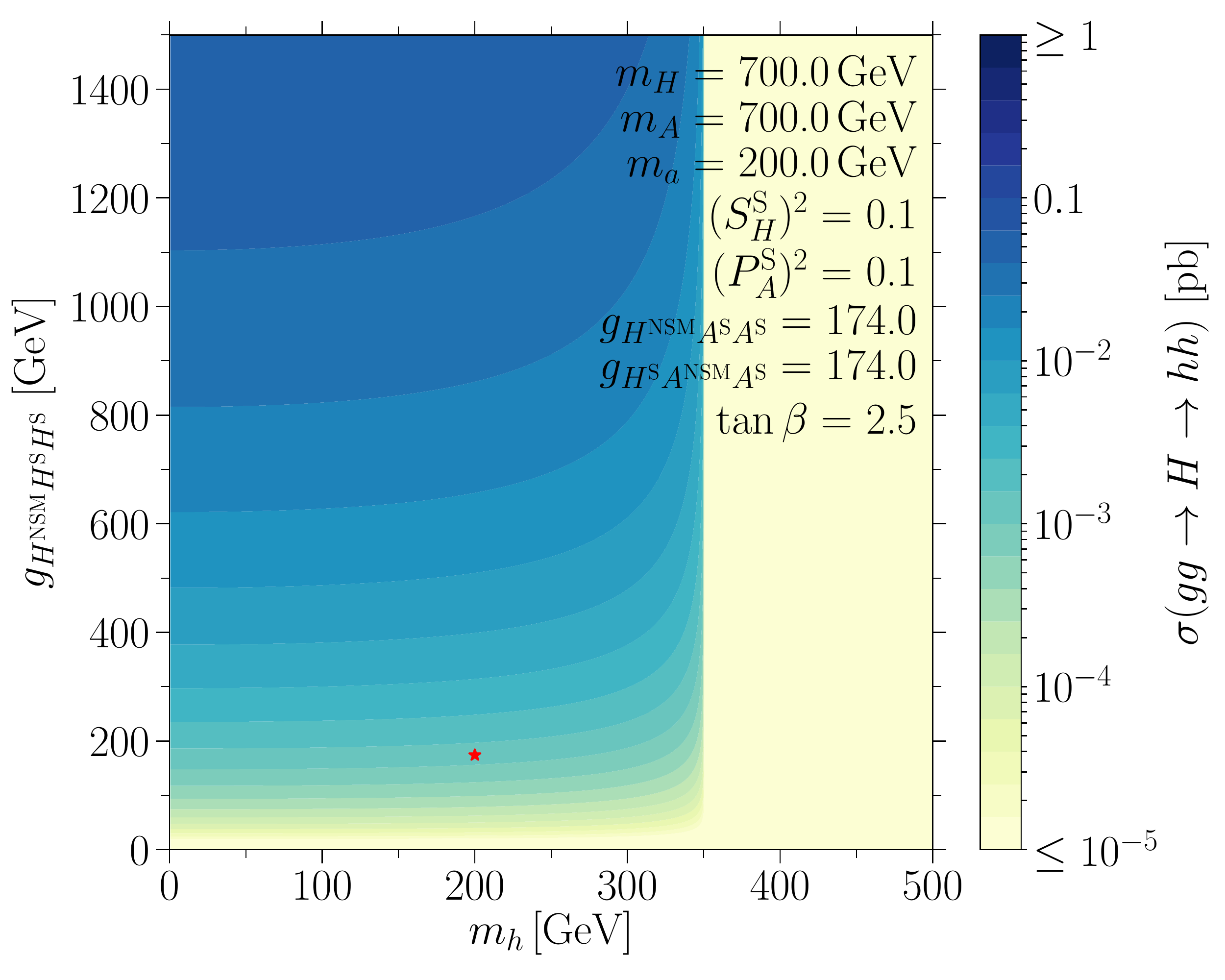}
      \caption{Cross sections as denoted by color bars for ($gg \to A \to ha$; first row), ($gg \to H \to aa$; second row), and ($gg \to H \to hh$; third row) in the planes of the most relevant masses and trilinear couplings. The red stars indicate the benchmark point from the third column in Table~\ref{tab:BP_h125S}.}
      \label{fig:HhhaaAah}
   \end{centering}
\end{figure}

\begin{figure}[t!]
   \begin{centering}
      \includegraphics[width = 2.5in]{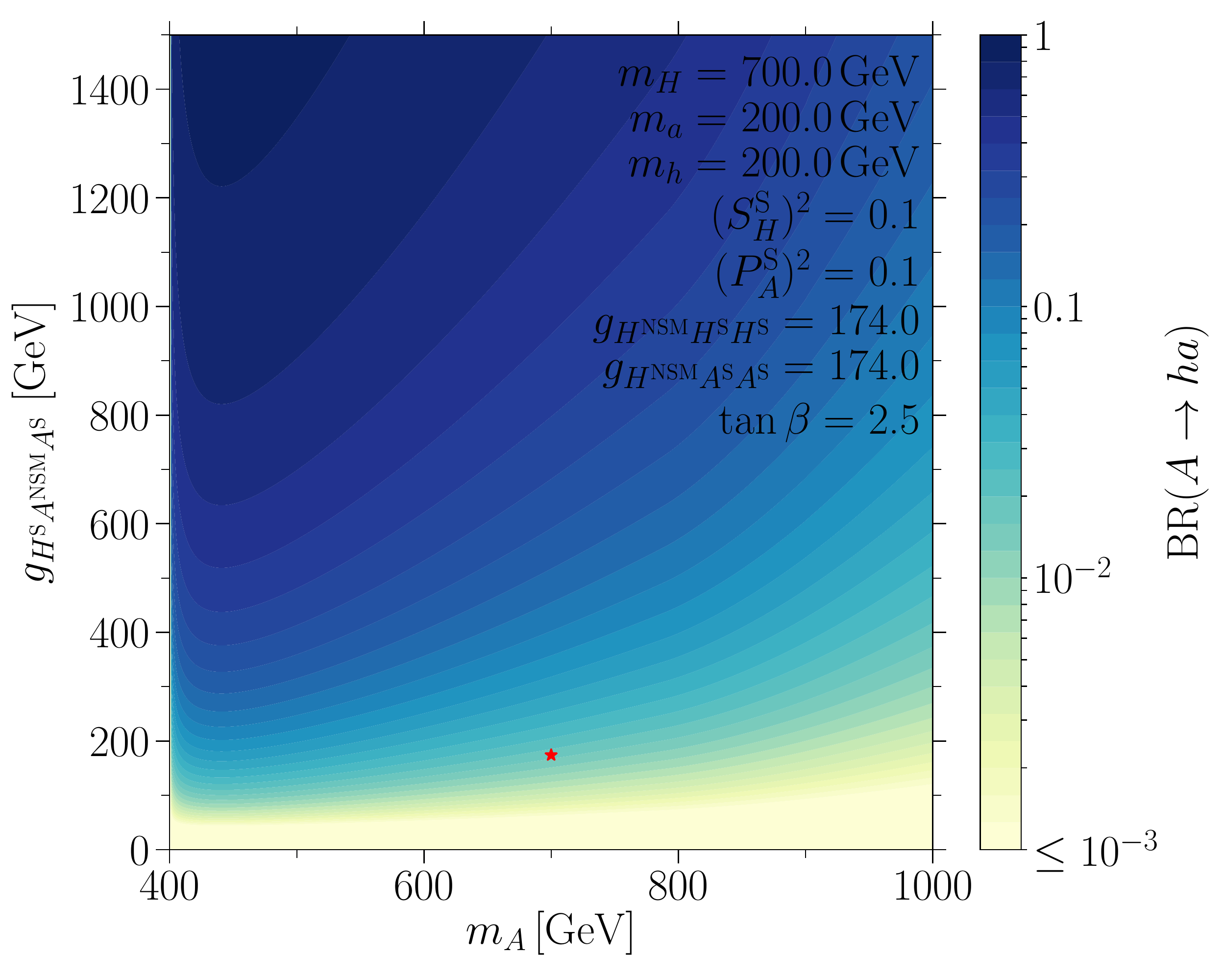}
      \hspace{.5in}
      \includegraphics[width = 2.5in]{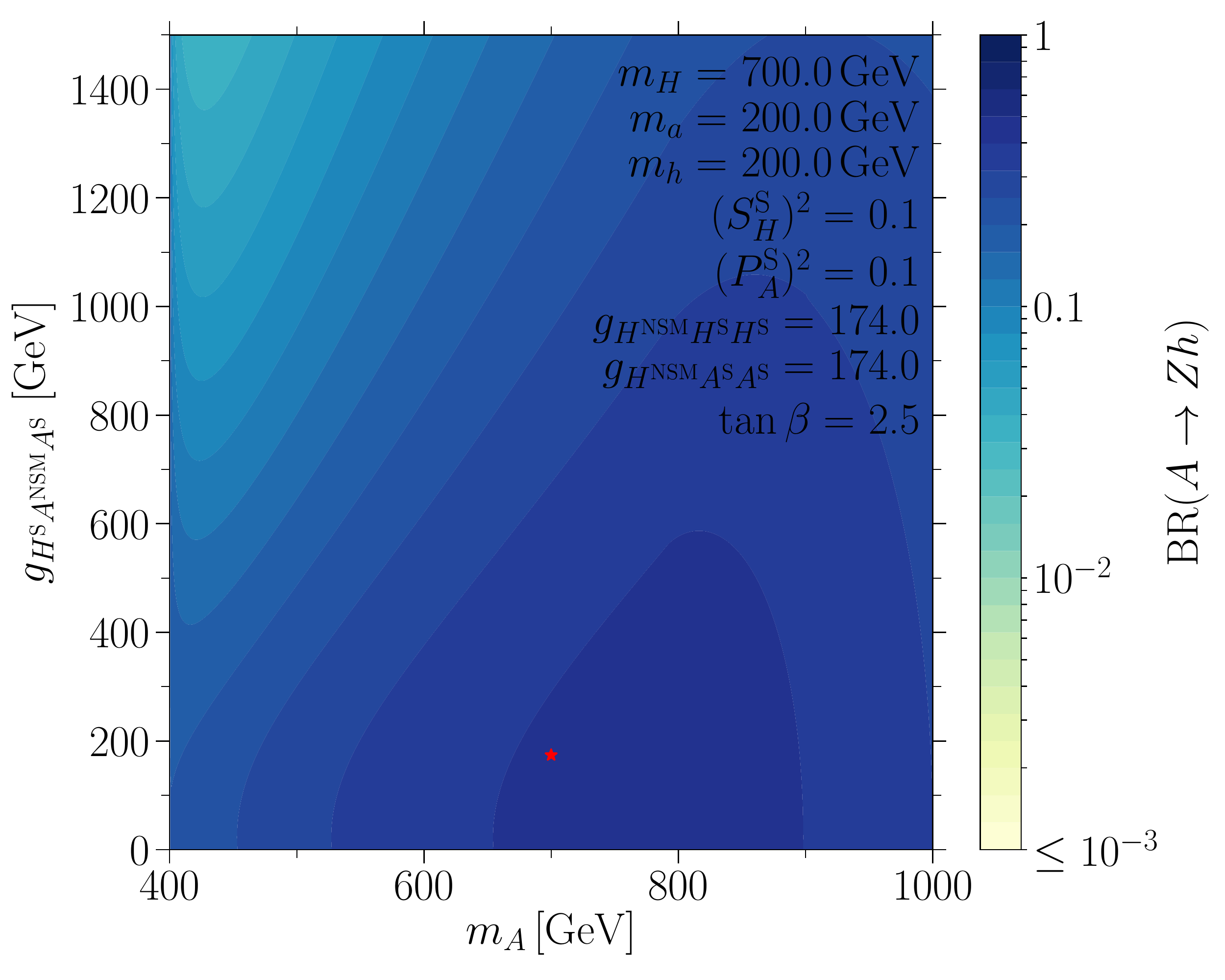}

      \includegraphics[width = 2.5in]{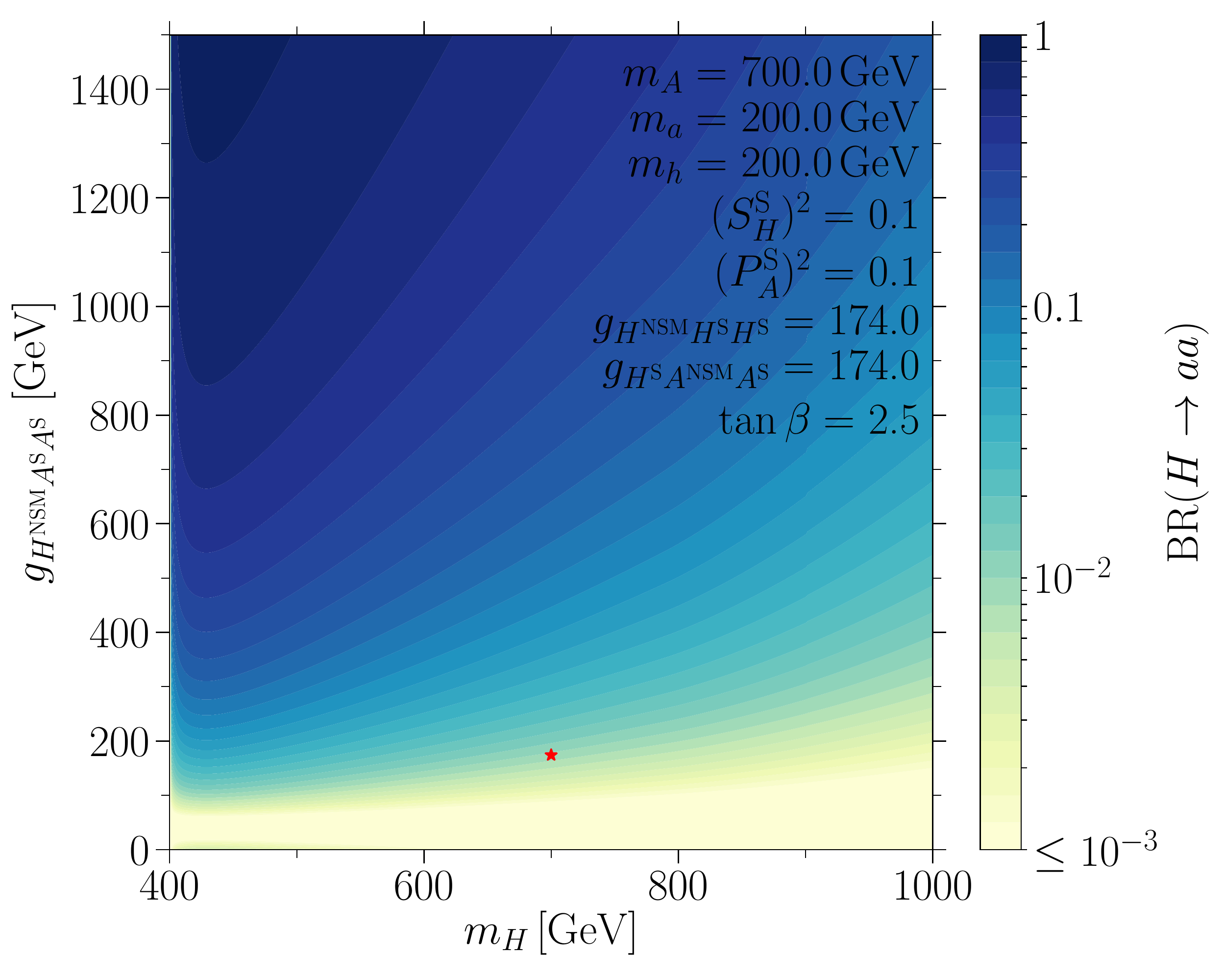}
      \hspace{.5in}
      \includegraphics[width = 2.5in]{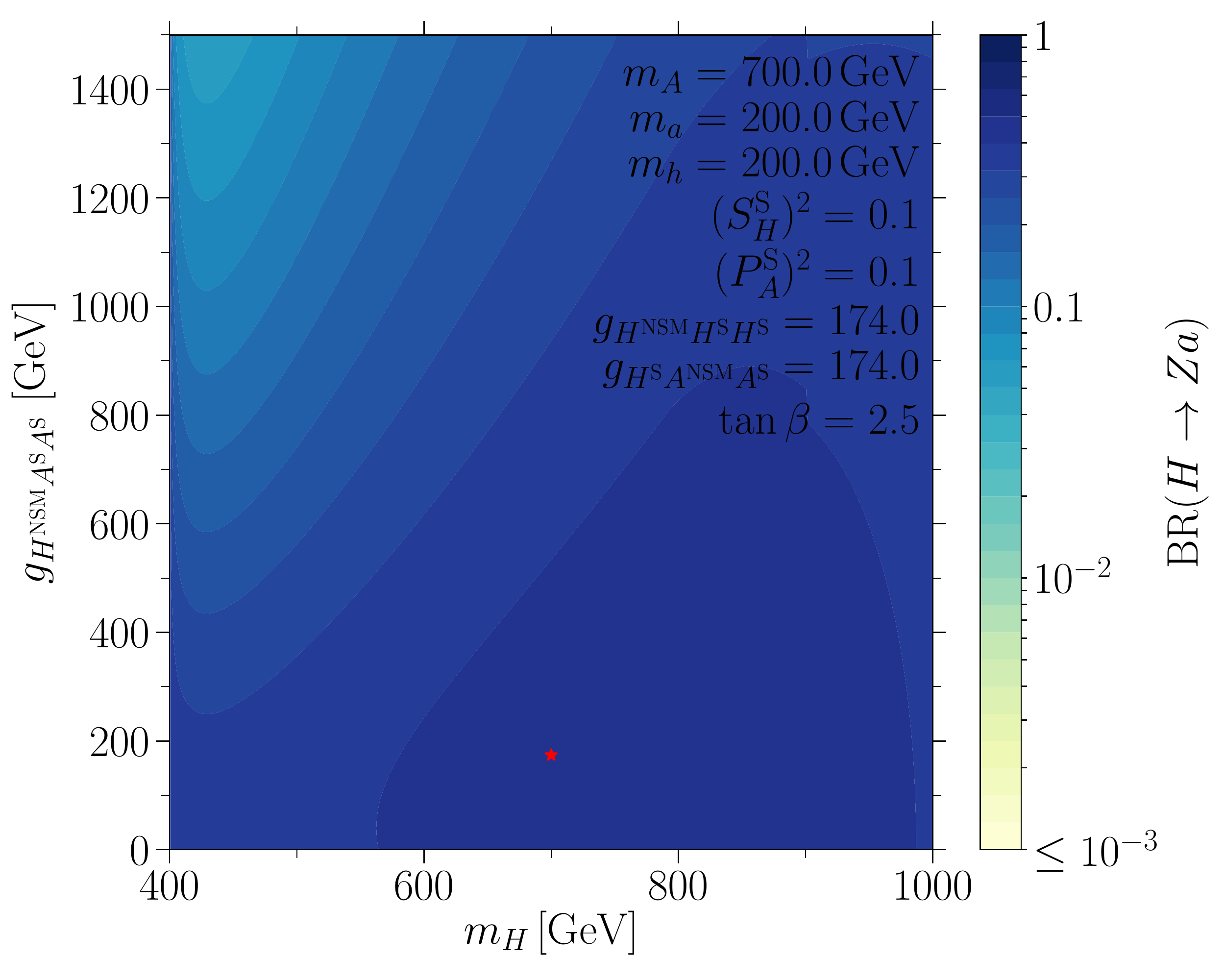}

      \includegraphics[width = 2.5in]{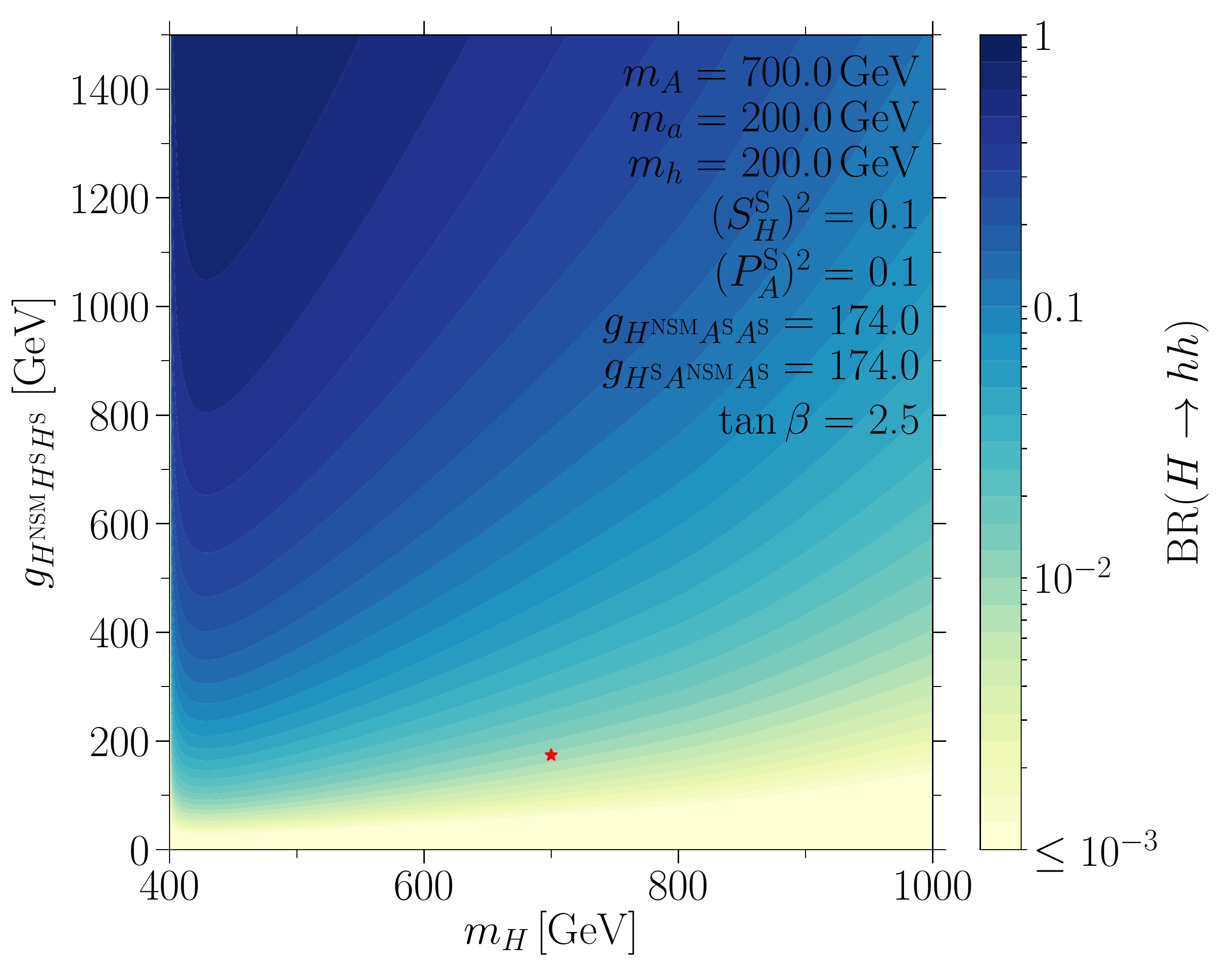}
      \hspace{.5in}
      \includegraphics[width = 2.5in]{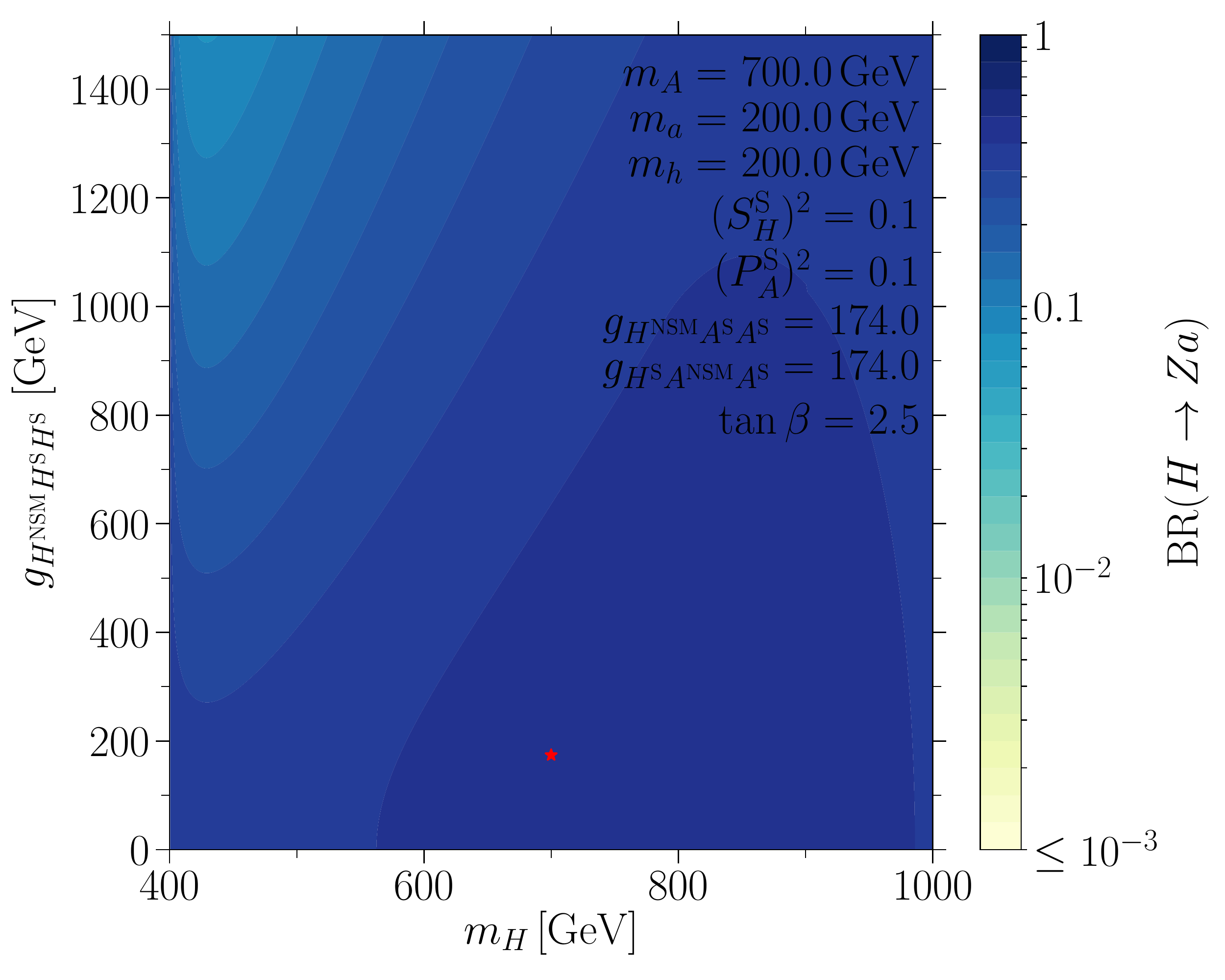}
      \caption{Most relevant branching ratios for ($A \to ha$; first row), ($H \to aa$; second row), and ($H \to hh$; third row) searches in the planes of the most relevant masses and trilinear couplings. The red stars indicate the benchmark point from the third column in Table~\ref{tab:BP_h125S}.}
      \label{fig:HhhaaAah_BRs}
   \end{centering}
\end{figure}

To the best of our knowledge, no projections for the sensitivity of resonant $hh$, $aa$, or $ha$ production exist. Hence, we present only the cross section for our benchmark scenario here. We have chosen conservative values for the trilinear couplings. We show the variation in the cross-section as a function of the most relevant couplings and masses for each of these channels in Fig.~\ref{fig:HhhaaAah}. The three rows in the figure denote the three different channels. We have not taken into account the branching ratios of the daughter states $h$ or $a$, since these are arbitrarily controlled by the possibility of decays into missing energy. 

Note that unlike ($H \to h_{125} h$) and ($A \to h_{125} a$) discussed previously, the decays ($H \to hh$), ($H \to aa$), and ($A \to ha$) considered here are essentially governed by the free trilinear couplings between the Higgs basis interaction states as well as the masses of the particles involved in each process. In Fig.~\ref{fig:HhhaaAah_BRs} we show the most relevant branching ratios for such processes. For each channel (i.e. the different rows), we show the branching ratio corresponding to the signal channel in the left panel, while the right panel shows the corresponding branching ratio into a $Z$ boson and an $h$ or $a$.

From Fig.~\ref{fig:HhhaaAah} we see that the production cross-section for these channels can easily be $\mathcal{O}(100)$\,fb, which may be explored at the LHC.

\FloatBarrier

\subsection{Max Misalignment Scenario: $H \to h_{125} h_{125}$}

\begin{table}[h!]
   \centering
   \begin{tabular}{c||c|c|c|c}
      \hline\hline
      $m_H$ [GeV] & \multicolumn{4}{c}{$700$} \\
      $m_A$ [GeV] & \multicolumn{4}{c}{$1000$} \\
      $m_h$ [GeV] & \multicolumn{4}{c}{$m_H-100 \rm ~GeV = 600$} \\
      $m_a$ [GeV] & \multicolumn{4}{c}{$950$} \\
      \hline
      $S_{h_{125}}^{\rm NSM}$ & $0.2$ & $0.2$ & $-0.2$ & $-0.2$ \\
      $S_{h_{125}}^{\rm S}$ & $0.5$ & $-0.5$ & $0.5$ & $-0.5$ \\
      $S_H^{\rm S}$ & \multicolumn{4}{c}{$0.3$} \\
      \hline
      $g_{H^{\rm SM} H^{\rm SM} H^{\rm SM}} = 3 \mathcal{M}^2_{S,11}/\sqrt{2}v$ [GeV] & $1800$ & $1400$ & $1400$ & $1800$ \\
      $g_{H^{\rm SM} H^{\rm SM} H^{\rm NSM}} = 3 \mathcal{M}^2_{S,12}/\sqrt{2}v$ [GeV] & $1500$ & $480$ & $-480$ & $-1500$ \\
      $g_{H^{\rm SM} H^{\rm SM} H^{\rm S}} = \mathcal{M}^2_{S,13}/\sqrt{2}v$ [GeV] & $340$ & $-680$ & $680$ & $-340$ \\
      $g_{H^{\rm SM} H^{\rm NSM} H^{\rm S}} = \mathcal{M}^2_{S,23}/\sqrt{2}v$ [GeV] & $-560$ & $-370$ & $-370$ & $-560$ \\
      \hline
      $\sigma(ggH)$ [pb] & $0.96$ & $0.018$ & $0.36$ & $0.097$ \\
      BR($H \to h_{125} h_{125}$) & $0.032$ & $0.35$ & $0.15$ & $0.016$ \\
      BR($H \to Z Z$) & $0.23$ & $0.20$ & $0.13$ & $0.31$ \\
      BR($H \to W W$) & $0.46$ & $0.40$ & $0.26$ & $0.63$ \\
      \hline\hline
   \end{tabular}
   \caption{Benchmark Scenario for ($gg \to H \to h_{125} h_{125}$) resonant $h_{125}$ pair production. Of the parameters not listed, $\tan\beta = 2.5$, and all free trilinear couplings, cf. Eq.~\eqref{eq:trilin} are set to 0. The relevant trilinear couplings are computed from the scalar squared mass matrix, $\mathcal{M}^2_{S}$, obtained from the given values of the masses and mixing angles.}
   \label{tab:BP_h125h125}
\end{table}

\begin{figure}
   \begin{centering}
      \includegraphics[width = 2.5in]{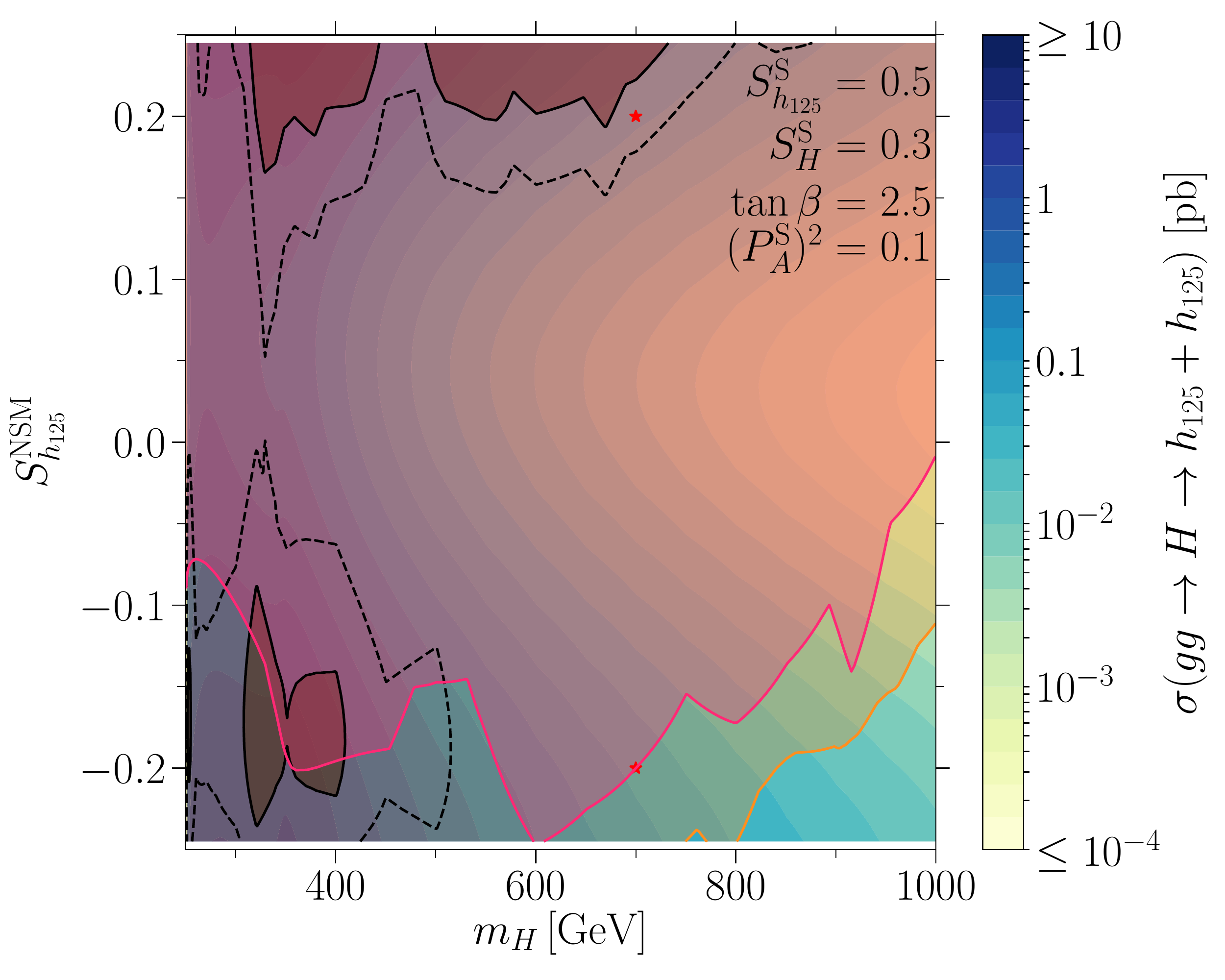}
      \hspace{.5in}
      \includegraphics[width = 2.5in]{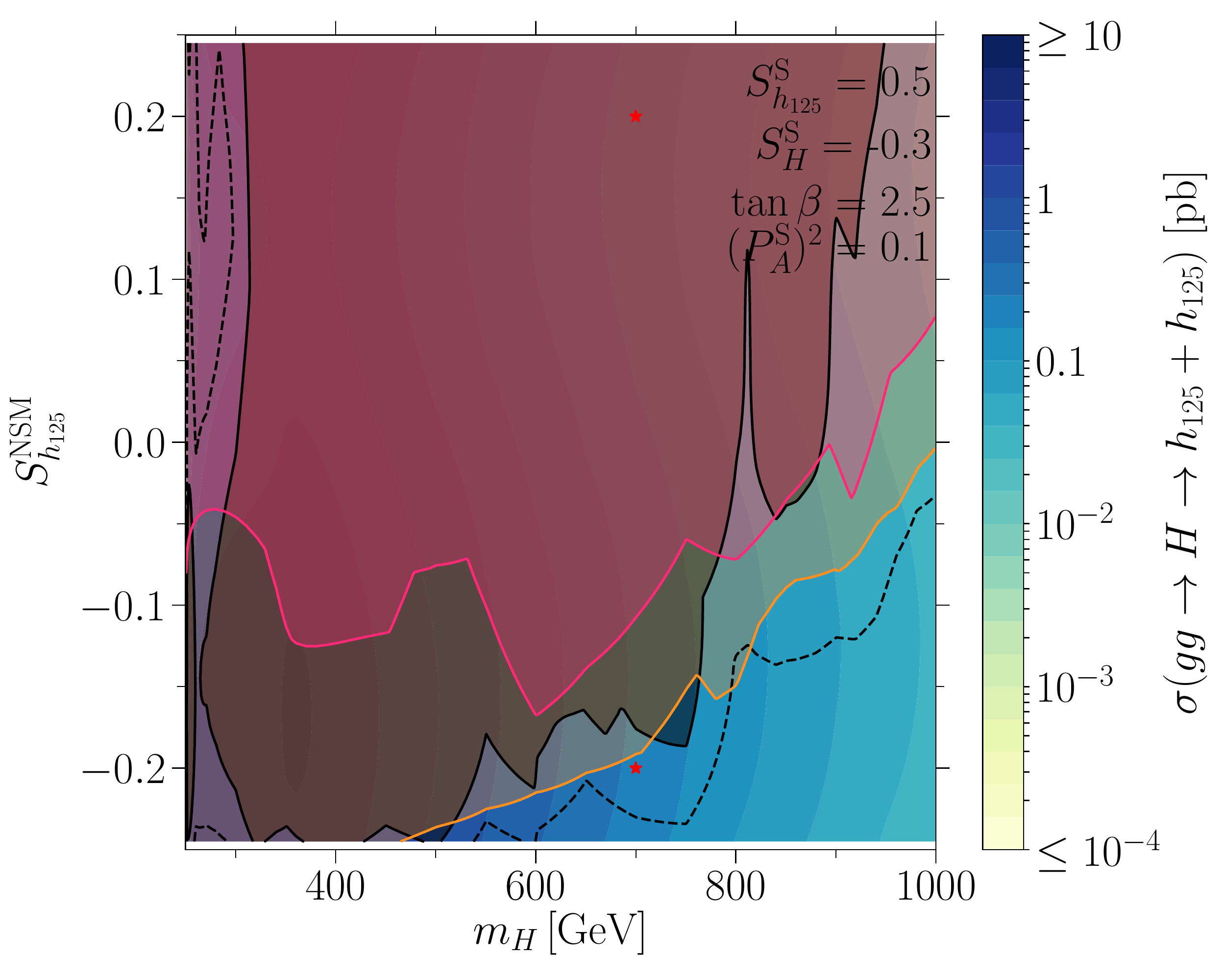}

      \includegraphics[width = 2.5in]{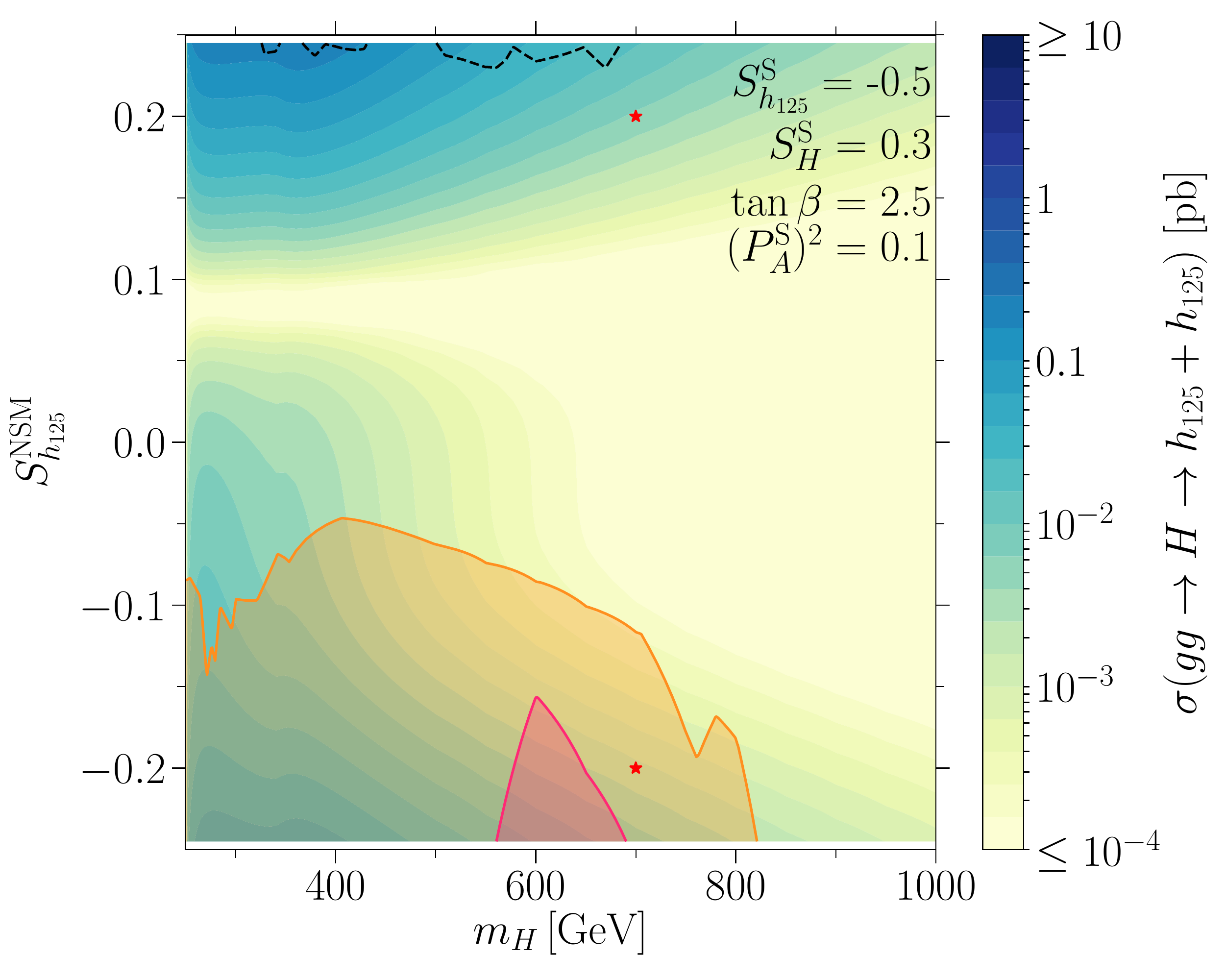}
      \hspace{.5in}
      \includegraphics[width = 2.5in]{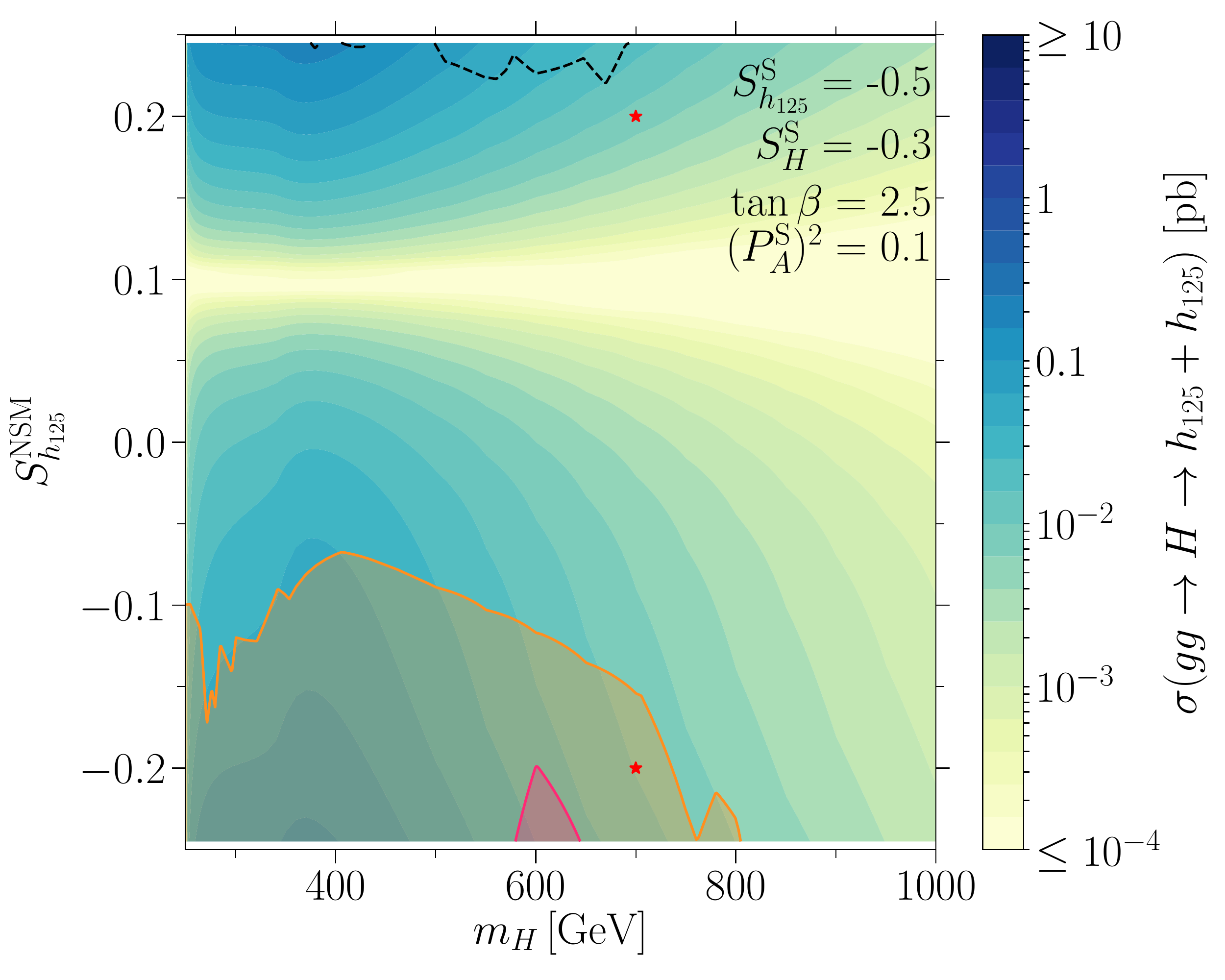}

      \includegraphics[width = 2.5in]{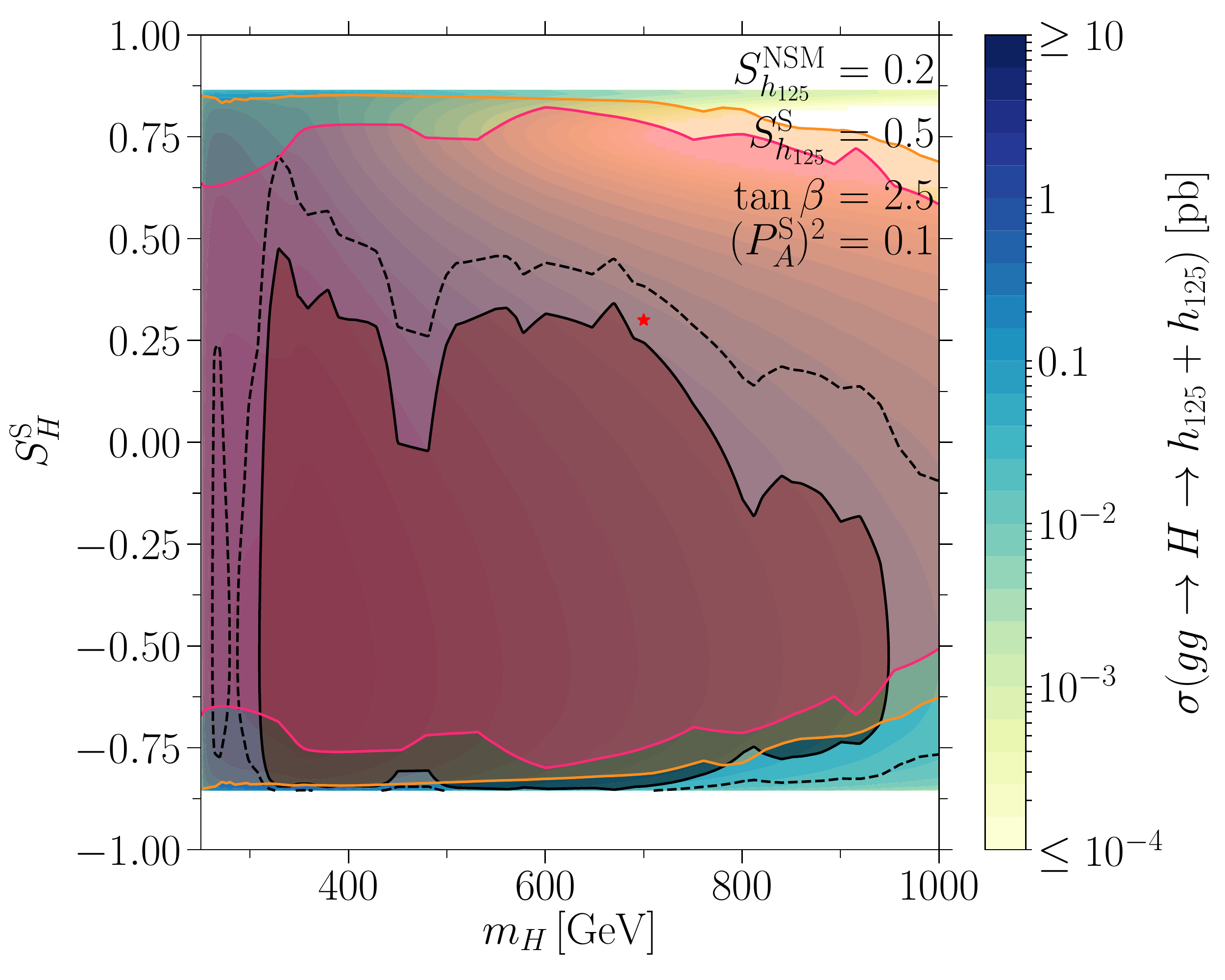}
      \hspace{.5in}
      \includegraphics[width = 2.5in]{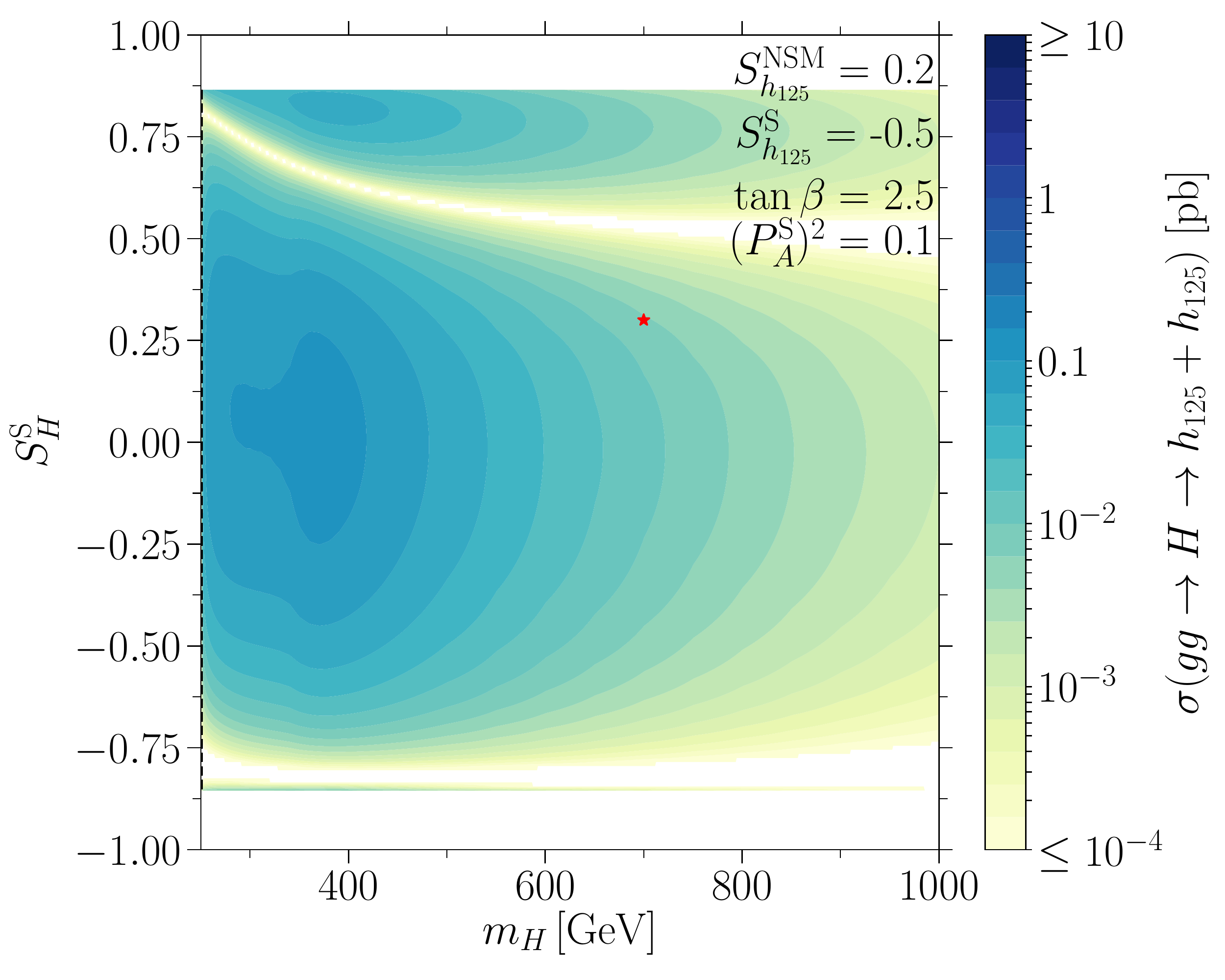}

      \includegraphics[width = 2.5in]{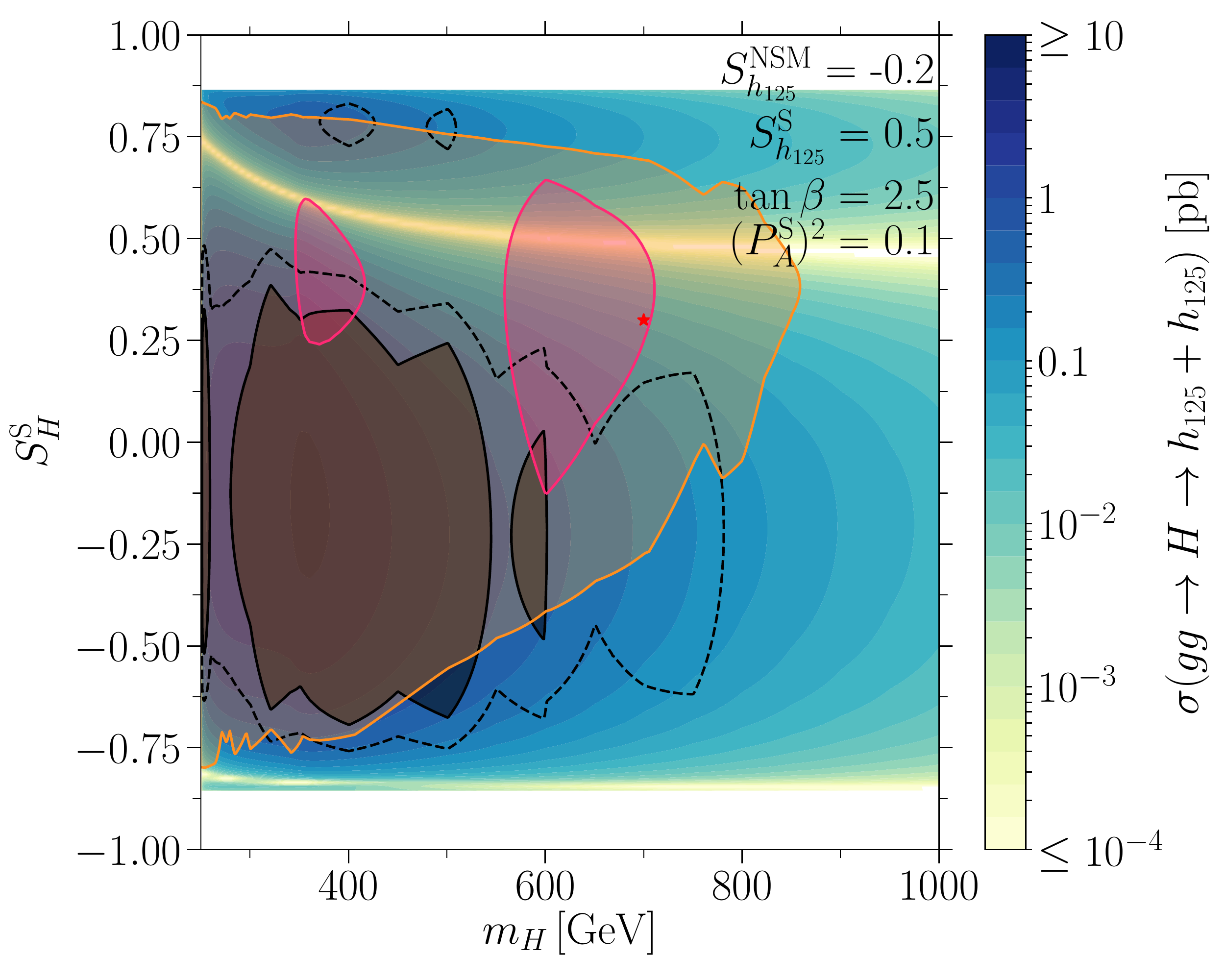}
      \hspace{.5in}
      \includegraphics[width = 2.5in]{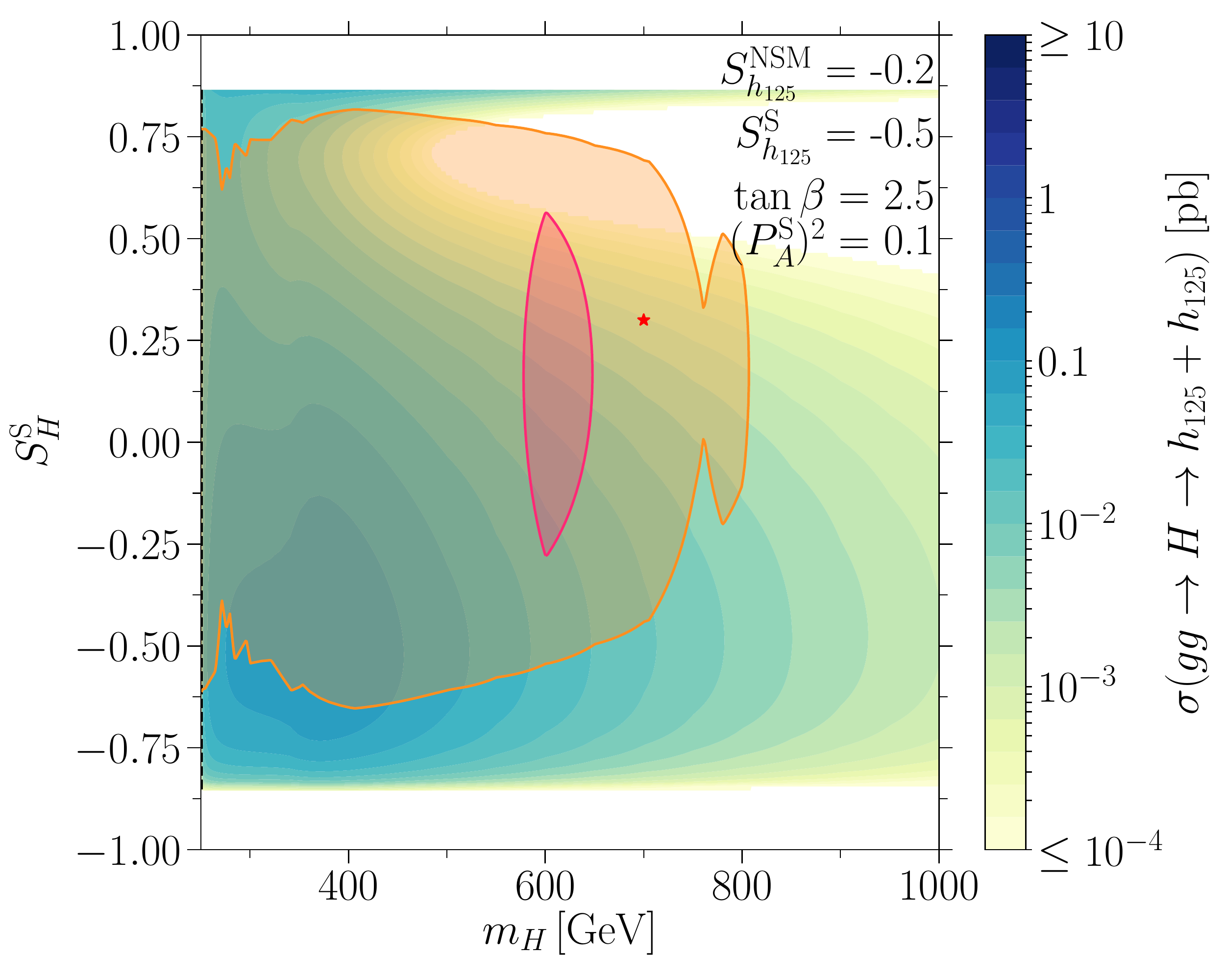}
      \caption{Production cross section and and current LHC limits for ($gg \to H \to h_{125} h_{125}$) in the plane of the heavy Higgs boson $m_H$ and various relevant mixing angles. The colored contours denote the total cross-section according to the color bar label. Gray shaded regions are excluded by current ($gg \to H \to h_{125} h_{125}$) LHC searches. Black dashed lines label cross sections which are a factor two smaller than the current sensitivity, and hence may be expected to be probed with 300\,fb$^{-1}$ of data. Red [orange] shaded area denote regions ruled out by current ($gg \to H \to WW$) [($gg \to H \to ZZ$)] LHC searches. The red stars indicate the benchmark points presented in Table~\ref{tab:BP_h125h125}.}
      \label{fig:h125h125}
   \end{centering}
\end{figure}

\begin{figure}
   \begin{centering}
      \includegraphics[width = 2.5in]{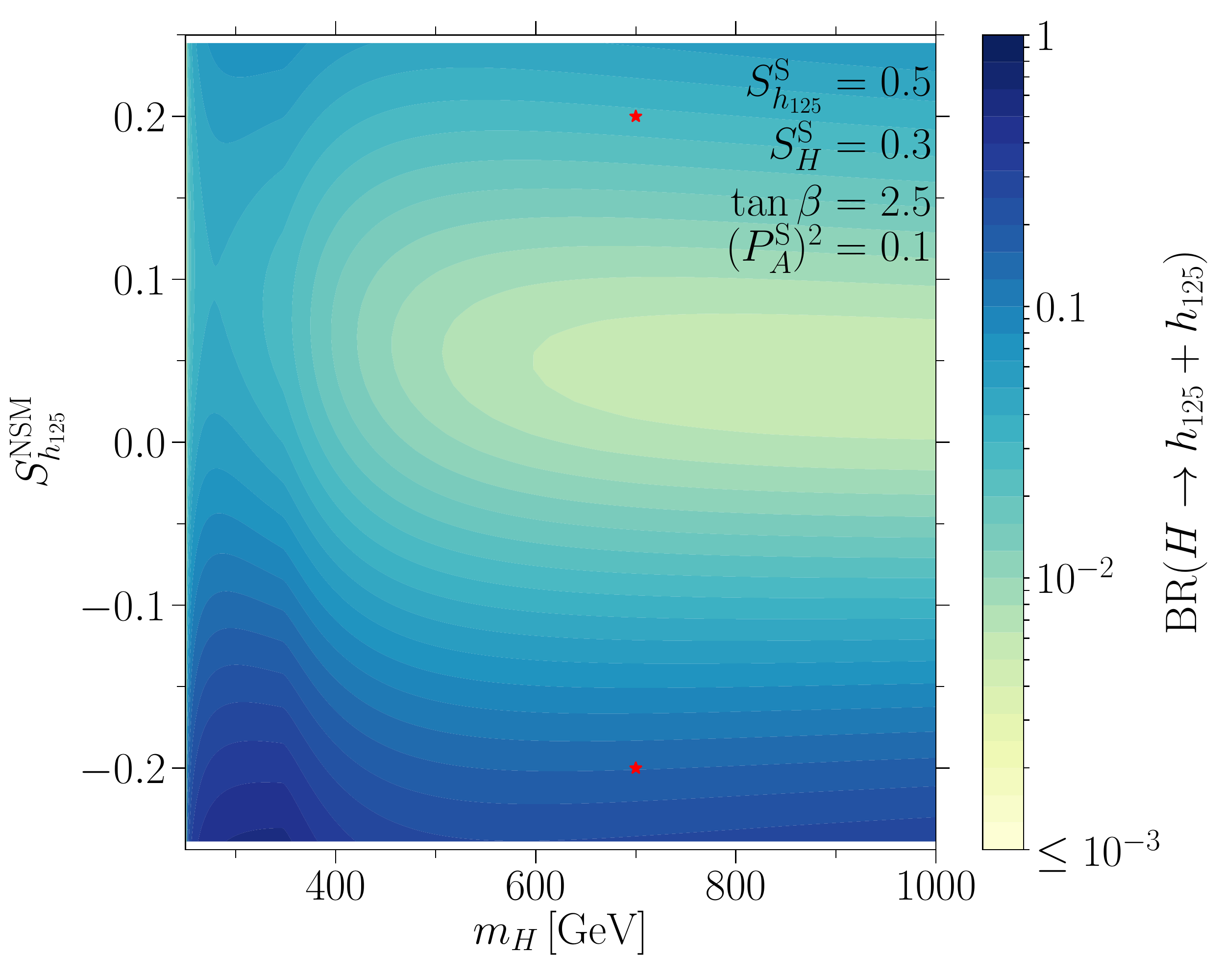}
      \hspace{.5in}
      \includegraphics[width = 2.5in]{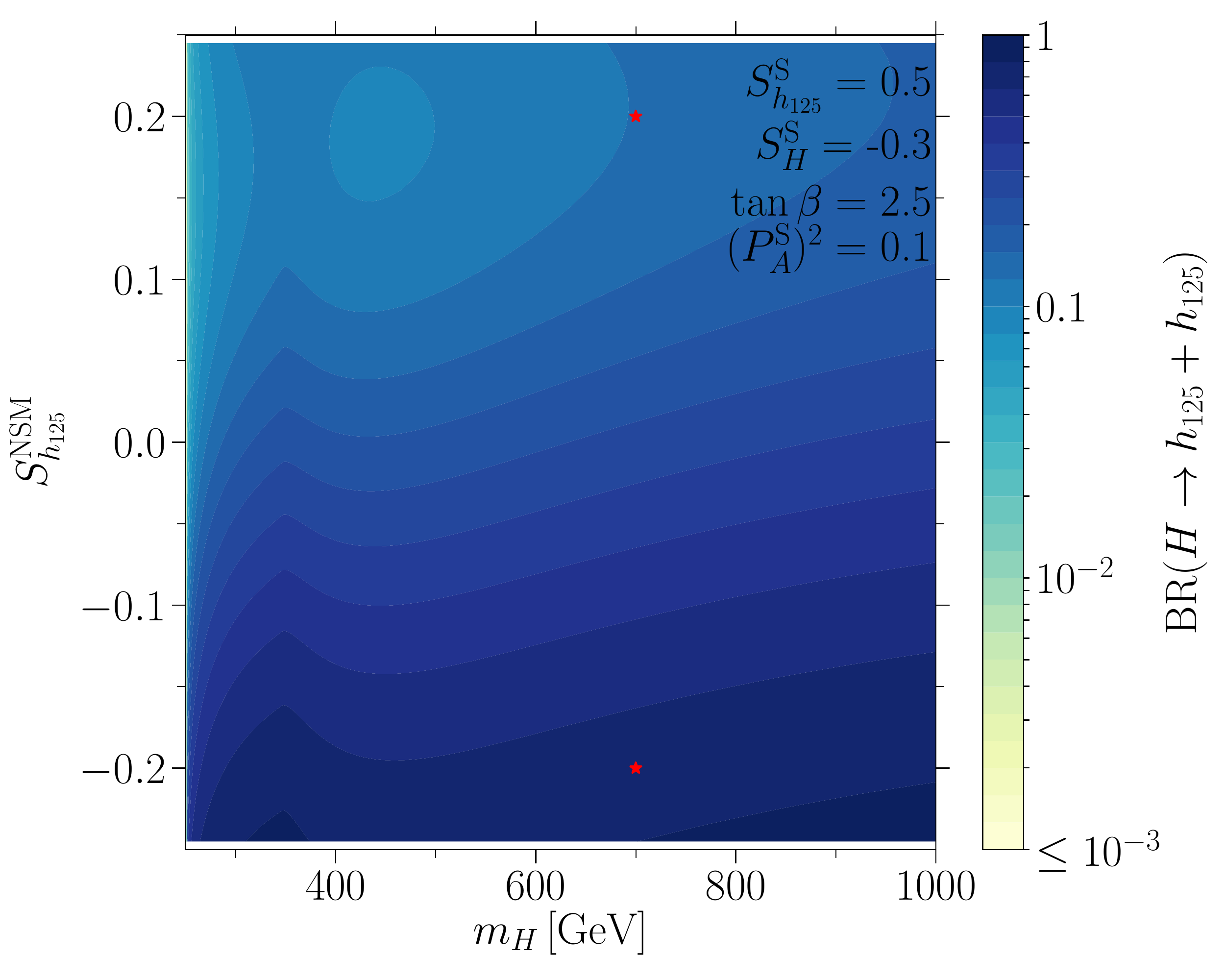}

      \includegraphics[width = 2.5in]{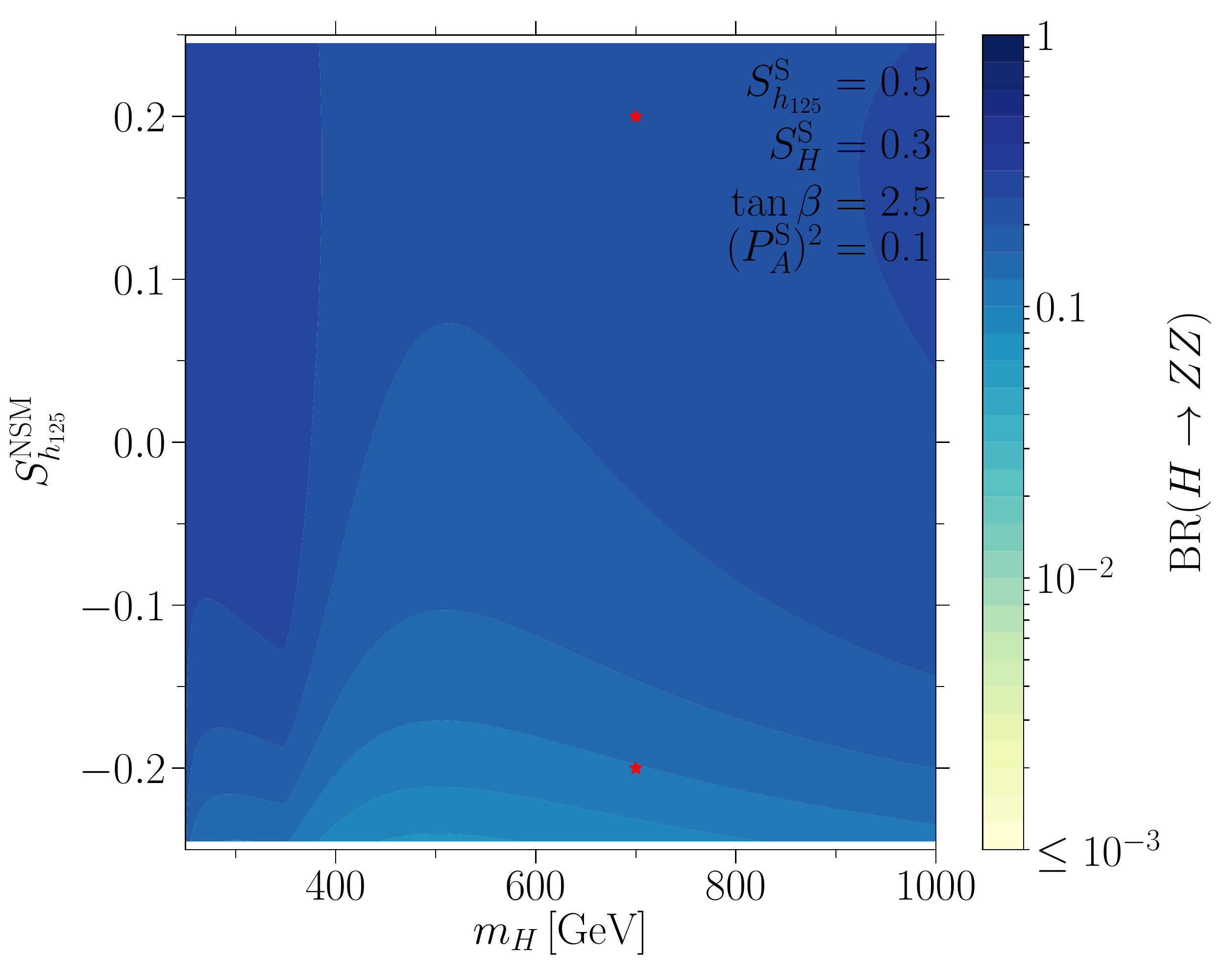}
      \hspace{.5in}
      \includegraphics[width = 2.5in]{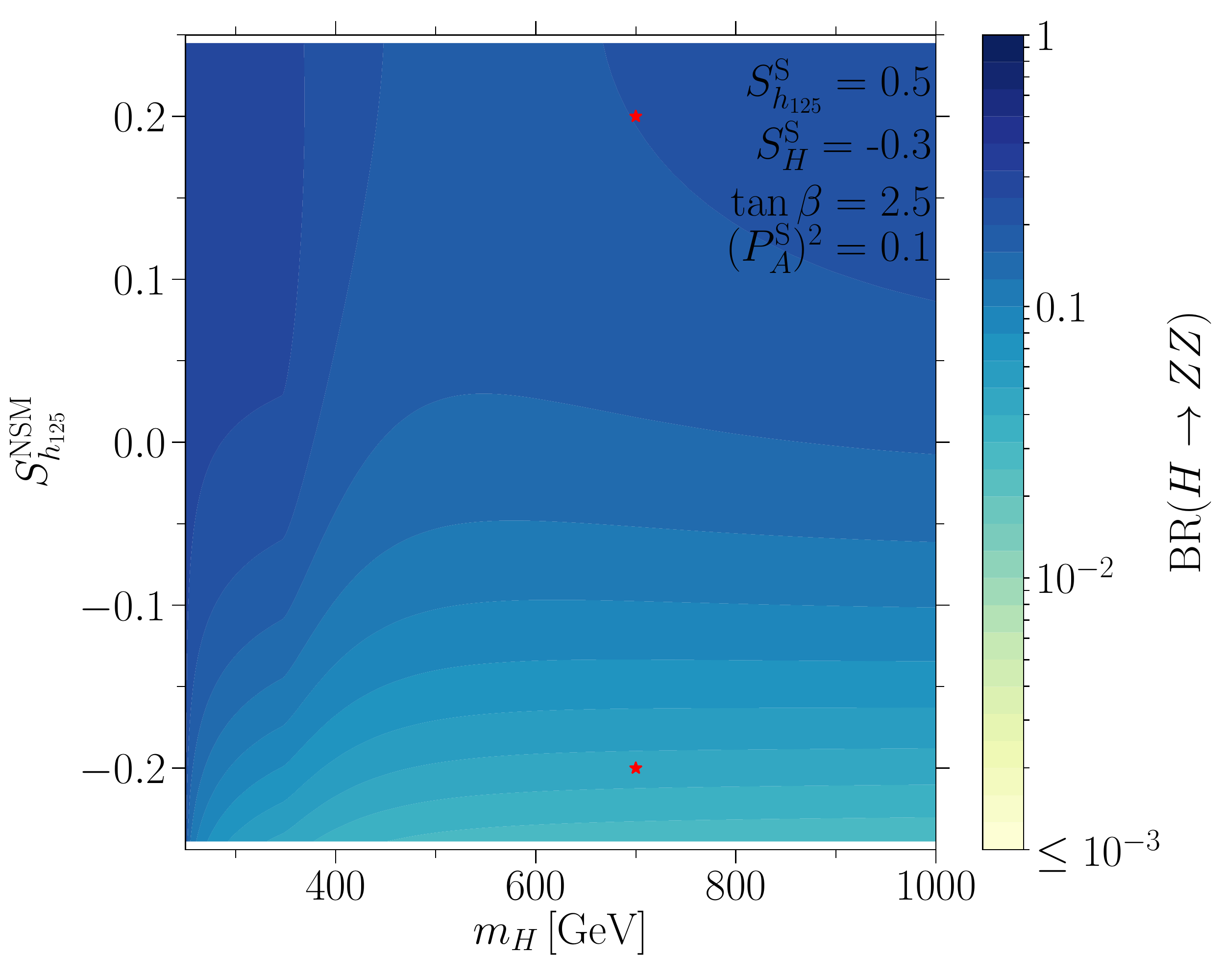}

      \includegraphics[width = 2.5in]{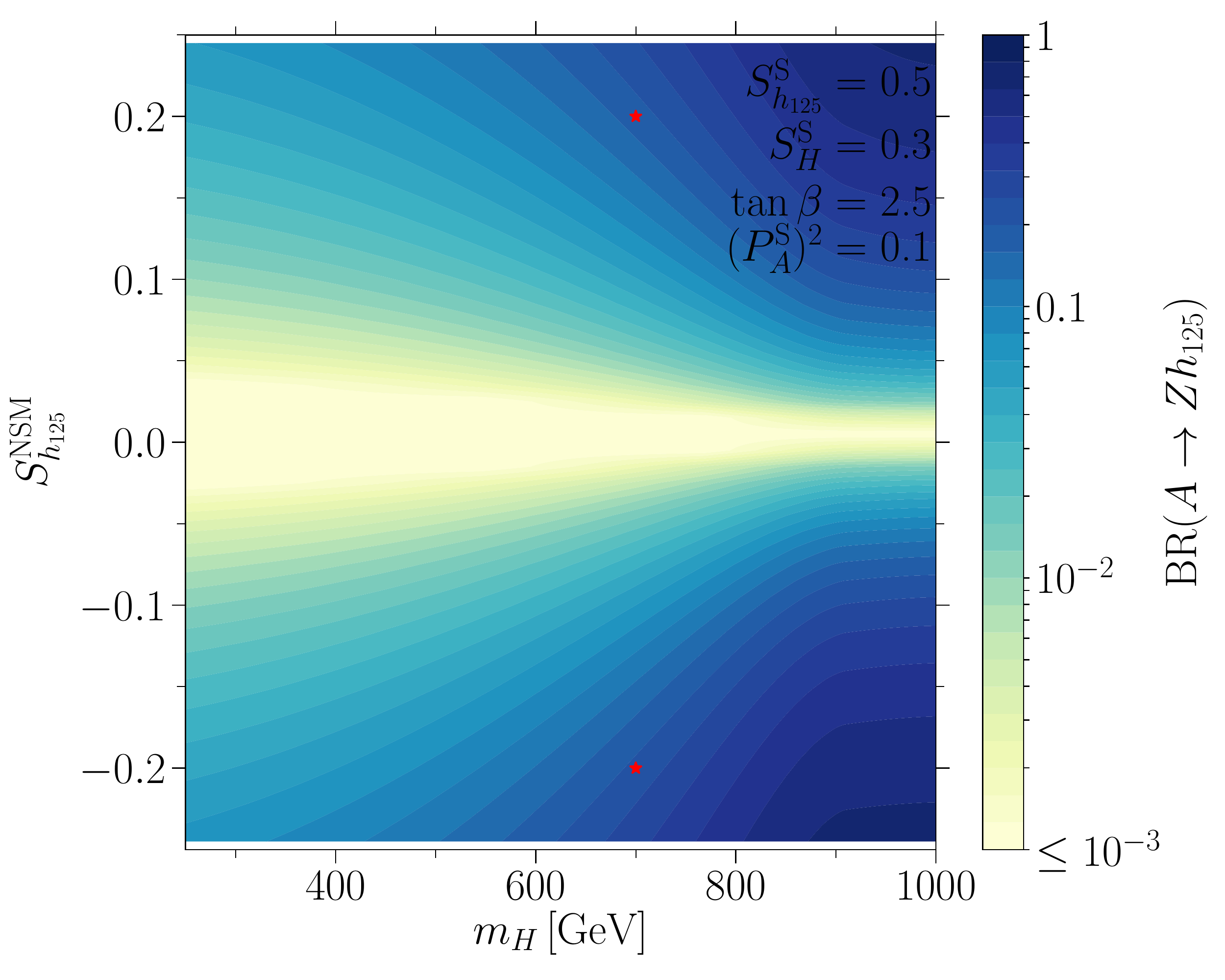}
      \hspace{.5in}
      \includegraphics[width = 2.5in]{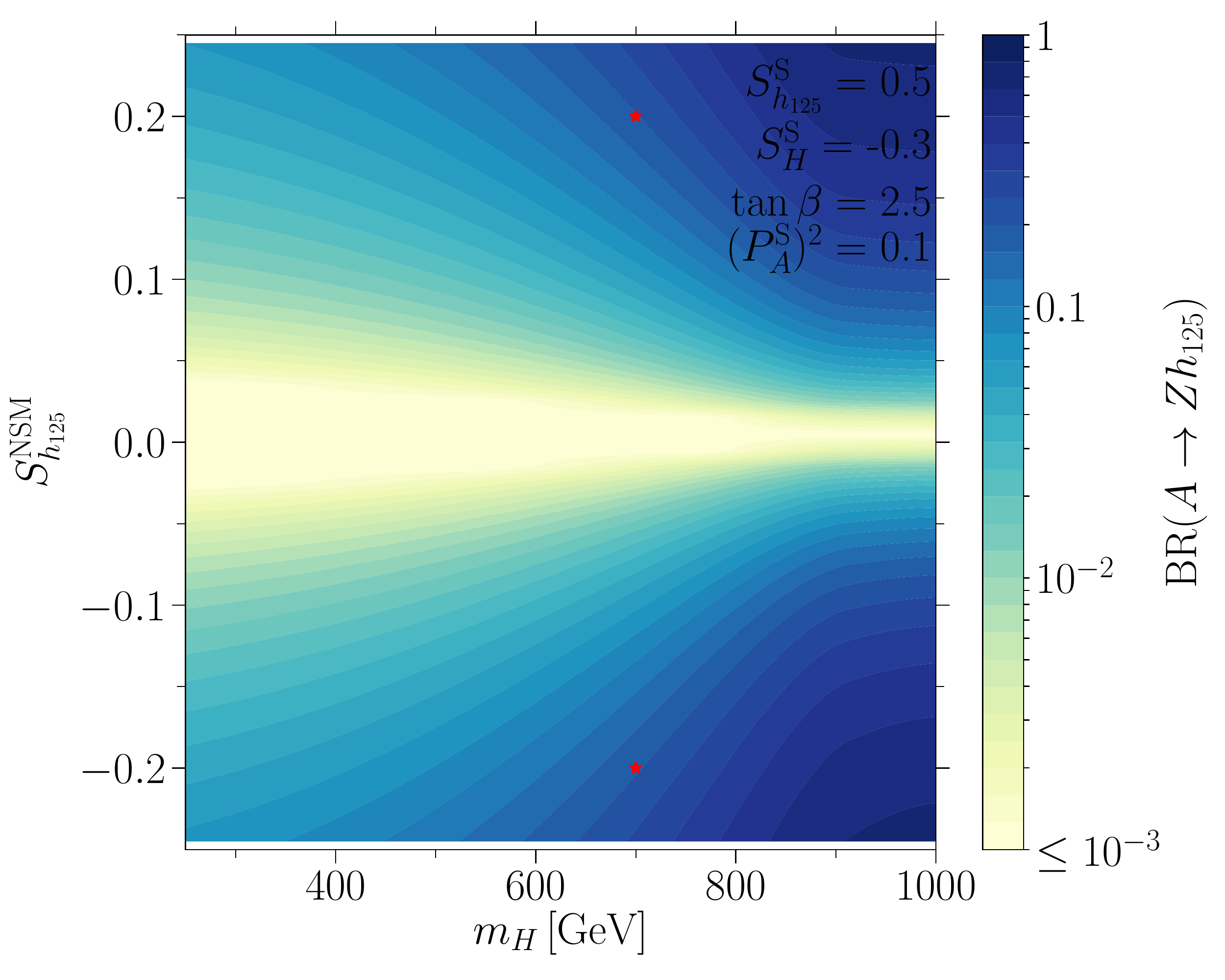}

      \includegraphics[width = 2.5in]{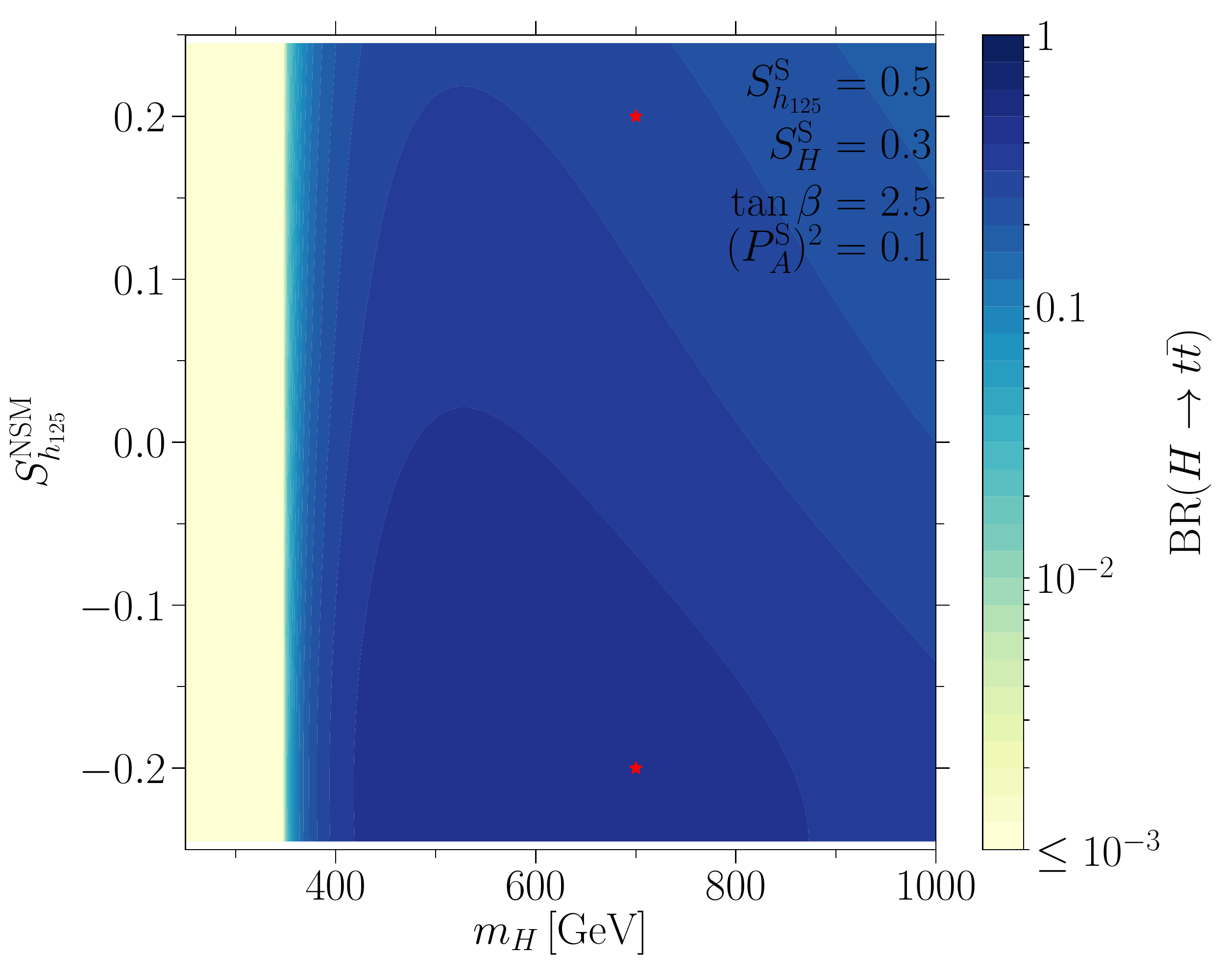}
      \hspace{.5in}
      \includegraphics[width = 2.5in]{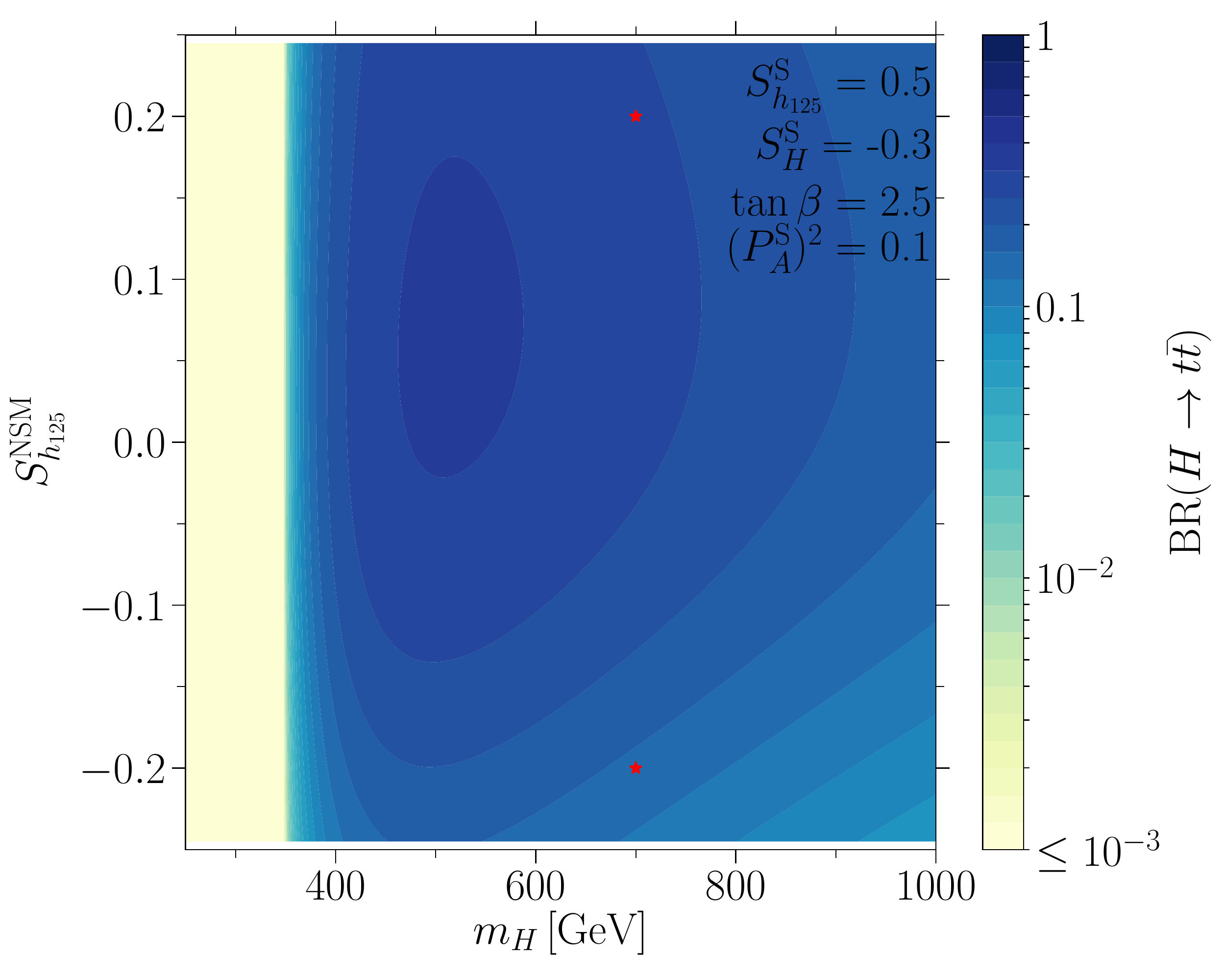}
      \caption{Most relevant branching ratios as labeled by the colorbar for ($gg \to H \to h_{125} h_{125}$) searches in the $m_H$--$S_{h_{125}}^{\rm NSM}$ plane. For all plots $S_{h_{125}}^{\rm S} = 0.5$. For plots in the left column $S_{h_{125}}^{\rm S} = 0.3$, while in the right column $S_{h_{125}}^{\rm S} = -0.3$. The red stars indicate the benchmark points presented in Table~\ref{tab:BP_h125h125}. Note that we do not show the ($H \to WW$) branching ratio, its scaling with the mass and the mixing angle is identical to BR$(H \to ZZ)$.}
      \label{fig:h125h125_BRs1}
   \end{centering}
\end{figure}

\begin{figure}
   \begin{centering}
      \includegraphics[width = 2.5in]{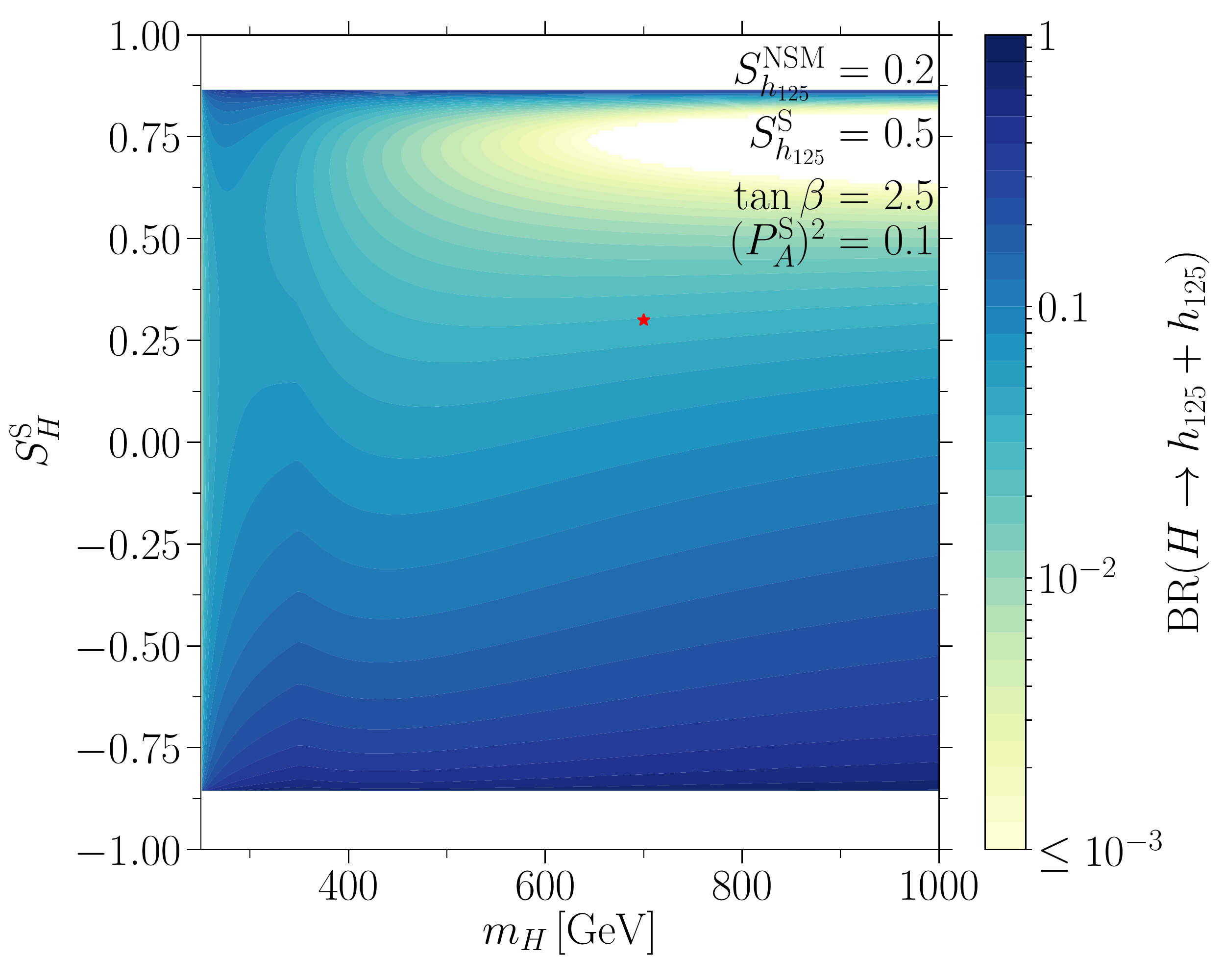}
      \hspace{.5in}
      \includegraphics[width = 2.5in]{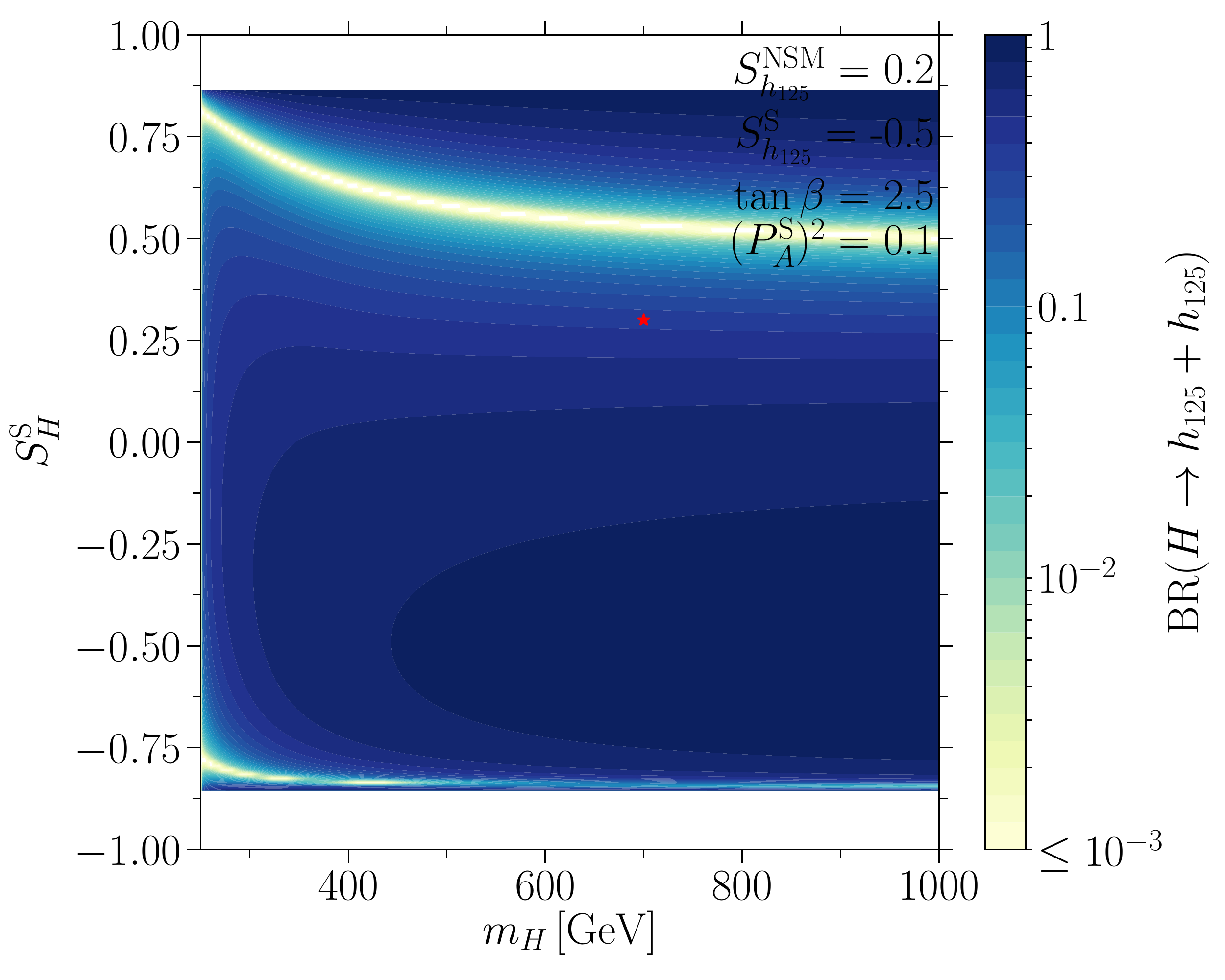}

      \includegraphics[width = 2.5in]{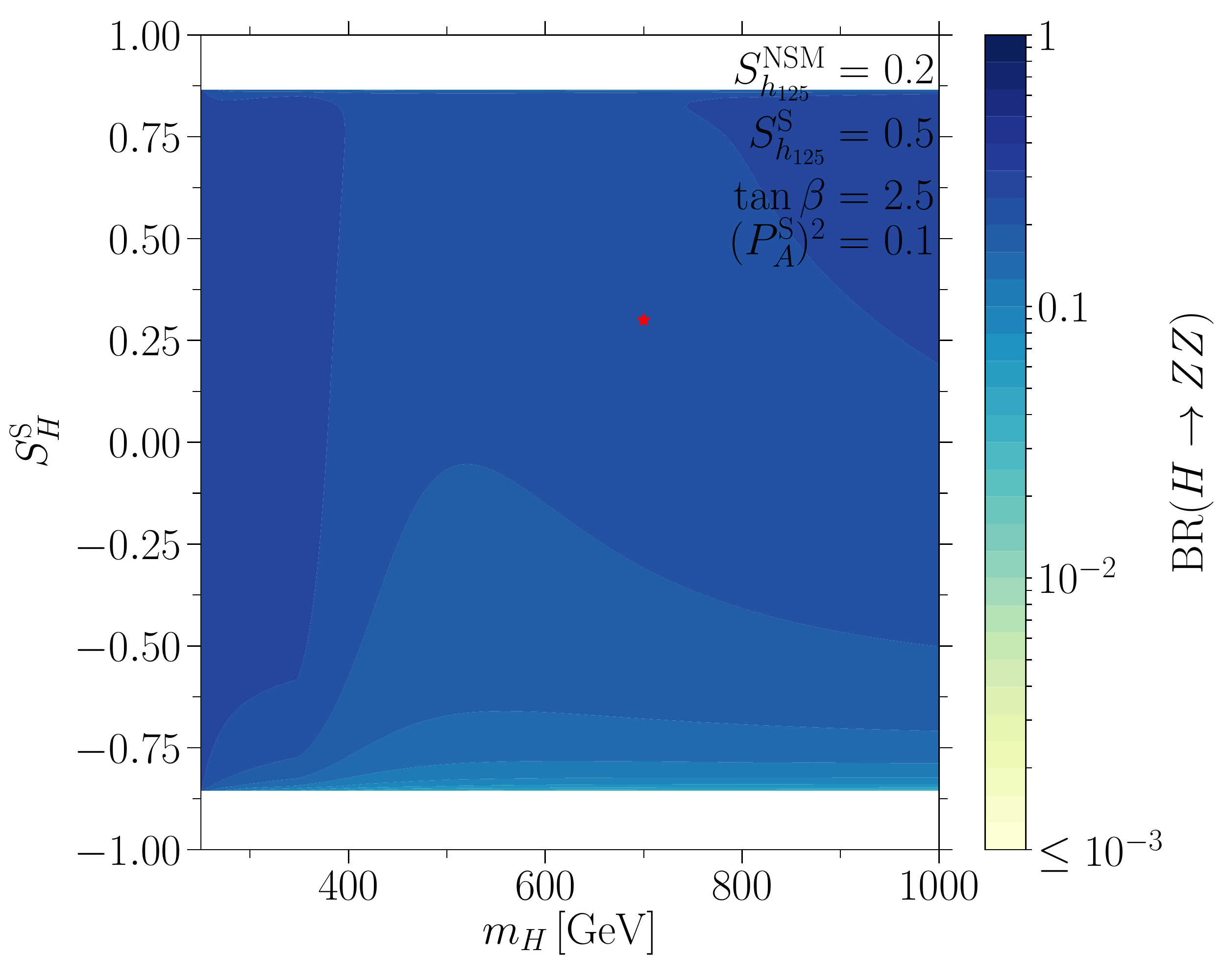}
      \hspace{.5in}
      \includegraphics[width = 2.5in]{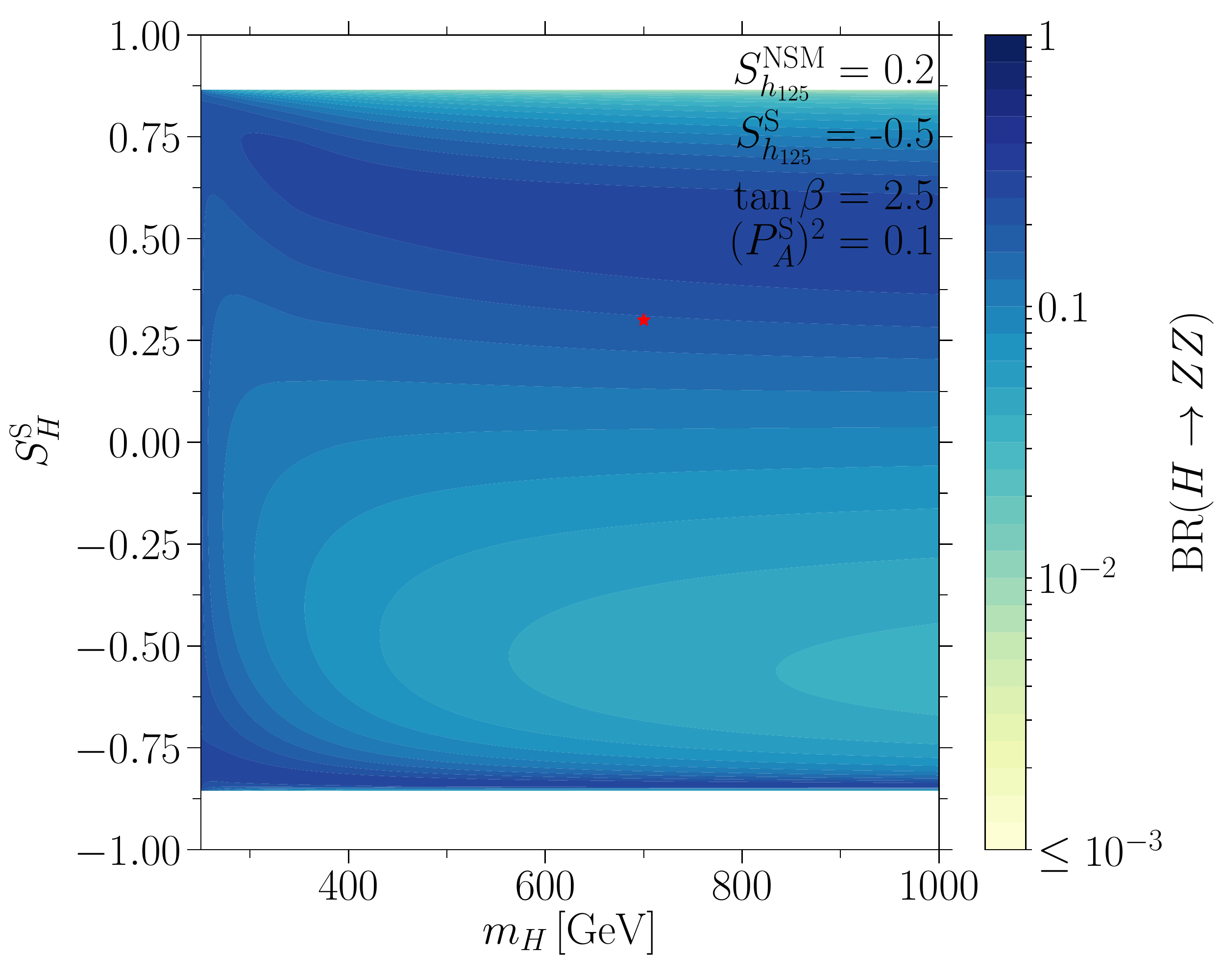}

      \includegraphics[width = 2.5in]{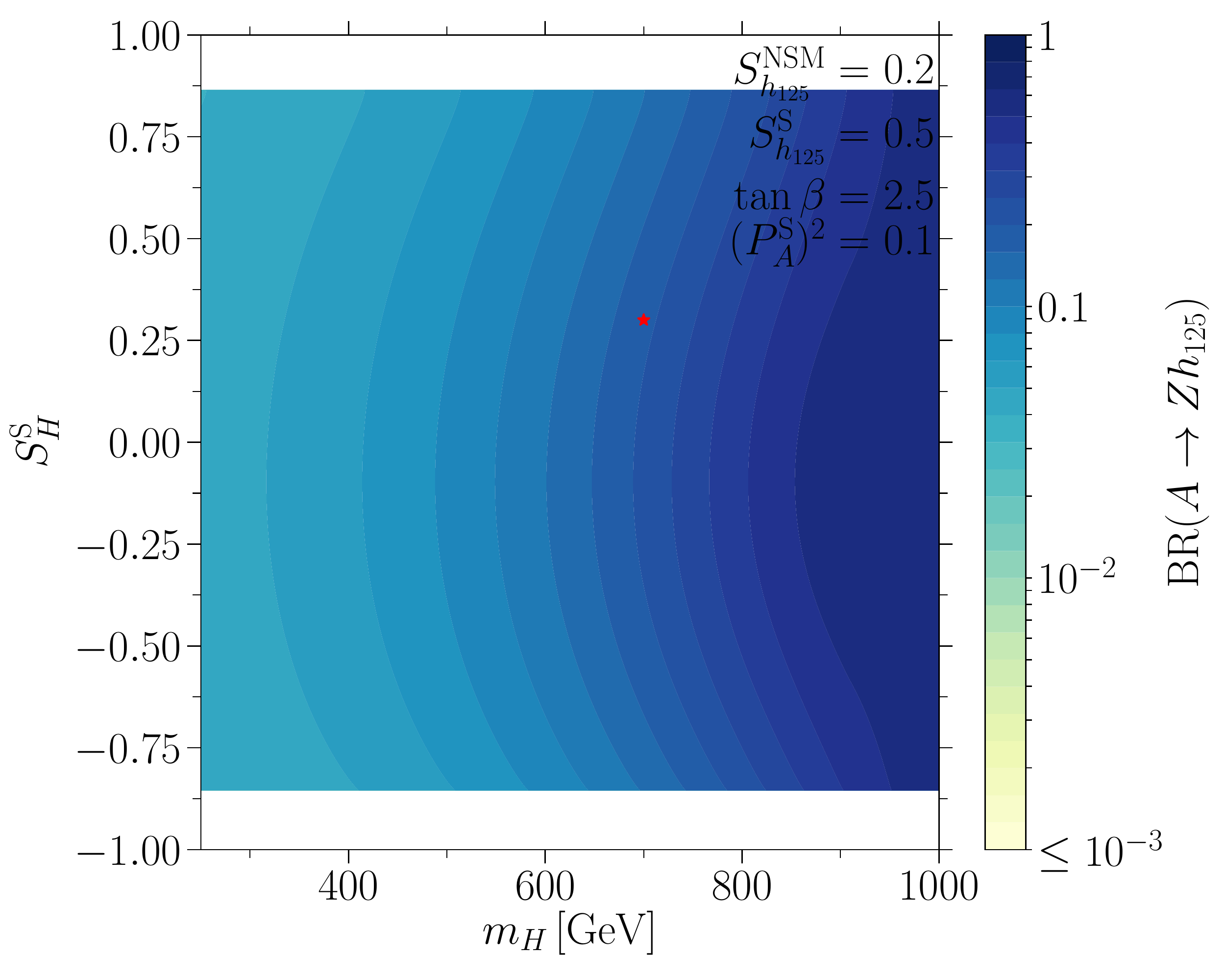}
      \hspace{.5in}
      \includegraphics[width = 2.5in]{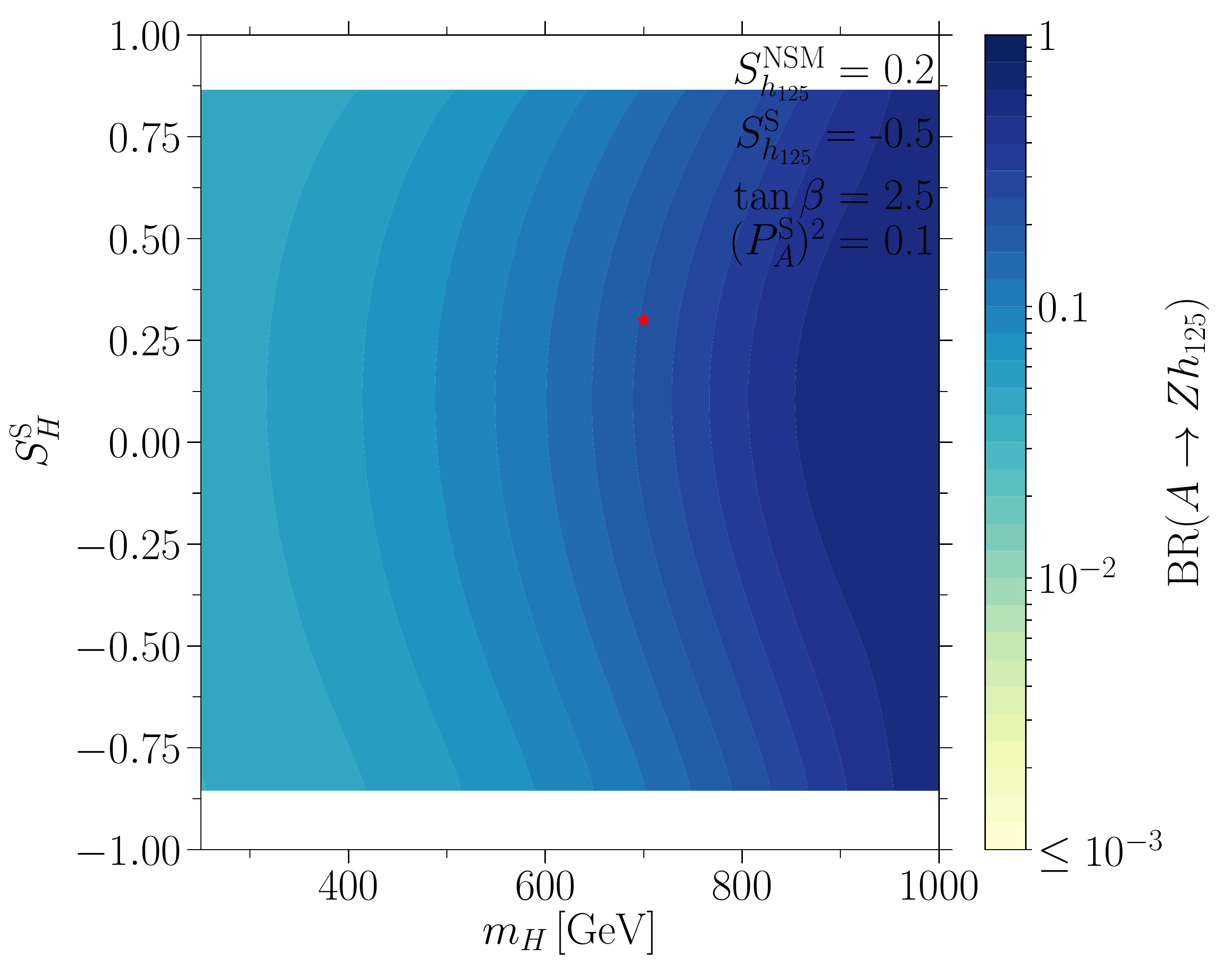}

      \includegraphics[width = 2.5in]{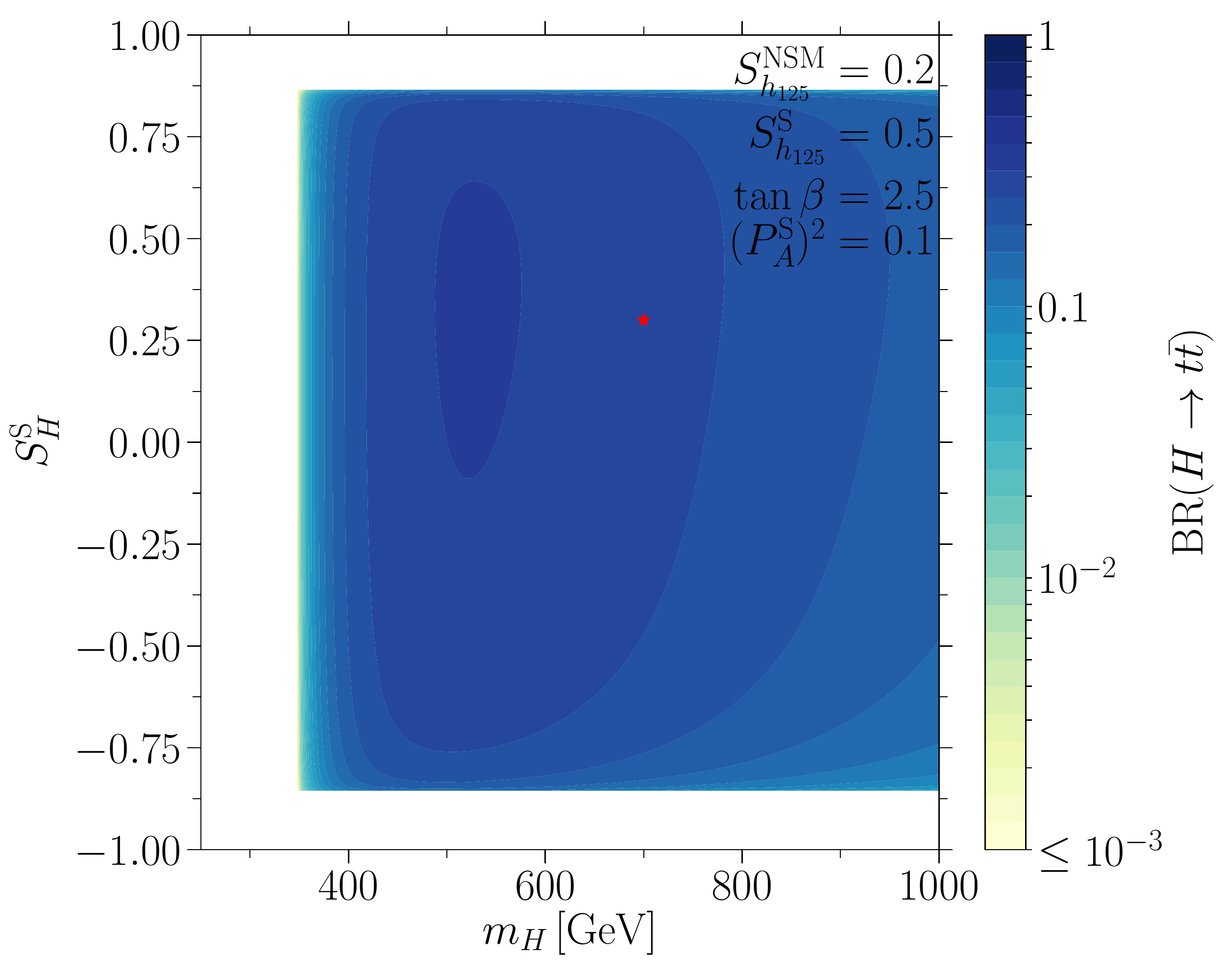}
      \hspace{.5in}
      \includegraphics[width = 2.5in]{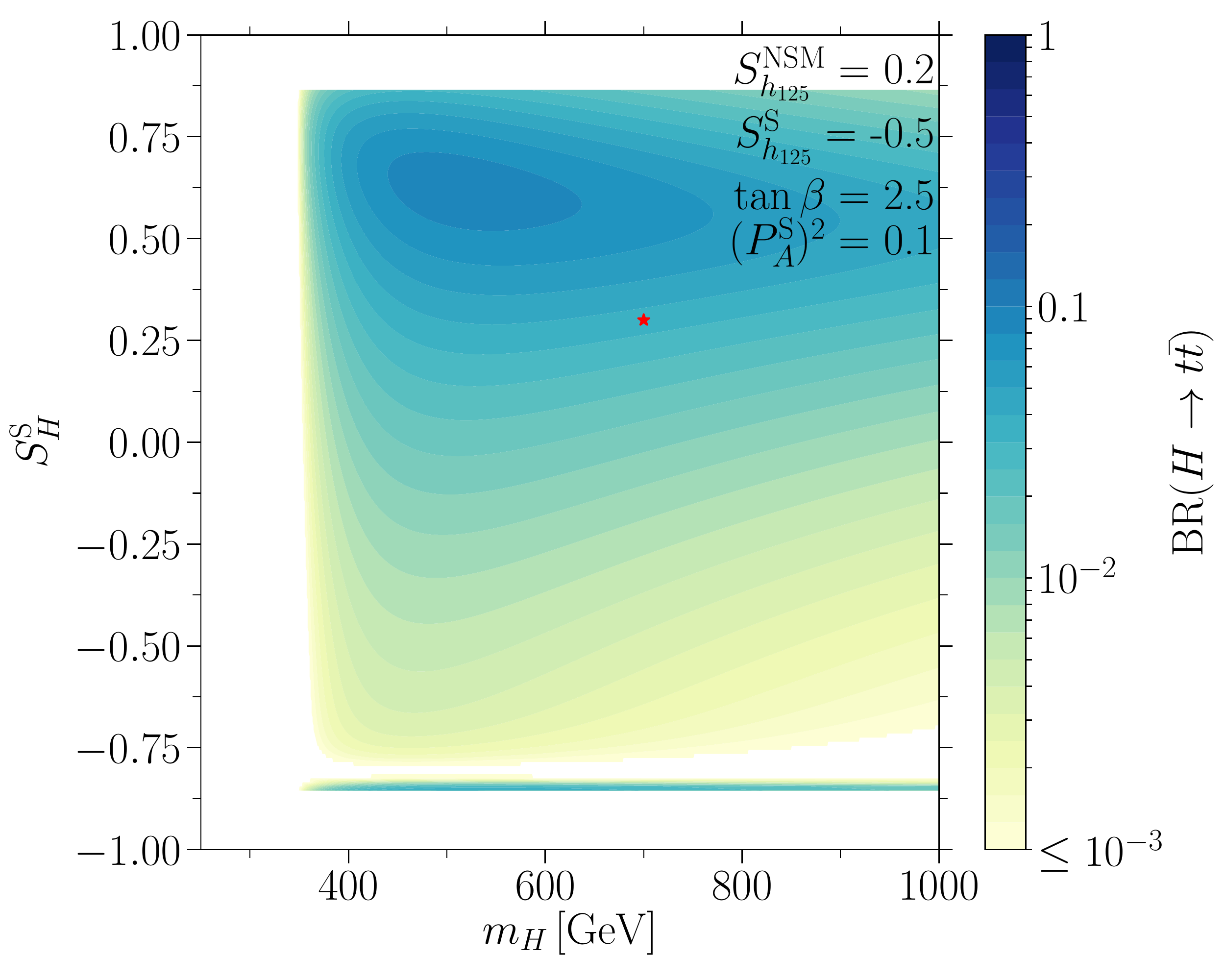}
      \caption{Most relevant branching ratios as labeled by the colorbar for ($gg \to H \to h_{125} h_{125}$) searches in the $m_H$--$S_H^{\rm S}$ plane. For all plots $S_{h_{125}}^{\rm NSM} = 0.2$. For plots in the left column $S_{h_{125}}^{\rm S} = 0.5$, while in the right column $S_{h_{125}}^{\rm S} = -0.5$. The red stars indicate the benchmark points presented in Table~\ref{tab:BP_h125h125}. Note that we do not show the ($H \to WW$) branching ratio, its scaling with the mass and the mixing angle is identical to BR$(H \to ZZ)$.}
      \label{fig:h125h125_BRs2}
   \end{centering}
\end{figure}

\begin{figure}
   \begin{centering}
      \includegraphics[width = 2.5in]{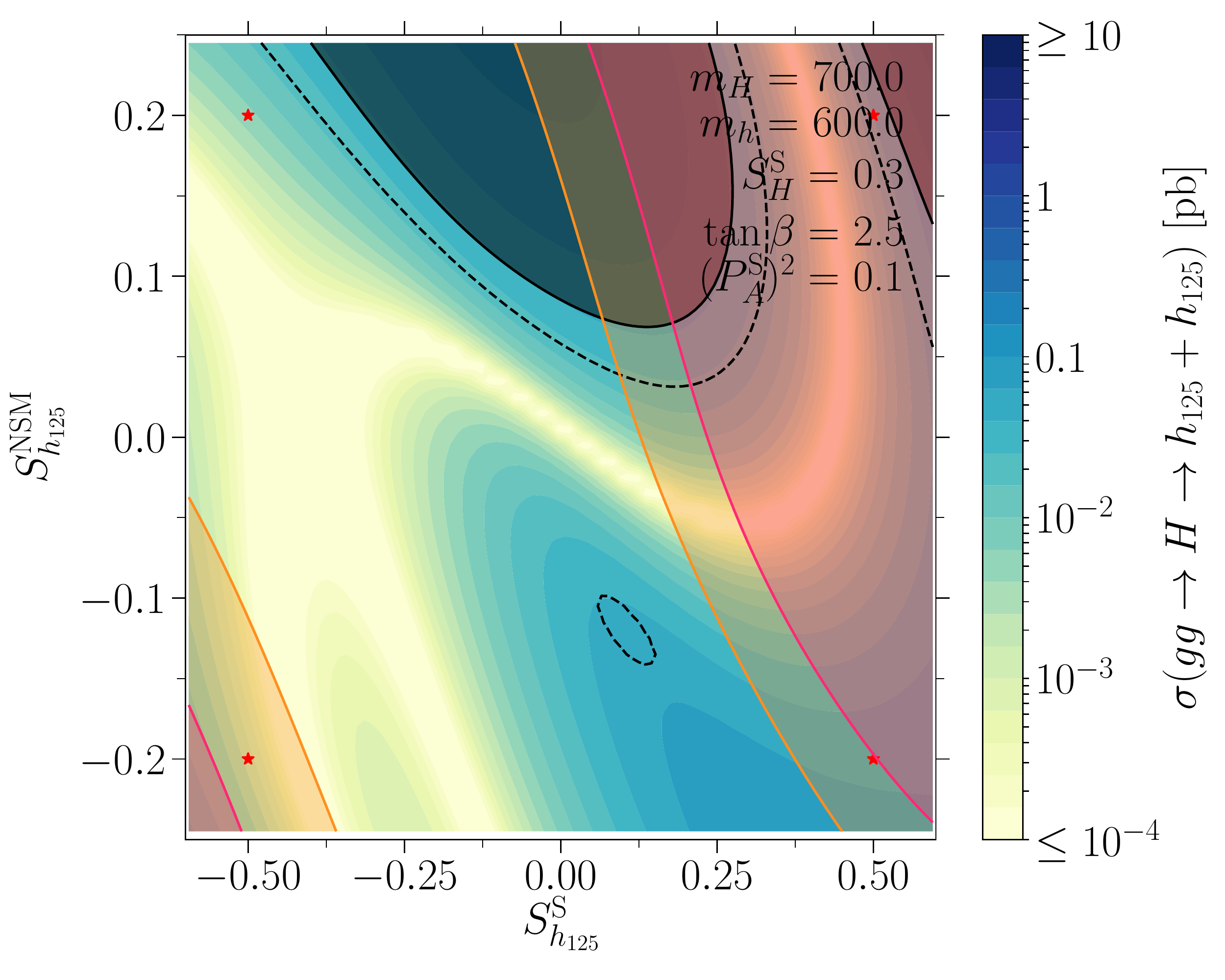}
      \hspace{.5in}
      \includegraphics[width = 2.5in]{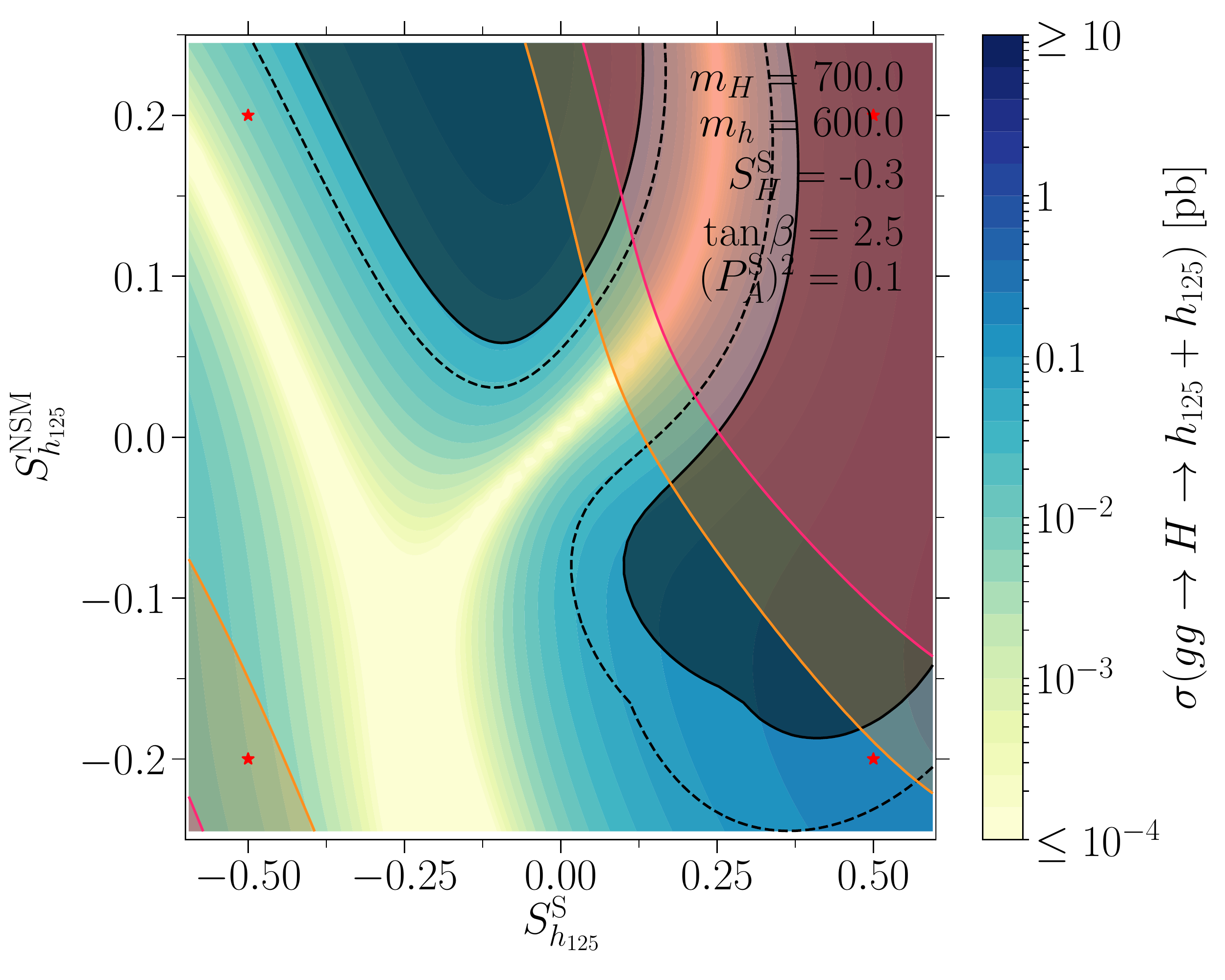}

      \includegraphics[width = 2.5in]{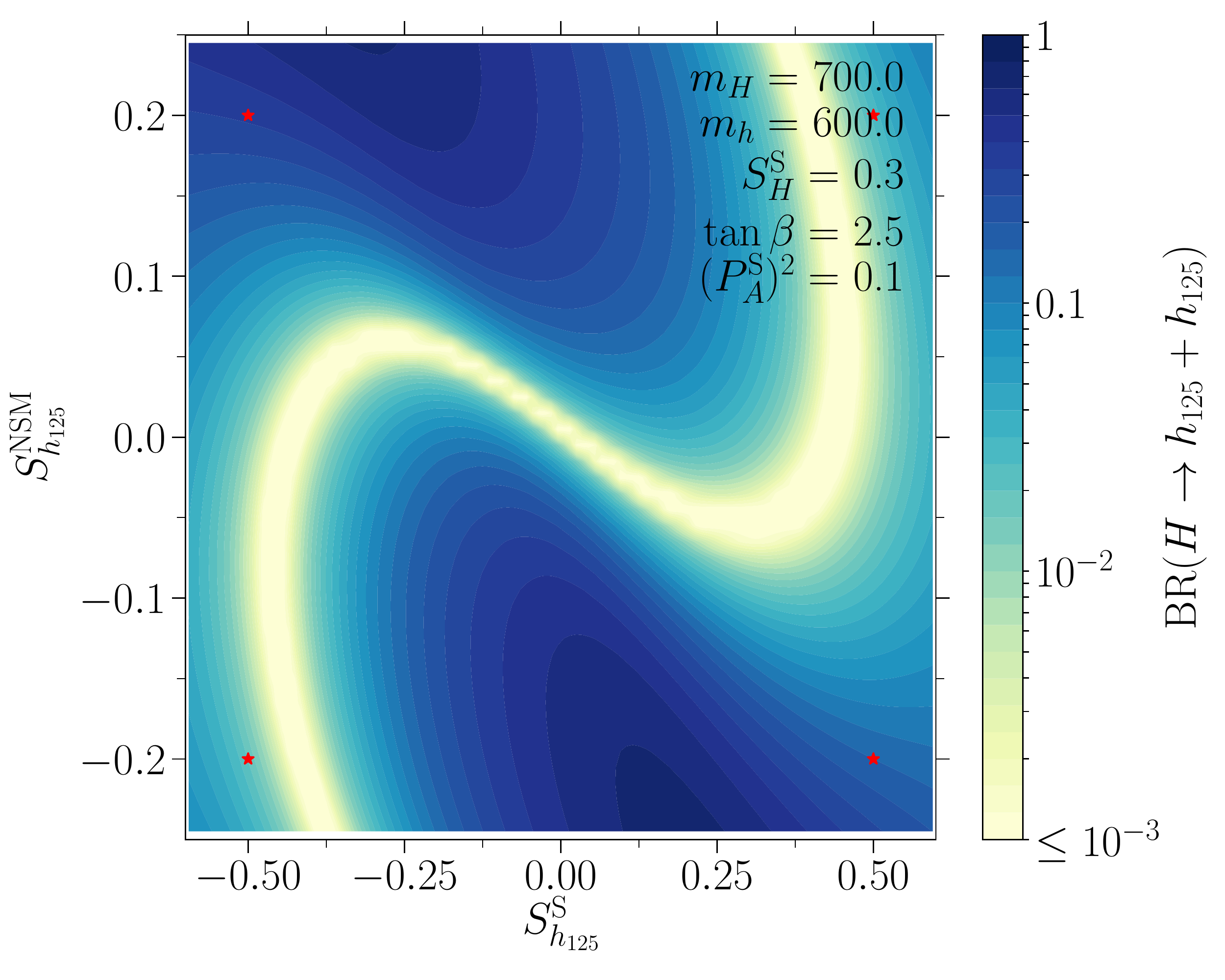}
      \hspace{.5in}
      \includegraphics[width = 2.5in]{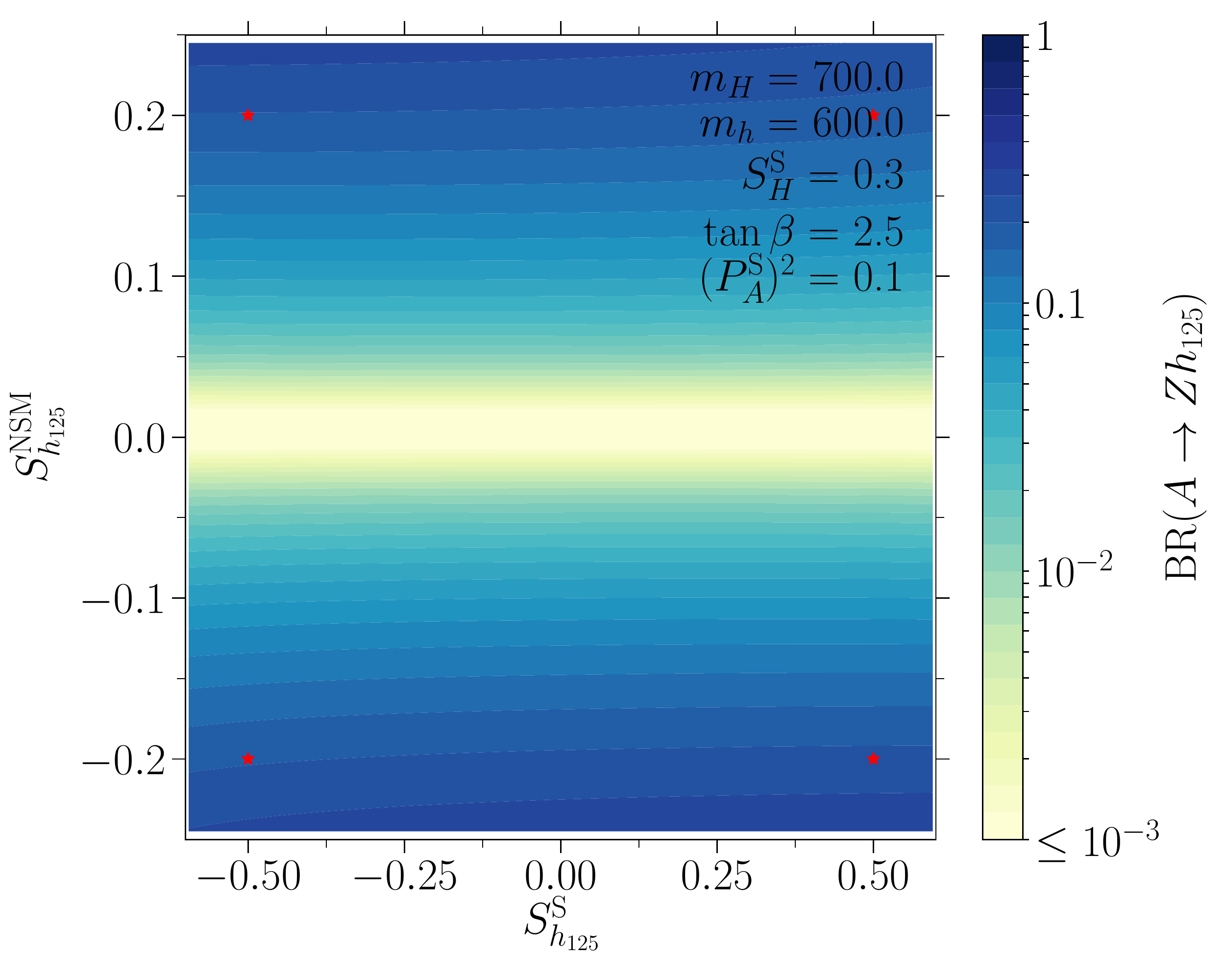}

      \includegraphics[width = 2.5in]{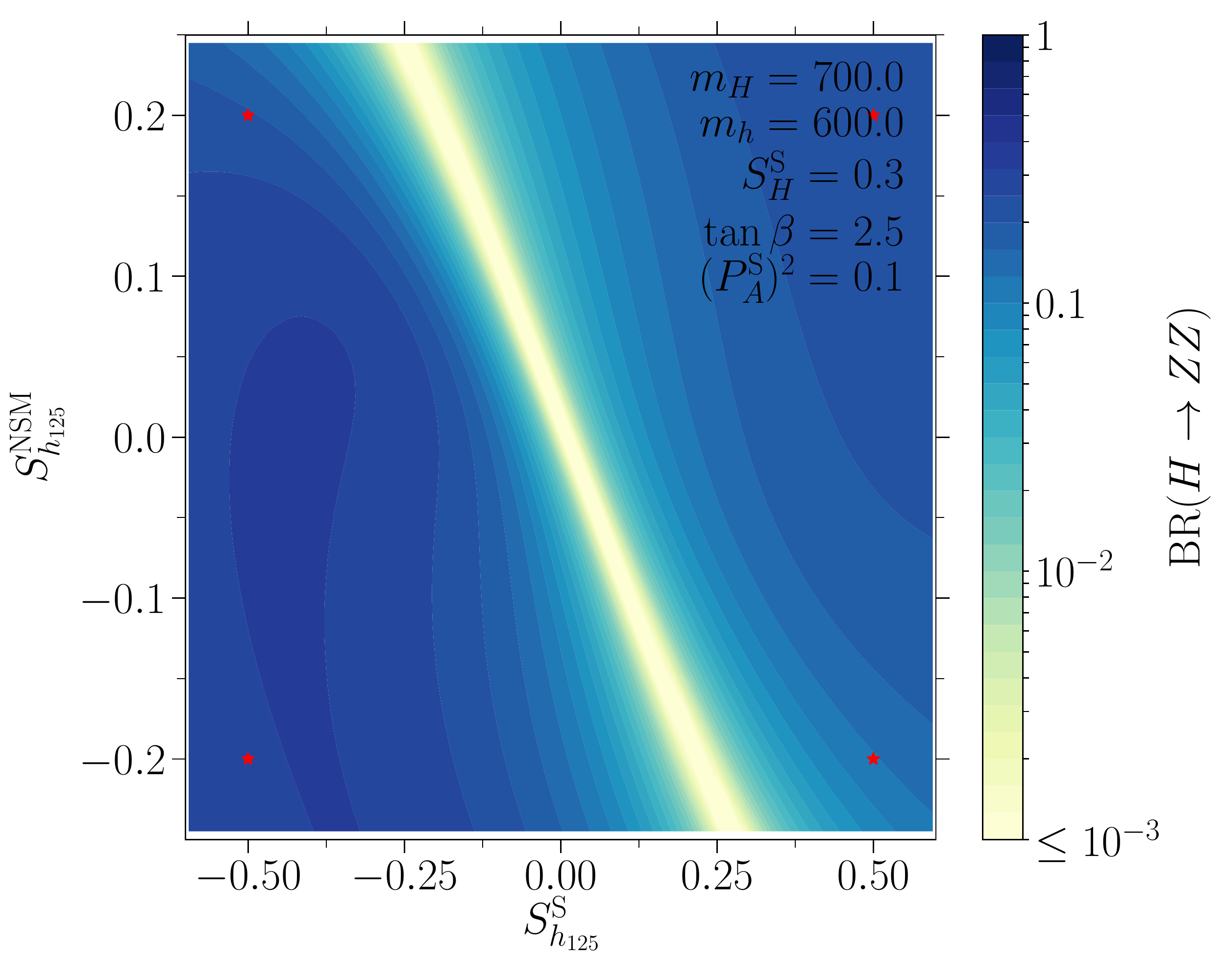}
      \hspace{.5in}
      \includegraphics[width = 2.5in]{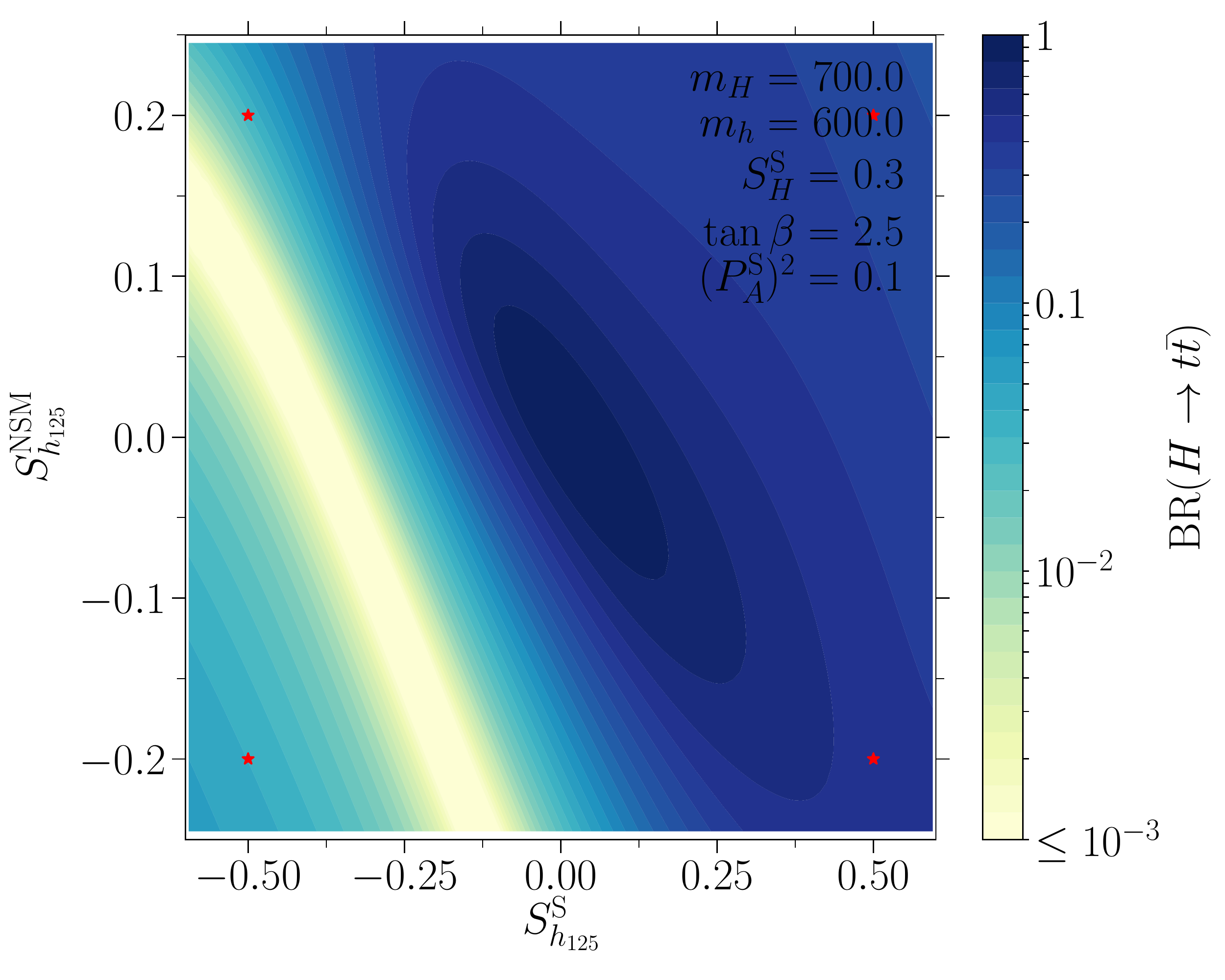}

      \includegraphics[width = 2.5in]{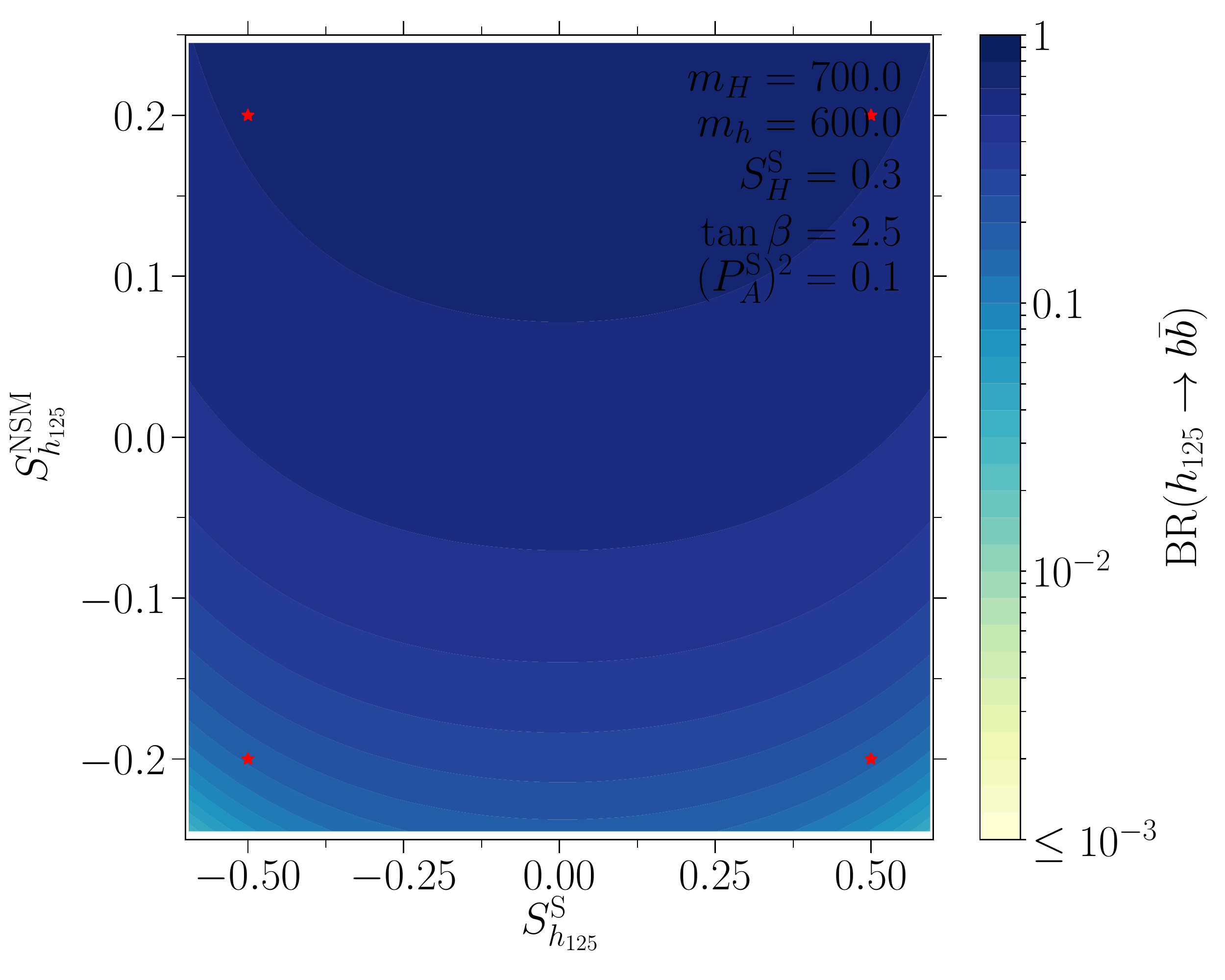}
      \hspace{.5in}
      \includegraphics[width = 2.5in]{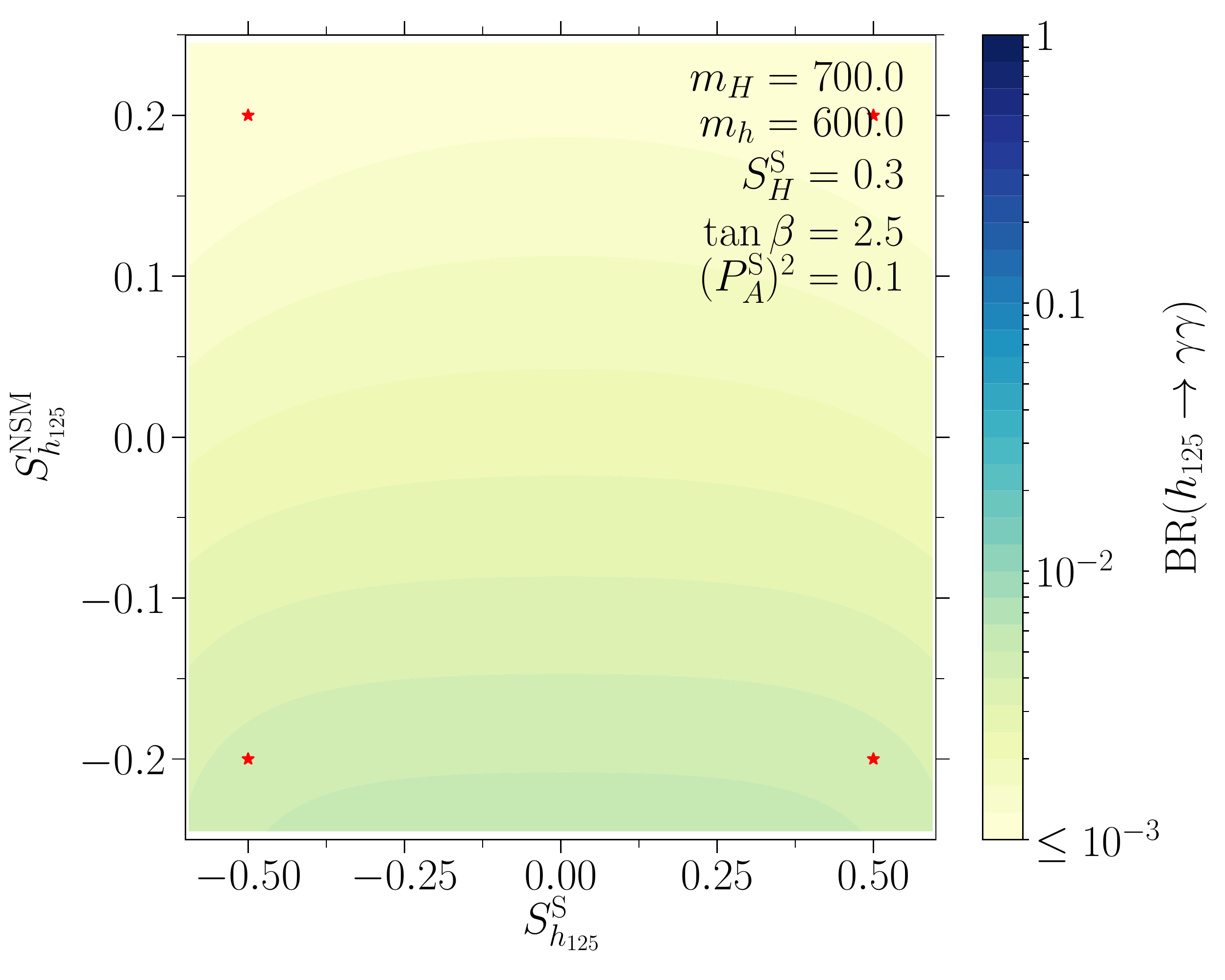}
      \caption{{\it Top row:} Production cross section and current LHC limits for ($gg \to H \to h_{125} h_{125}$; gray), ($gg \to H \to WW$; red), ($gg \to H \to ZZ$; orange) in the $S_{h_{125}}^{\rm S}$--$S_{h_{125}}^{\rm NSM}$ plane. {\it Rows 2--4:} Most relevant branching ratios of $H$ and $h_{125}$ as indicated by the colorbar in the $S_{h_{125}}^{\rm S}$--$S_{h_{125}}^{\rm NSM}$ plane for $S_H^{\rm S} = 0.3$, as in the left panel of the top row. The red stars indicate the benchmark points presented in Table~\ref{tab:BP_h125h125}. Note that while we do not show the ($h_{125} \to \tau\tau$) branching ratio, its scaling with the mixing angles is identical to that of BR($h_{125} \to b\bar{b}$; bottom left panel), since both of them are down-type fermions.}
      \label{fig:h125h125_mixAngles}
   \end{centering}
\end{figure}

In this section, we present benchmark scenarios for resonant pair production of SM-like Higgs bosons in the 2HDM+S. Note that here we compare the production cross section to {\it current} exclusion limits from the 13\,TeV LHC in the $4b$ final state taken from Refs.~\cite{ATLAS-CONF-2016-017,CMS-PAS-HIG-16-002,ATLAS-CONF-2016-049,CMS-PAS-HIG-17-009} and in the $2b2\gamma$ final state taken from Refs.~\cite{CMS-PAS-HIG-16-032,ATLAS-CONF-2016-004,CMS-PAS-HIG-17-008}. Projections for 300\,fb$^{-1}$ may be obtained by rescaling the limit with the increased luminosity. In the CP-conserving 2HDM+S, the only resonant production mechanisms at the LHC are ($pp \to H \to h_{125} h_{125}$) and ($pp \to h \to h_{125} h_{125}$). We focus on the former process, benchmark scenarios for the latter channel can be obtained by interchanging the values of the relevant mixing angles. The SM production of $h_{125}$ pairs (including interference effects) is not expected to be relevant in the region of parameter space considered here.

As discussed in Sec.~\ref{sec:general}, the branching ratios of $H$ and $h$ into pairs of SM-like Higgs states are suppressed by the proximity to alignment. For our benchmark scenario, we consider the most optimistic case of approximately maximal misalignment allowed by current LHC constraints on the couplings of the observed SM-like state~\cite{Khachatryan:2016vau,Sirunyan:2017exp,Sirunyan:2018egh,Sirunyan:2018koj,ATLAS-CONF-2017-047}. Approximately, these measurements allow a $H^{\rm NSM}$ component of $(S_{h_{125}}^{\rm NSM})^2 \lesssim 0.05$ and a $H^{\rm S}$ component of $(S_{h_{125}}^{\rm S})^2 \lesssim 0.3$. In Table~\ref{tab:BP_h125h125} we show the parameters for our benchmark points. In the interest of maximizing the ($H\to h_{125} h_{125})$ cross section, we have fixed the masses of $h$ and $a$ such that ($H \to hh$), ($H \to aa$), and ($H \to h_{125} h$) decays are kinematically forbidden. All free trilinear couplings between the Higgs basis interaction states (i.e. those not related to mixing angles and masses) are set to 0. The dominant effect of the mixing angle $S_H^{\rm S}$ is an overall suppression of the gluon fusion production cross section of $H$. One expects that the mixing angle $S_{h_{125}}^{\rm S}$ would have minimal effect since its main impact is to suppress the total $h_{125}$ width, however, since the coupling $g_{h_{125}h_{125}H}$ is a sum of small factors, signs can play a very relevant role. Hence, we present benchmark scenarios for the different possible combinations of the signs of the relevant mixing angles. The mass of the heavy Higgs boson is chosen to be $m_H =700\,$GeV for reasons similar to the ones for the previous benchmarks. The mass of $h$ for the benchmark points is chosen to be 600\,GeV such that decays of $H$ into the other Higgs channels discussed in the previous sections are forbidden. When varying $m_H$ in the plots, $m_h$ will be varied as ($m_h = m_H - 100\,$GeV). Note that the choice of $m_h$ has considerable impact on the relevant trilinear couplings which are computed from the masses and mixing angles. Our choice of the masses of the pseudoscalar states ensures that ($H \to Z a$) decays (and similar channels) are kinematically forbidden, but otherwise plays no role for $(H \to h_{125} h_{125}$). Thus, we fix the masses to $m_A = 1\,$TeV and $m_a = 950\,$GeV.

Fig.~\ref{fig:h125h125} demonstrates how the resonant $h_{125}$ pair production cross section changes as a function of the mass of the non SM-like Higgs boson $m_H$ and the relevant mixing angles. Gray regions are already excluded by current LHC data from $(gg \to H \to h_{125} h_{125})$ searches, taken from Refs.~\cite{ATLAS-CONF-2016-017,CMS-PAS-HIG-16-002,ATLAS-CONF-2016-049,CMS-PAS-HIG-17-009,CMS-PAS-HIG-16-032,ATLAS-CONF-2016-004,CMS-PAS-HIG-17-008}. The dashed lines indicate where the cross section is a factor of two smaller than current limits, which may be expected to be probed with 300\,fb$^{-1}$ of data. We also indicate regions of parameter space ruled out by current searches in the $gg \to H \to ZZ$ channel~\cite{ATLAS-CONF-2015-071,ATLAS-CONF-2016-012,ATLAS-CONF-2016-016,CMS-PAS-HIG-16-001,CMS-PAS-HIG-16-033,ATLAS-CONF-2016-056,ATLAS-CONF-2016-079,Aaboud:2017rel,Sirunyan:2018qlb} by the orange shaded regions, and regions ruled out by $(gg \to H \to WW)$ searches~\cite{ATLAS-CONF-2016-021,CMS-PAS-HIG-16-023,ATLAS-CONF-2016-074,ATLAS-CONF-2016-062,Aaboud:2017gsl,Aaboud:2017fgj,CMS-PAS-HIG-17-033} by the red shaded regions. The red stars indicate the benchmark points in Table~\ref{tab:BP_h125h125}. Depending on the relative signs of the mixing angles, the benchmark points may either be already marginally excluded~(top and bottom panels on the right), or, may be challenging to probe even with high luminosity. It is interesting to note that for some regions of parameter space, bounds from resonant $h_{125}$ pair production give stronger constraints on the mixing angles of $h_{125}$ then the precision measurements of the production cross sections and branching ratios of $h_{125}$. We also stress that decays of $(H\to ZZ/W^+W^-)$ are only proportional to the SM component of $H$ and hence directly to $S_{h_{125}}^{\rm NSM}$. Therefore more robust constraints on the misalignment of $h_{125}$ could be obtained by combining such searches. We show the most relevant branching ratios of $H$ in the $m_H$--$S_{h_{125}}^{\rm NSM}$ plane in Fig.~\ref{fig:h125h125_BRs1}, and in the $m_H$--$S_H^{\rm S}$ plane in Fig.~\ref{fig:h125h125_BRs2}.

In Fig.~\ref{fig:h125h125_mixAngles} we show the production cross section together with the constraints from current $(gg \to H \to h_{125} h_{125})$, $(gg \to H \to ZZ)$, and $(gg \to H \to WW)$ searches together with the most relevant branching ratios of $H$ and $h_{125}$ in the plane of the $H^{\rm S}$ and $H^{\rm NSM}$ component of the SM-like mass eigenstate $h_{125}$.

\FloatBarrier

\section{NMSSM Benchmarks} \label{sec:NMSSM}
As mentioned earlier, the 2HDM+S is a generalized version of the Next-to-Supersymmetric Standard Model's (NMSSM's) Higgs sector. In Ref.~\cite{Baum:2018zhf} we presented the mapping of both the general and the $Z_3$ NMSSM to the 2HDM+S. The NMSSM benchmarks presented below are a subset of $Z_3$ NMSSM benchmarks recently published in Ref.~\cite{Baum:2019uzg}. We reproduce the relevant information here.

In Table~\ref{tab:BP_params_masses} we present the NMSSM parameters and mass spectra, in Table~\ref{tab:BP_signal_strength} the signal strengths, and in Table~\ref{tab:BP_xsec_BR} the most relevant production cross sections and branching ratios for two Benchmark Points, BP1 and BP2. A description of the most important features of the benchmark points can be found below. The benchmark points were chosen as examples of points which are simultaneously within the projected reach of resonant double Higgs search channels with 3000\,fb$^{-1}$ and difficult to detect with conventional search strategies.
\begin{itemize}
   \item BP1: Mono-Higgs
   \item BP2: Higgs+visible
\end{itemize}

The benchmark points presented here feature Higgs mass eigenstates approximately aligned with the Higgs basis interaction eigenstates. In particular, they show very small doublet-doublet mixing $|S_{h_{125}}^{\rm NSM}| < 0.01$ as required by the observed phenomenology of the 125\,GeV SM-like state. The doublet-singlet mixing can take somewhat larger values; among the two benchmark points we find the largest mixing angle for BP1 ($S_{h_{125}}^{\rm S} = 0.117$).

This proximity to the alignment limit is ensured by the values of $\lambda$ and $\kappa/\lambda$ close to what is dictated by the alignment conditions, cf. Refs.~\cite{Carena:2015moc,Baum:2017gbj,Baum:2019uzg}, and is found to be a generic feature of the allowed NMSSM parameter space we scanned. Note also that for both benchmark points, all non-SM states have masses larger than $m_{h_{125}}/2$. Therefore, $h_{125}$ can only decay into pairs of SM particles. Together with the approximate alignment of $h_{125}$ with $H^{\rm SM}$, this ensures compatibility of the $h_{125}$ phenomenology with LHC observations.

For both benchmark points, the lighter non SM-like CP-even state $h$ and the lighter CP-odd state $a$ are mostly singlet-like, while the heavier states $H$ and $A$ are dominantly composed of the non SM-like doublet interaction states $H^{\rm NSM}$ and $A^{\rm NSM}$, respectively. 

The mass spectra for the benchmark points are chosen such that the non SM-like doublet-like states $H$ and $A$ are heavy enough to be difficult to detect in conventional searches ($\{m_A, m_H\} \gtrsim 350\,$GeV) but light enough such that they are readily produced at the LHC. Hence both BP1 and BP2 feature masses of the doublet-like states of $m_A \sim m_H \sim 700\,$GeV. In order to allow for sufficiently large mass gaps necessary for resonant double Higgs production, the mass of the singlet-like pseudo-scalar states has been chosen considerably lighter than the mass of the doublet-like states, $m_a \sim 200\,$GeV for BP1, and $m_a \sim 160\,$GeV for BP2. Further, while BP1 features similar singlet masses $h$ and $a$, BP2 has much larger mass splittings. The corresponding singlet-like scalar masses are $m_h \sim 165\,$GeV for BP1, and $m_h \sim 560\,$GeV for BP2. Regarding the lightest neutralino, BP1 features $m_{\chi_1}\sim 100\,$GeV, whereas BP2 features a much heavier neutralino $m_{\chi_1}\sim 500\,$GeV.

\begin{table}
	\begin{center}
	\begin{tabular}{c|cc}
		\hline\hline 
		& BP1 & BP2 \\
		& Mono-Higgs & Higgs+visible \\
		\hline
		$\lambda$ & $0.602$ & $0.602$ \\
		$\kappa$ & $-0.281$ & $0.347$ \\
		$\tan\beta$ & $2.73$ & $1.40$ \\
		$\mu$ [GeV] & $-193$ & $-466$ \\
		$A_\lambda$ [GeV] & $-784$ & $-270$ \\
		$A_\kappa$ [GeV] & $-200$ & $26.3$ \\
		$M_A$ [GeV] & $639$ & $732$ \\
		$M_{Q_3}$ [TeV] & $7.66$ & $7.78$ \\
		\hline
		$m_{h_{125}}$ [GeV] & $127$ & $128$ \\
		$m_h$ [GeV] & $165$ & $561$ \\
		$m_H$ [GeV] & $648$ & $750$ \\
		$m_a$ [GeV] & $205$ & $168$ \\
		$m_A$ [GeV] & $662$ & $749$ \\
		\hline
		$(S_{h_{125}}^{\rm S})^2$ & $1.34 \times 10^{-2}$ & $3.96 \times 10^{-3}$ \\
		$(S_h^{\rm S})^2$ & $0.972$ & $0.986$ \\
		$(S_H^{\rm S})^2$ & $1.41 \times 10^{-2}$ & $9.78 \times 10^{-3}$ \\
		$(P_A^{\rm S})^2$ & $5.92 \times 10^{-2}$ & $3.93 \times 10^{-3}$ \\
		\hline
		$m_{\chi_1}$ [GeV] & $102$ & $486$ \\
		$m_{\chi_2}$ [GeV] & $212$ & $494$ \\
		$m_{\chi_3}$ [GeV] & $292$ & $572$ \\
		\hline\hline
		\end{tabular}
	\caption{NMSSM parameters and mass spectra for our benchmark points. BP1: Mono-Higgs, BP2: Higgs+visible. The first block from the top shows the parameters used as input parameters in \texttt{NMSSMTools} $\{\lambda, \kappa, \tan\beta, \mu, A_\lambda, A_\kappa, M_{Q_3}\}$ where the first 6 parameters are those appearing in the scalar potential, and $M_{Q_3} = M_{U_3}$ is the stop mass parameter which controls the radiative corrections to the scalar mass matrices. For the convenience of the reader we also record the value of $M_A$. The remaining parameters are fixed to $M_1 = M_2 = 1\,$TeV, $M_3 = 2\,$TeV, $A_t = \mu \cot\beta$, $A_b = \mu \tan\beta$, and all sfermion mass parameters (except $M_{Q_3} = M_{U_3}$) are fixed to 3\,TeV. The second block shows the mass spectrum of the Higgs sector, and the third block values of the singlet components of the non SM-like Higgs bosons. In particular, these blocks contain the masses of the CP-odd states $a$ and $A$ and the mixing angle in the CP-odd sector $P_A^{\rm S}$. In the fourth block we record the masses of the three lightest neutralinos. Since we set the bino and wino mass parameters to $M_1 = M_2 = 1\,$TeV, the two heaviest neutralinos $\chi_4$ and $\chi_5$ are bino- and wino-like with masses $m_{\chi_4} \approx m_{\chi_5} \approx 1\,$TeV, while the three lightest neutralinos, $\chi_1$, $\chi_2$, and $\chi_3$, are Higgsino- and singlino-like.}
	\label{tab:BP_params_masses}
	\end{center}
\end{table}

\begin{table}
	\begin{center}
	\begin{tabular}{c|cc}
		\hline\hline 
		& BP1 & BP2 \\
		& Mono-Higgs & Higgs+visible V\\
		\hline
		$\max\left[\mu_{\rm Proj.}^{300\,{\rm fb}^{-1}}({\rm Resonant~Higgs\;production})\right]$ & $1.04$ & $0.442$ \\
		$\max\left[\mu_{\rm Curr.\,Lim.}^{<37\,{\rm fb}^{-1}}({\rm conventional})\right]$ & $8.76 \times 10^{-3}$ & $8.03\times 10^{-3}$ \\
		\hline
		Mono-Higgs Channels &&\\
		$\mu_{\rm Proj.}^{300\,{\rm fb}^{-1}}(gg \to H \to h_{125} h \to \gamma\gamma \chi_1 \chi_1)$ & -- & -- \\
		$\mu_{\rm Proj.}^{300\,{\rm fb}^{-1}}(gg \to A \to h_{125} a \to \gamma\gamma \chi_1 \chi_1)$ & $1.04$ & --\\
		\hline
		Higgs+visible Channels &&\\
		$\mu_{\rm Proj.}^{300\,{\rm fb}^{-1}}(gg \to H \to h_{125} h \to b\bar{b}b\bar{b})$ & $9.93 \times 10^{-2}$ & $5.85 \times 10^{-6}$ \\
		$\mu_{\rm Proj.}^{300\,{\rm fb}^{-1}}(gg \to A \to h_{125} a \to b\bar{b}b\bar{b})$ & $2.83 \times 10^{-2}$ & $0.442$ \\
		\hline\hline
		\end{tabular}
	\caption{LHC signal strengths for the benchmark points BP1 and BP2 defined in Table~\ref{tab:BP_params_masses}. In the first two rows we record the signal strength projected at the LHC for $\mathcal{L} = 300\,{\rm fb}^{-1}$ of data in the dominant resonant double Higgs production channel, $\max\left[\mu_{\rm Proj.}^{300\,{\rm fb}^{-1}}({\rm Resonant~Higgs\;production})\right]$, and the largest signal strength in the conventional channels. $\max\left[\mu_{\rm Curr.\,Lim.}^{<37\,{\rm fb}^{-1}}({\rm conventional})\right]$. See Ref.~\cite{Baum:2019uzg} for details.	In the remaining rows, we record the projected signal strength at the LHC for $\mathcal{L} = 300\,{\rm fb}^{-1}$ of data in the respective final states arising through resonant double Higgs productions.}
	\label{tab:BP_signal_strength}
	\end{center}
\end{table}

\begin{table}
	\begin{center}
	{\fontsize{11pt}{12pt}\selectfont
	\begin{tabular}{c|cc}
		\hline\hline 
		& BP1 & BP2 \\
		& Mono-Higgs & Higgs+visible \\
		\hline
		$\sigma(gg \to h)$ [pb] & $7.10 \times 10^{-2}$ & $1.7 \times 10^{-4}$ \\
		${\rm BR}(h \to \tau^+ \tau^-)$ & $1.74 \times 10^{-2}$ & $3.06 \times 10^{-5}$ \\
		${\rm BR}(h \to b\bar{b})$ & $0.151$ & $2.15 \times 10^{-4}$ \\
		${\rm BR}(h \to t\bar{t})$ & -- & $9.34 \times 10^{-4}$ \\
		${\rm BR}(h \to \gamma\gamma)$ & $4.32 \times 10^{-5}$ & $1.31 \times 10^{-6}$ \\
		${\rm BR}(h \to ZZ)$ & $1.77 \times 10^{-2}$ & $6.16 \times 10^{-2}$ \\
		${\rm BR}(h \to W^+ W^-)$ & $0.812$ & $0.128$ \\
		${\rm BR}(h \to \chi_1 \chi_1)$ & -- & -- \\
		\hline
		$\sigma(gg \to H)$ [pb] & $0.134$ & $0.239$ \\
		${\rm BR}(H \to \tau^+ \tau^-)$ & $8.66 \times 10^{-4}$ & $1.82 \times 10^{-4}$ \\
		${\rm BR}(H \to b\bar{b})$ & $6.02 \times 10^{-3}$ & $1.41 \times 10^{-3}$ \\ 
		${\rm BR}(H \to t\bar{t})$ & $0.281$ & $0.961$ \\
		${\rm BR}(H \to \gamma\gamma)$ & $2.29 \times 10^{-6}$ \\
		${\rm BR}(H \to ZZ)$ & $7.31 \times 10^{-5}$ & $6.51 \times 10^{-4}$ \\
		${\rm BR}(H \to W^+ W^-)$ & $1.50 \times 10^{-4}$ & $1.33 \times 10^{-3}$ \\
		${\rm BR}(H \to \chi_1 \chi_1)$ & $6.66 \times 10^{-2}$ & -- \\
		${\rm BR}(H \to \chi_1 \chi_2)$ & $0.107$ & -- \\
		${\rm BR}(H \to \chi_2 \chi_3)$ & $0.110$ & -- \\
		${\rm BR}(H \to hh)$ & $2.46 \times 10^{-3}$ & -- \\
		${\rm BR}(H \to hh_{125})$ & $0.102$ & $6.08 \times 10^{-3}$ \\
		${\rm BR}(H \to h_{125}h_{125})$ & $1.73 \times 10^{-3}$ & $8.56 \times 10^{-4}$ \\
		${\rm BR}(H \to aa)$ & $2.47 \times 10^{-3}$ & $1.01 \times 10^{-4}$ \\
		${\rm BR}(H \to Za)$ & $0.308$ & $2.69 \times 10^{-2}$ \\
		\hline
		$\sigma(gg \to a)$ [pb] & $0.195$ & $8.36 \times 10^{-2}$ \\
		${\rm BR}(a \to \tau^+ \tau^-)$ & $1.58 \times 10^{-3}$ & $9.05 \times 10^{-2}$ \\
		${\rm BR}(a \to b\bar{b})$ & $1.32 \times 10^{-2}$ & $0.797$ \\
		${\rm BR}(a \to \gamma\gamma)$ & $6.20 \times 10^{-6}$ & $5.60 \times 10^{-3}$ \\
		${\rm BR}(a \to \chi_1 \chi_1)$ & $0.985$ & --\\
		\hline
		$\sigma(gg \to A)$ [pb] & $0.175$ & $0.336$ \\
		${\rm BR}(A \to \tau^+ \tau^-)$ & $8.37 \times 10^{-4}$ & $1.60 \times 10^{-4}$ \\
		${\rm BR}(A \to b\bar{b})$ & $5.84 \times 10^{-3}$ & $1.18 \times 10^{-3}$ \\
		${\rm BR}(A \to t\bar{t})$ & $0.350$ & $0.973$ \\
		${\rm BR}(A \to \gamma\gamma)$ & $4.39 \times 10^{-6}$ & $7.95 \times 10^{-6}$ \\
		${\rm BR}(A \to \chi_1 \chi_1)$ & $0.102$ & -- \\
		${\rm BR}(A \to \chi_3 \chi_3)$ & $0.112$ & -- \\
		${\rm BR}(A \to ha)$ & $3.31 \times 10^{-3}$ & $2.69 \times 10^{-4}$ \\
		${\rm BR}(A \to h_{125}a)$ & $0.304$ & $1.88 \times 10^{-2}$ \\
		${\rm BR}(A \to Zh)$ & $8.40 \times 10^{-2}$ & $4.00 \times 10^{-3}$ \\
		${\rm BR}(A \to Zh_{125})$ & $5.61 \times 10^{-4}$ & $2.86 \times 10^{-4}$ \\
		\hline\hline
	\end{tabular}
	}
	\caption{Gluon fusion production cross sections at the $\sqrt{s} = 13\,$TeV LHC, $\sigma(gg \to \Phi)$, as well as the most relevant branching ratios for the non SM-like Higgs bosons $\Phi = \{h, H, a, A\}$ for the benchmark points BP1 and BP4 defined in Table~\ref{tab:BP_params_masses}.}
	\label{tab:BP_xsec_BR}
	\end{center}
\end{table}

Regarding the branching ratios important for resonant double Higgs production, we first note that the branching ratio of heavy Higgs bosons into pairs of SM-like Higgs bosons or a SM-like Higgs and a $Z$ boson is suppressed due to the proximity to alignment as discussed earlier, see also Refs.~\cite{Carena:2015moc, Baum:2017gbj, Baum:2018zhf}. For all benchmark points, we find
\begin{align*}
	{\rm BR}(H \to h_{125} h_{125}) &\ll \{ {\rm BR}(H \to h_{125} h), {\rm BR}(A \to h_{125} a)\} , \\
	{\rm BR}(A \to Z h_{125}) &\ll \{ {\rm BR}(A \to Z h), {\rm BR}(H \to Z a)\} .
\end{align*}

Additionally we note that branching ratios of the heavy non-SM like doublets into either $h_{125}$ or $Z$ and an additional singlet like state are generally comparable. This leads to multiple channels that may be probed at the LHC for each benchmark point, as discussed in detail below.

\subsubsection*{BP1 - Mono-Higgs}

This benchmark point features a Higgs spectrum with comparable masses of the singlet-like states $a$ and $h$, $m_a = 205\,$GeV and $m_h = 165\,$GeV. The heavier states $A$ and $H$ are mostly composed of $A^{\rm NSM}$ and $H^{\rm NSM}$, respectively, and are approximately mass degenerate with $m_A \approx m_H \approx 650\,$GeV. The Higgsino mass parameter has a value of $\mu = -193\,$GeV, and $\kappa = -0.281$, leading to $2|\kappa|/\lambda = 0.93$. Thus, the lightest neutralino $\chi_1$ is mostly singlino-like but has sizable Higgsino components. Its mass is $m_{\chi_1} = 102\,$GeV, allowing for ($a \to \chi_1 \chi_1$) decays but not for ($h \to \chi_1 \chi_1$) decays. The second-lightest neutralino $\chi_2$ is dominantly Higgsino-like with a mass of $m_{\chi_2} = 212\,{\rm GeV} \approx |\mu|$, while $\chi_3$ is mostly Higgsino-like but has a sizable singlino component and a mass of $m_{\chi_3} = 292\,$GeV. 

Due to their singlet-like nature, the direct production cross sections of $a$ and $h$ are much smaller than those of a SM Higgs boson of the same mass, rendering them beyond the reach of conventional search channels at the LHC which rely on direct production of $a$ or $h$. The dominant decay modes of $h$ are into pairs of $b$-quarks, and, facilitated by its (small) doublet component, into pairs of $W$-bosons. The singlet-like pseudo-scalar on the other hand is kinematically allowed to decay into pairs of neutralinos ($a \to \chi_1 \chi_1$). Because $\chi_1$ has sizable singlino as well as Higgsino components, such decays proceed via both of the NMSSM's large couplings $\lambda$ and $\kappa$, rendering the corresponding branching ratio large, ${\rm BR}(a \to \chi_1 \chi_1) = 0.985$.

The heavier (doublet-like) CP-even state $H$ mostly decays into pairs of top quarks, neutralinos, and, most relevant for resonant double Higgs production, via ($H \to h h_{125}$) and ($H \to Z a$). The cross section ($gg \to H \to h h_{125}$) is not large enough for it to be within reach of the Higgs+visible search modes. However, facilitated by the sizable branching ratios of $(H \to Z a)$ and ($a \to \chi_1 \chi_1$), this benchmark point is within the projected reach of mono-$Z$ searches, $\mu_{\rm Proj.}^{3000\,{\rm fb}^{-1}}(gg \to H \to Z a \to \ell^+ \ell^- \chi_1 \chi_1) = 1.73$ at the LHC with 3000\,fb$^{-1}$ of data~\cite{Baum:2019uzg}.

The heavier (doublet-like) CP-odd state $A$ mostly decays into pairs of top quarks, neutralinos, and through the ($A \to h_{125} a$) channel. The sizable branching ratio of the latter decay mode, ${\rm BR}(A \to Z h_{125}) = 0.304$, together with the large branching ratio corresponding to $(a \to \chi_1 \chi_1)$ decays leads to a large projected signal strength in mono-Higgs searches via the corresponding decay chain, $\mu_{\rm Proj.}^{300\,{\rm fb}^{-1}}(gg \to A \to h_{125} a \to \gamma\gamma \chi_1 \chi_1) = 1.04$. 

Neither $H$ nor $A$ have large branching ratios into pairs of SM states except into pairs of top quarks, rendering them very difficult to detect by conventional searches.

\subsubsection*{BP2 - Higgs+visible}
Benchmark point BP2 features a Higgs spectrum with a large split between the masses of the singlet-like states $a$ and $h$, $m_a = 168\,$GeV and $m_h = 561\,$GeV. The heavier doublet-like states $A$ and $H$ are almost mass degenerate, $m_A \approx m_H \approx 750\,$GeV. The Higgsino mass parameter takes much larger absolute value than for BP1, $\mu = -466\,$GeV. Further, $\kappa$ also has a larger absolute value than for BP1, $\kappa = 0.347$, leading to $2|\kappa|/\lambda = 1.15$. Thus, the two lightest neutralinos, $\chi_1$ and $\chi_2$, are mostly Higgsino like and approximately mass degenerate, $m_{\chi_1} = 486\,$GeV and $m_{\chi_2} = 494\,$GeV. The third-lightest neutralino, $\chi_3$, is mostly composed of the singlino and has a mass of $m_{\chi_3} = 572\,$GeV. Note that because $|2|\kappa|/\lambda-1|$ is larger than for BP1, the Higgsino and singlino mass parameters are split further for BP2 than for BP1, leading to much smaller singlino-Higgsino mixing. Further, because of the relatively large masses of the neutralinos, none of the Higgs bosons are kinematically allowed to decay into pairs of neutralinos.

Similar to BP1, the large singlet components of $a$ and $h$ lead to direct production cross sections at the LHC much smaller than those of SM Higgs bosons of the same mass. Thus, they are out of reach of conventional search strategies. The dominant decay mode of the CP-even state $h$ is into pairs of $W$-bosons and pairs of the much lighter singlet-like CP-odd state, ${\rm BR}(h \to aa) = 0.740$. The CP-odd state $a$ decays mostly into pairs of $b$-quarks with a branching ratio of ${\rm BR}(a \to b\bar{b}) = 0.797$. It also has a sizable branching ratio into $\tau$-leptons, ${\rm BR}(a \to \tau^+ \tau^-) = 0.0905$.

The heavier (doublet-like) CP-even state $H$ predominantly decays into pairs of top quarks. Because of the small value of $\tan\beta$ compared to BP1, ($H \to h_{125} h$) decays, which are mostly controlled by the ($H^{\rm SM} H^{\rm NSM} H^{\rm S}$) coupling, are suppressed. The largest branching ratio of $H$ relevant for non standard Higgs decays is ${\rm BR}(H \to Z a) = 0.0269$. However, this branching ratio is not sufficiently large to put BP2 within reach of $Z$+visible searches where the projected signal strength with 3000\,fb$^{-1}$ of data is only $\mu_{\rm Proj.}^{3000\,{\rm fb}^{-1}}(gg \to H \to Z a \to \ell^+ \ell^- \chi_1 \chi_1) = 0.136$~\cite{Baum:2019uzg}.

Similar to $H$, the CP-odd doublet-like state $A$ mostly decays into pairs of top quarks. The largest branching ratio relevant for resonant double Higgs production is ${\rm BR}(A \to h_{125} a) = 0.0188$. This decay mode is mostly controlled by the $(H^{\rm SM} A^{\rm NSM} A^{\rm S})$ coupling, which becomes largest for values of $\tan\beta = 1$, but is suppressed for ${\rm sgn}(\kappa) = +1$. Nonetheless, together with the large branching ratio of $a$ into pairs of $b$-quarks, this branching ratio is sufficiently large to render BP2 within reach of Higgs+visible searches with a projected signal strength $\mu_{\rm Proj.}^{3000\,{\rm fb}^{-1}}(gg \to A \to h_{125} a \to b\bar{b}b\bar{b}) =1.22$ with 3000\,fb$^{-1}$ of data. With 300\,fb$^{-1}$ of data, the projected signal strength is $\mu_{\rm Proj.}^{300\,{\rm fb}^{-1}}(gg \to A \to h_{125} a \to b\bar{b}b\bar{b}) = 0.442$

Since both $H$ and $A$ decay dominantly into pairs of top quarks, BP2 is very challenging to discover with conventional search strategies. 

\section{Conclusions}

The 2HDM+S provides a well motivated framework for analyzing the complex phenomenology which may result from more complete models such as the NMSSM. The mapping from both the general and the $Z_3$ NMSSM to the 2HDM+S parameters is given in Ref.~\cite{Baum:2018zhf}. 

We have presented benchmark scenarios in the 2HDM+S optimized for three sets of signatures which may be probed with 300\,fb$^{-1}$ of data at the LHC:
\begin{itemize}
   \item $A\to h_{125} a$\,, $H\to h_{125} h$\,,
   \item $A\to a h$\,, $H\to hh$\,, $H\to aa$\,, and 
   \item $H\to h_{125}h_{125}$.
\end{itemize}
We have also presented benchmark points in the NMSSM previously published in Ref.~\cite{Baum:2019uzg}, which while being optimized for ($A\to h_{125} a, ~H\to h_{125} h)$ show the complementarity of the different possible channels, in particular decays involving a $Z$ boson in the final state.

\acknowledgments
We thank the conveners of the LHCHXSWG-HH working group for encouraging our work on this note. We also thank the members of the working group for useful discussions which lead to the present form of the note. 
SB and acknowledges support by the Vetenskapsr\r{a}det (Swedish Research Council) through contract No. 638-2013-8993 and the Oskar Klein Centre for Cosmoparticle Physics. NRS is supported by DoE grant DE-SC0007983 and Wayne State University. 

\bibliographystyle{JHEP.bst}
\bibliography{theBib}

\end{document}